\documentclass{ar-1col}
\usepackage{rotating}
\usepackage[ruled,lined,algonl]{algorithm2e}
\usepackage{amsmath,amssymb}
\usepackage{natbib}
\usepackage{url}
\usepackage{graphicx}
\usepackage[total={30.5pc,47pc},centering]{geometry}
\usepackage{hyperref}

\jname{Annu. Rev. Astron. Astrophys.}
\jvol{55}
\jyear{2017}
\doi{10.1146/((please add article doi))}

\def\bsym{\mathbf}
\def\beq{\begin{equation}}   \def\eeq{\end{equation}}
\def\be{\begin{eqnarray}}   \def\ee{\end{eqnarray}}
\def\ben{\begin{equation}\begin{aligned}} \def\een{\end{aligned}\end{equation}} 
\def\sec#1{Section~\ref{sec:#1}}
\def\fig#1{Figure~\ref{fig:#1}} 

\def\equ#1{Equation~(\ref{equ:#1})}

\def\alg#1{Algorithm~\ref{alg:#1}}

\definecolor{grey}{rgb}{0.35,0.35,0.35}

\newcommand\mnras{MNRAS}
\newcommand\apj{ApJ}
\newcommand\prd{PRD}

\newcommand\apjl{ApJL}
\newcommand\aap{A\&A}

\newcommand\araa{ARA\&A}

\newcommand\pasp{PASP}
\newcommand\aapr{AAPR}
\newcommand\nar{NAR}
\newcommand\aj{AJ}

\begin{document}

\markboth{Sanjib Sharma}{MCMC and Bayesian Data Analysis}
\title {Markov Chain Monte Carlo Methods for Bayesian Data Analysis in Astronomy}
\author{Sanjib Sharma$^1$
\affil{$^1$Sydney Institute for Astronomy, School of
  Physics, University of Sydney, NSW 2006, Australia, email:
  sanjib.sharma@sydney.edu.au}
{\tt Draft version. To appear in Annual Review of Astronomy and Astrophysics.}
}
\begin{abstract}
Markov Chain Monte Carlo based Bayesian data analysis has now become 
the method of choice for analyzing and interpreting 
data in almost all disciplines of science. 
In astronomy, over the last decade,   
we have also seen a steady increase in the number of papers 
that employ Monte Carlo based Bayesian analysis. 
New, efficient Monte Carlo based methods are continuously 
being developed and explored. 
In this review, we first explain the basics 
of Bayesian theory and discuss how to set up 
data analysis problems within this framework. 
Next, we provide an overview of various 
Monte Carlo based methods for performing 
Bayesian data analysis.   
Finally, we discuss advanced ideas that 
enable us to tackle complex problems and thus 
hold great promise for the future.   
We also distribute downloadable computer software 
(\url{https://github.com/sanjibs/bmcmc/}) that implements  
some of the algorithms and examples discussed here.
\end{abstract}
\maketitle

\tableofcontents

\section{Introduction}
Markov Chain Monte Carlo (MCMC) and Bayesian Statistics  
are two independent disciplines, the former being a method to sample from a
distribution while the latter is a theory to interpret
observed data. When these two disciplines are combined
together, the effect is so dramatic and powerful that it has
revolutionized data analysis in almost all disciplines of
science, and astronomy is no exception. 
This review explores the power of this combination.

What is so special about MCMC based Bayesian data analysis?
The usefulness of Bayesian methods in science and astronomy 
is easy to understand. 
In many situations,  it is easy to predict the
outcome given a cause. But in science, most often, we are faced with the
opposite question. Given the outcome of an experiment what 
are the causes, or what is the probability of a cause as 
compared to some other cause? If we have some prior
information, how does that help us? This opposite problem 
is more difficult to solve. The power of Bayesian theory lies 
in the fact that it provides 
a unified framework to quantitatively answer such 
questions. Hence it has become indispensable for science. 
As opposed to deductive logic, Bayesian theory provides a framework 
for plausible reasoning, a concept which is more powerful and 
general, an idea championed by \citet{jaynes2003probability}
in his book. 

The question  now is how does one solve a problem 
that has been set up using Bayesian theory. 
This mostly involves computing the
probability distribution function (pdf) of some parameters given the data
and is written as $p(\theta|D)$. Here, $\theta$ need not be a 
single parameter; in general, it represents a set of parameters. 
Usually here and elsewhere, such functions 
do not have analytical solutions and so we need methods to 
numerically evaluate the distribution. This is where 
MCMC methods come to the rescue. 
They provide an efficient and easy way to sample 
points from any given distribution which is analogous 
to evaluating the distribution.   

Bayesian data analysis \citep{jeffreys193961} and
Markov Chain Monte Carlo \citep{metropolis1953equation}
techniques have existed for more than 50 years.
Their tremendous increase in popularity over the last decade 
is due to an increase in computational power which has made 
it affordable to do such computations.

The simplest and the most widely used MCMC algorithm is the ``random 
walk'' Metropolis algorithm (\sec{mhalgo}). However, the efficiency of this  
algorithm depends upon the ``proposal distribution'' which the user 
has to supply. This means that there is some problem-specific 
fine tuning to be done by the user. The problem to find
a suitable proposal distribution becomes worse as the 
dimensionality of the space over which the sampling is done 
increases. Correlations and degeneracies between the 
coordinates further exacerbate the problem. 
Many algorithms have been proposed to solve this 
problem and it remains an active area of research. 
Some algorithms work better for specific problems 
and under special conditions but algorithms that work 
well in general are in high demand. 
Multimodal distributions pose additional problems for 
MCMC algorithms. In such situations, an MCMC chain can easily 
get stuck at a local density maximum. To overcome this, 
algorithms like simulated tempering and parallel tempering
have been proposed (\sec{partemp}). Hence discussion of efficient 
MCMC algorithms is one focus of this review.

Given its general applicability, the Bayesian framework can be used 
in almost any field of astronomy. Hence, it is not possible 
to discuss all its applications. 
However, there are many examples where either alternatives 
do not exist or are inferior. The aim of this review 
is to specifically discuss such cases where 
Bayesian-MCMC methods have enjoyed great success. 
The Bayesian framework by itself is very simple. The difficult 
part when attempting to solve a problem is to express the 
problem within this framework and then to choose the appropriate 
MCMC method to solve it. The best way to master this is by 
studying a diverse set of applications, and we aim to 
provide this in our review (\sec{casestudy}). 
Finally, we also discuss a few advanced topics 
like non-parametric models 
and hierarchical Bayesian models (\sec{bhm}) which are not yet 
main stream in astronomy but are very powerful and allow 
one to solve complex problems. 

To summarize, the review has three main aims. The first is to explain 
the basics of Bayesian theory 
using simple familiar problems, e.g., fitting 
a straight line to a set of data points with errors in both 
coordinates and in the presence of outliers. This is targeted 
at readers who are new to the topic. The second goal is to 
provide a concise overview of
recent developments. This will benefit people who are familiar 
with Bayesian data analysis but are interested in learning 
more. The final aim is to discuss emerging ideas that   
hold great promise in future. 
We also develop and distribute
downloadable software (available at
\url{https://github.com/sanjibs/bmcmc/} or by running the 
command \verb|pip install bmcmc|)
 implementing 
some of the algorithms and examples that we discuss.

\begin{figure}
\centering \includegraphics[width=0.95\textwidth]{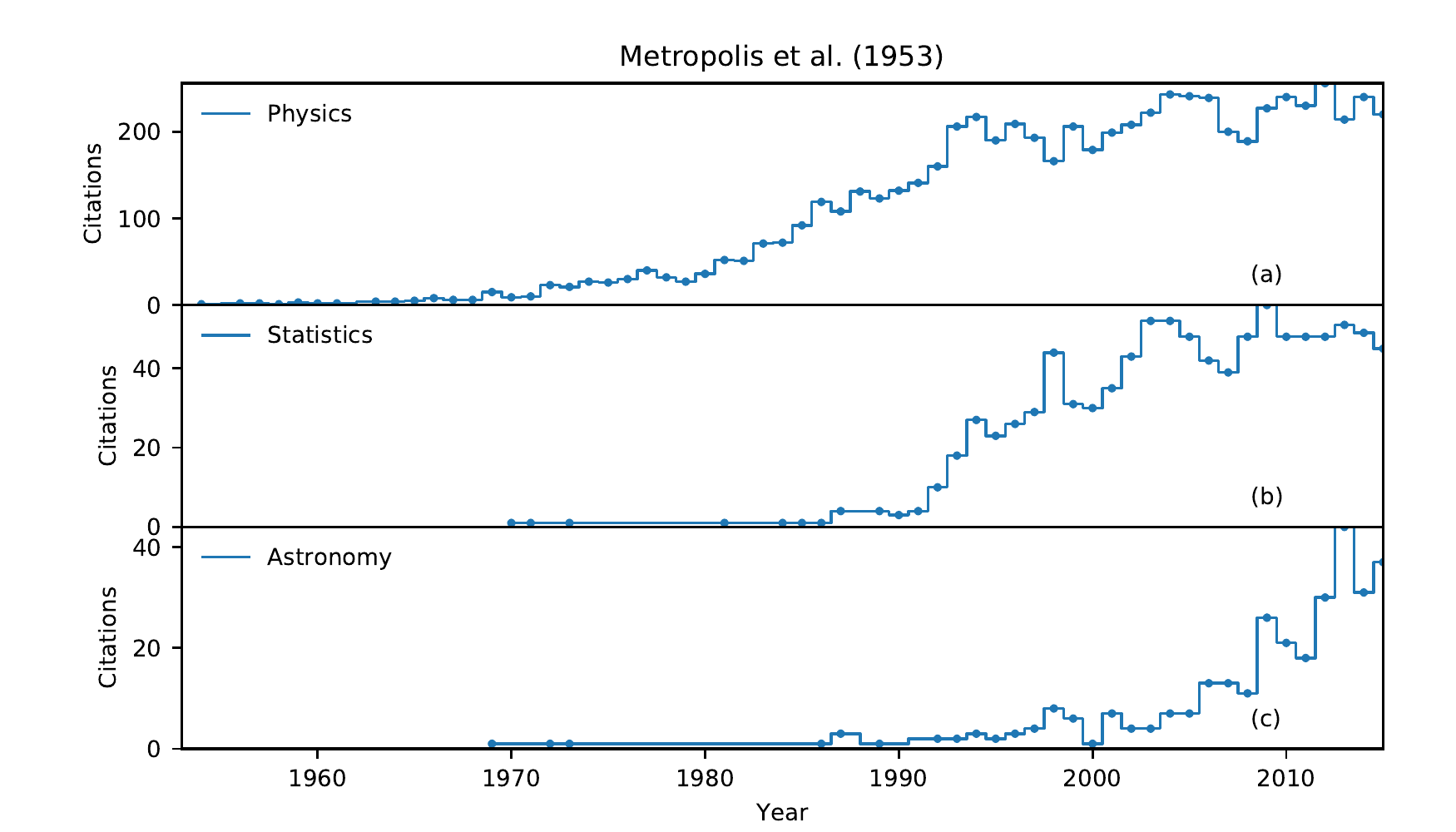}
\caption{Citation history for the
  \citet{metropolis1953equation} paper for three different
  subject areas. 
\label{fig:metropolis_citations}}
\end{figure}

\subsection{Rise of MCMC based Bayesian methods in astronomy and science} 
The emergence of Bayesian statistics has a long and 
interesting history dating back to 1763 when 
Thomas Bayes laid down the basic ideas 
of his new probability theory \citep[][published 
posthumously by Richard Price]{mr1763essay}. 
It was rediscovered 
independently by Laplace \citep{de1774memoire} 
and used in a wide variety of contexts, e.g., celestial 
mechanics, population statistics, reliability, and
jurisprudence. However, after that it was largely ignored. 
A few scientists like, 
Bruno de Finetti and Harold Jeffreys kept the 
Bayesian theory alive in the first half of the 20th
century. Harold Jeffreys published 
the book Theory of Probability \citep{jeffreys193961}, 
which for a long time remained the 
main reference for using the Bayes theorem. 
The Bayes theorem was used in the Second World War at
Bletchley Park, United Kingdom, for cracking the German Enigma code,
but its use remained classified for many years afterwards.
From 1950 onwards, the tide turned towards 
Bayesian methods. However, the lack of proper tools to do 
Bayesian inference remained a challenge. The frequentist 
methods in comparison were simpler to implement which made them more 
popular. Recent statement by the American Statistical
  Association, \citep{wasserstein2016asa} warning on the 
misuse of P values is another example of the superiority of 
the Bayesian methods of hypothesis testing.  

Interestingly, efficient methods like MCMC 
to sample distributions had been invented by 1954
in the context of solving problems in statistical
mechanics  \citep{metropolis1953equation}. (The  brand name Monte Carlo was coined by
\citet{metropolis1949monte} where they discussed 
a stochastic method making use of random numbers to 
solve a class of problems in mathematical physics 
which are difficult to solve due to the large number 
of dimensions.)  
Such problems typically involve $N$ interacting particles.  
A single configuration $\omega$ of such a system is fully specified by 
giving the position and velocity of all the particles; 
i.e., $\omega$ can be defined by a point in
$\mathcal{R}^{2N}$ space, also known as the
configuration space $\Omega$. The total energy is
a function of the configuration $E(\omega)$. For a system in
equilibrium, the probability of a configuration is given by 
$\exp(-E(\omega)/kT)$, where $k$ is the Boltzmann constant and 
$T$ is the temperature of the system. 
Computing any thermodynamic property of the system, e.g., 
pressure or energy typically involves computing integrals 
of the form 
\be
\bar{F}=\frac{\int F(\omega) \exp(-E(\omega)/kT) {\rm d} \omega}{Z} 
\ee
for which $Z=\int \exp(-E(\omega)/kT) d \omega$ is known 
as the partition function. The integrals over $\omega$ are 
in most cases analytically and computationally intractable.  
The idea of Metropolis and colleagues was to start with an arbitrary 
configuration of $N$ particles and then move each particle  
by a random walk. If $\Delta E<0$, the move is always accepted,
otherwise, it is accepted stochastically with 
probability $\exp(-\Delta E/kT)$, which is the 
ratio of the probability of the new configuration 
with respect to the old.
The method ends 
up choosing a configuration $\omega$ sampled 
from $\exp(-E(\omega)/kT)$. 
The method immediately became popular in the 
statistical physics community. 

However, the fact 
that the same method can be used for sampling 
an arbitrary pdf $p(\omega)$ 
by simply replacing $E(\omega)/kT $ 
with $\ln(p(\omega))$ had to wait till the 
important paper by \citet{hastings1970monte}. 
He generalized the work of Metropolis and colleagues and derived the
essential condition for the acceptance ratio 
that a Markov chain ought to satisfy in order 
to sample the target distribution. 
The generalized algorithm is   
now known as the Metropolis-Hastings (MH) algorithm. 
Later Hastings's student 
Peskun showed  that, 
among the available choices, the one by Metropolis 
and colleagues was the most efficient
\citep{peskun1973optimum}. 
Despite its introduction 
to the statistical community, the ideas remained 
dormant till 1980. 

Around 1980 things suddenly changed and a few influential algorithms 
appeared. 
Simulated annealing was presented 
by \citet{kirkpatrick1983optimization} 
to solve combinatorial optimization problems 
using the MH algorithm in conjunction with 
ideas of annealing from solid state physics. 
It is especially useful for situations where we have 
multiple maxima and applies to any setting 
where we have to minimize an objective function $C(\omega)$.  
This is done by sampling $\exp(-C(\omega)/T)$, with
progressively decreasing $T$ to allow annealing and 
selection of a globally optimum solution.  
A year later 
\citet{geman1984stochastic} introduced what we currently 
know as ``Gibbs sampling'' in the context of image
restoration. 
This was the first proper use of MCMC techniques to solve a problem 
set up in a Bayesian framework, in the sense that simulating from conditional
distributions is the same as simulating from the joint
distribution. However, there exists earlier 
work related to Gibbs sampling; 
the Hammersley-Clifford theorem which was developed in 
the early 1970s and the work by \citet{besag1974spatial}.

At about this time, one of the most influential
methods of the 20th-century emerged $-$
the expectation maximization (EM) 
algorithm by \citet{dempster1977maximum}. 
This provided a way to deal with missing data 
and hidden variables and vastly 
increased the range of problems that can be 
addressed by Bayesian methods. 
The EM algorithm is deterministic and 
has some sensitivity to the starting configuration. 
To address this, stochastic versions were 
developed \citep{celeux1985sem} quickly followed 
by the data augmentation (DA) 
algorithm \citep{tanner1987calculation}. 

\begin{figure}
\centering 
\begin{minipage}{6cm}
\centering 
\includegraphics[width=6.0cm]{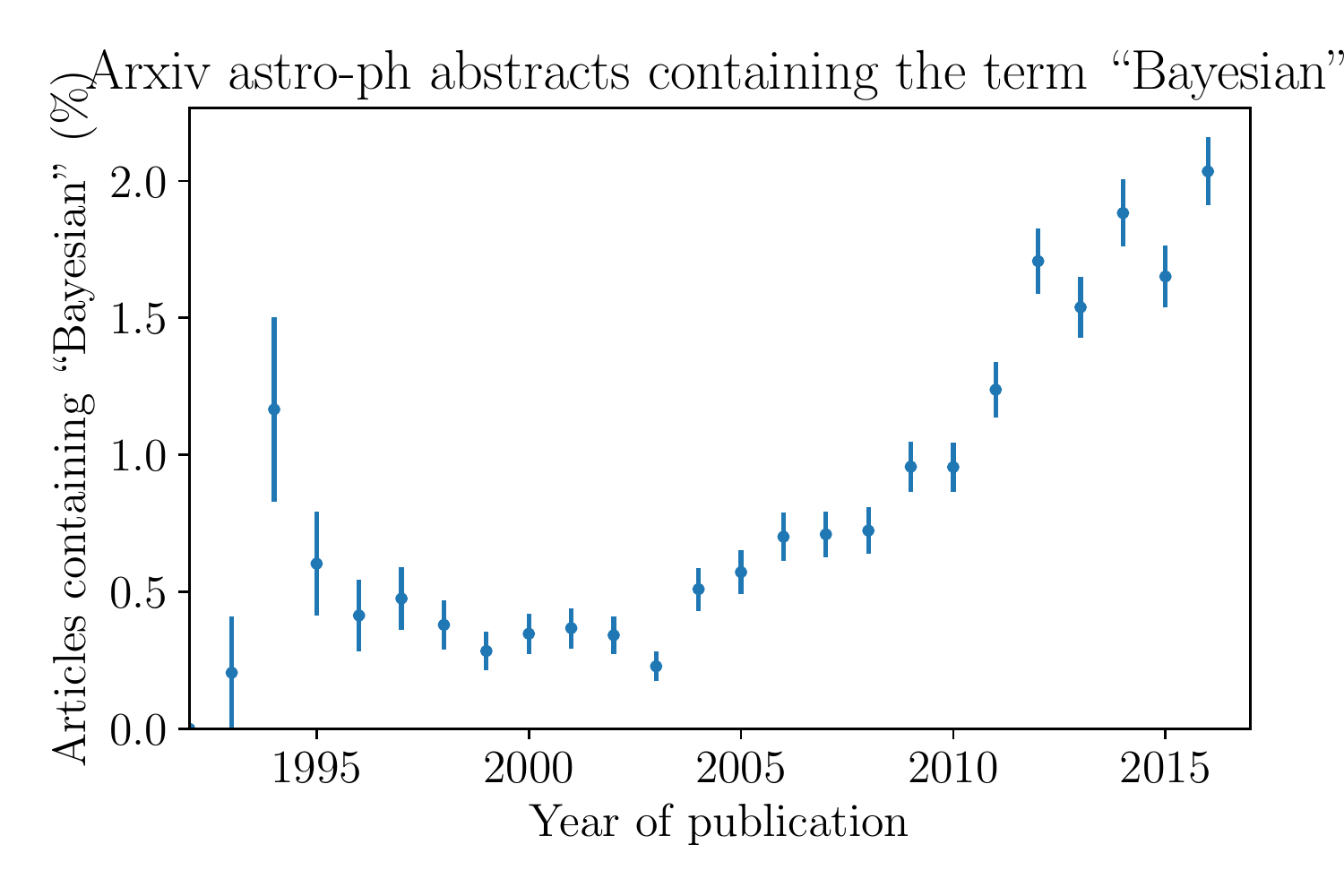}
\end{minipage}
\centering 
\begin{minipage}{6cm}
\centering 
\includegraphics[width=6.0cm]{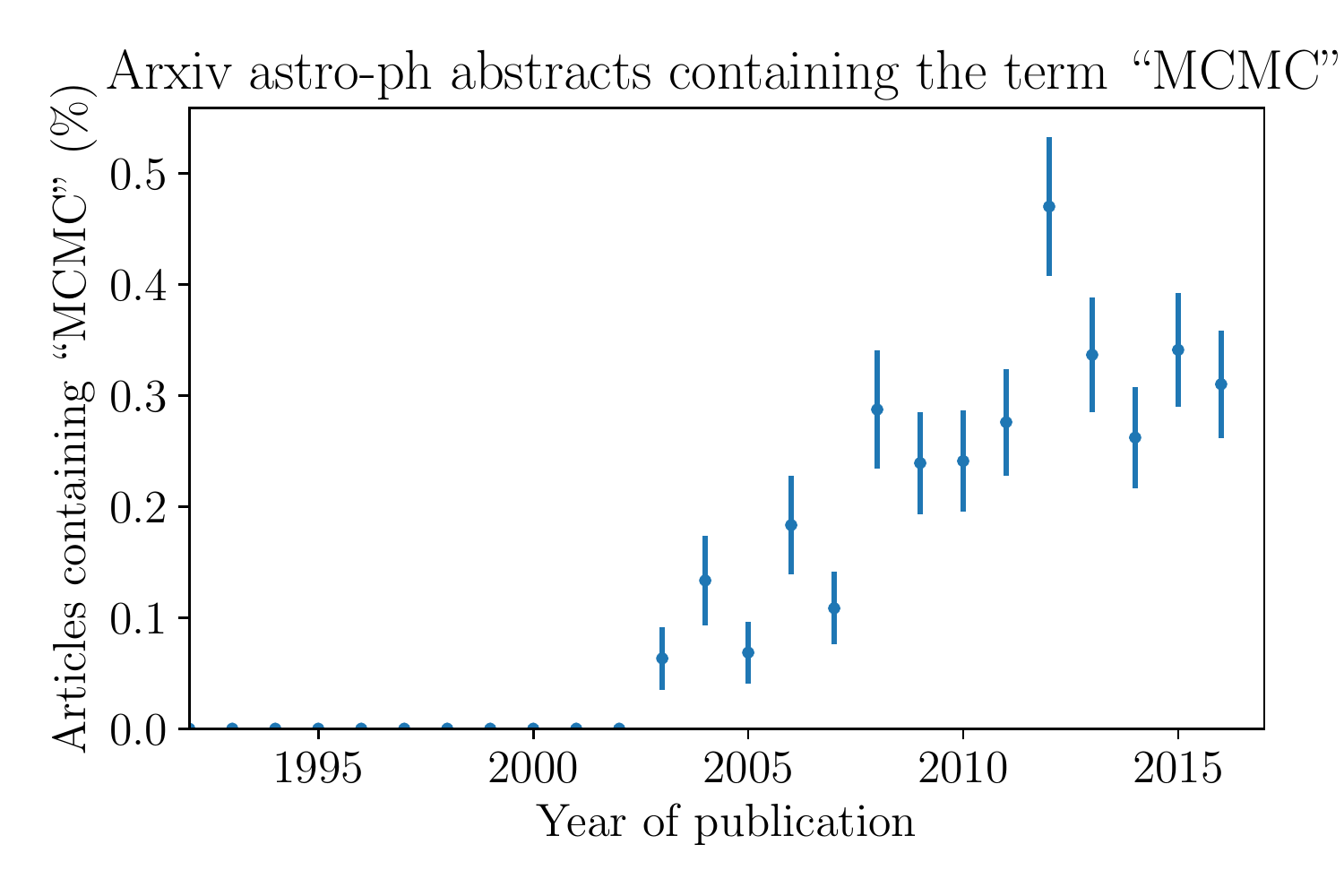}
\end{minipage}
\caption{Percentage of articles in Arxiv astro-ph abstracts
  containing the word Bayesian (left) and MCMC
  (right). Computed using the code {\it arxiv.py}, courtesy Dustin Lang.}
\label{fig:arxiv_stats}
\end{figure}

The watershed moment in the field of statistics is largely 
credited to the paper by \citet{gelfand1990sampling} that  
unified the ideas of Gibbs sampling, DA and 
the EM algorithm \citep{tanner2010data,robert2011short}. It firmly 
established that Gibbs sampling and Metropolis-Hastings based
MCMC algorithms can be used to solve a wide class of
problems that fall into the category of hierarchical
Bayesian models. The citation history of the famous 
\citet{metropolis1953equation} paper shown in \fig{metropolis_citations} 
corroborates the historical narrations on this topic. 
In physics, the MH 
algorithm was well known in the period 1970-1990, but
this was not so in statistics or astronomy. 
In astronomy, 
a watershed moment can be seen in 2002; this is visible 
more clearly in \fig{arxiv_stats} where we track the usage of the words {\it MCMC} and {\it Bayesian}. 

But prior to 2002, the Bayesian-MCMC technique was not unknown to 
the astronomy community. We can see its use in
\citet{saha1994unfolding} who applied it to extract galaxy kinematics from   
absorption line spectra. 
Further seeds were planted down the line by 
\citet{1998PhRvD..58h2001C} while studying gravitational
wave radiation, and then by \citet{2001CQGra..18.2677C} 
and \citet{2001ApJ...563L..95K} in
the context of cosmological parameter estimation using 
cosmic microwave background data. Inspired by these papers, 
\citet{2002PhRvD..66j3511L}
more than any other paper 
seems to have galvanized the astronomy community 
in the use of Bayesian and MCMC techniques.  
They laid out in detail the Bayesian-MCMC framework, applied it to one of the most 
important data sets of the time (cosmic background
radiation) and used it to address a significant scientific
question $-$ the fundamental parameters of our universe. Additionally, 
they made their MCMC code publicly available, which
was instrumental in lowering the barrier for new entrants 
to the field.

\section{Bayesian Data Analysis}
In this section we briefly review the basics of the Bayesian
theory. We start with the Bayes theorem and then use it to 
set up the problem of fitting a model to data. This is 
followed by a discussion of the  
role of priors in Bayesian analysis.
Next, the Bayesian solution of fitting a straight line is
discussed in detail to illustrate the ideas discussed.
Finally, we show how to perform model selection. 
To further explore the topics discussed here, 
many excellent resources are available.  
A stimulating discussion on Bayesian theory can be 
found in \citet{jaynes2003probability}. 
\citet{sivia2006data} and \citet{gregory2005bayesian} 
are excellent textbooks with a good emphasis on 
applications in science. \citet{2010arXiv1008.4686H} 
provides lucid tutorial on fitting models to data.
A fascinating discussion 
on Bayesian versus frequentist approaches to solving 
problems can be found in \citet{loredo1990laplace}. 
A review with emphasis on cosmology is given by 
\citet{2008ConPh..49...71T}.

\subsection{Bayes' Theorem}
\citet{cox1946probability} showed that the rules of Bayesian
probability theory can be derived from
just two basic rules: 
\be
& p(H|I)+p(\bar{H}|I) = 1  &\textrm{ Sum Rule},\\
& p(H,D|I) = p(H|D,I)p(D|I)=p(D|H,I)p(H|I) &\textrm{ Product Rule}.
\ee
Here $H$ stands for some proposition being true 
and $D$ stands for some other proposition being true,
and
$\bar{H}$ means the proposition $H$ is false.  
So the sum rule just states that the probability of a proposition
being true plus the probability of it being false is unity. 
The product rule expresses the joint probability of two
propositions being true in terms of conditional
probabilities, one being true given the other is true. 
The vertical bar $|$ is a conditioning symbol and means
`given'. $I$ denotes relevant background information that is
used to construct the probabilities.  
The product rule leads to the Bayes Theorem
\be
p(H|D,I)=\frac{p(D|H,I)p(H|I)}{p(D|I)},\quad 
{\rm Posterior=\frac{Likelihood \times Prior}{Evidence}}, 
\ee 
where we identify $H$ with the hypothesis and $D$ with the
data. 
The $p(D|H,I)$ is the probability of observing the data 
 $D$ if the hypothesis is true and is known as the {\it likelihood}. 
The quantity $p(H|I)$ is 
the {\it prior} and specifies our prior knowledge of $H$
being true. The 
$p(H|D,I)$, known as {\it posterior}, expresses our updated
belief about the truth of the hypothesis in
light of the data $D$. The quantity $p(D|I)$ is a constant and
serves the purpose of normalizing $\int p(H|D,I)\:{\rm d}H$
to 1. It is known as the {\it evidence}. 

Another important result that can be derived from the sum
rule and the product rule is the marginalization
equation,  
\be 
p(X|I)=\int p(X,Y|I)dY=\sum_i p(X,Y_i|I).
\ee
First let us write the sum rule in an alternate form.  
Instead of considering just $Y$ and $\bar{Y}$, we consider
a set of possibilities $\{Y_i\}$ that are mutually exclusive.
\be
\sum p(Y_i|I)=1, &\textrm{ Extended Sum Rule}.
\ee 
Now, making use of the product rule and the sum rule we get
\be
\sum_i p(X,Y_i|I) &= \sum_i p(Y_i|X,I)p(X|I) & \textrm{  using product rule} \\ 
&= p(X|I) \sum_i p(Y_i|X,I) = p(X|I) & \textrm{  using sum rule}.
\ee 

\subsection{Fitting a model to data}
Typically, we have some data and we want to use it for
scientific inference. One of the most effective approaches to  
dealing with such problems is to develop a model
that describes how the data were created. Let $\theta$ be the
set of parameters of the model and $x^t$ a data
point generated by the model according to $f(x^t|\theta)$. 
The observed data points $x$ can have some measurement 
errors, described by a parameter $\sigma_x$. The probability
of the observed value is then given by $p(x|x^t,\sigma_x)$, which could be
$\mathcal{N}(x|x^t,\sigma_x^2)$ for Gaussian errors; hereafter,
$\mathcal{N}(.|\mu,\sigma^2)$ refers to a normal distribution
with mean $\mu$ and variance $\sigma^2$. The
probability of observed data point $x$ given a model 
and an error is then  
\be
p(x|\theta,\sigma_x)=\int f(x^t|\theta)p(x|x^t,\sigma_x) dx^t.
\ee   
We have integrated over true values $x^t$ which are
unknown. 

If we have reason to believe that there are
outliers in the data, e.g., a fraction of points are not
described by the model, we can supplement a background model
$f_b(x^t|\theta_b)$ with probability $P_b$ and parameters
$\theta_b$ \citep{press1997understanding,2010arXiv1008.4686H}.  The probability of the observed data points 
can then be written as,  
\be
p(x|\theta,\theta_b,P_b,\sigma_x) & = & \int
\left[P_b f_b(x^t|\theta_b)+(1-P_b)f(x^t|\theta)\right]p(x|x^t,\sigma_x)
dx^t \\
& = & p(x|\theta_b)P_b +p(x|\theta)(1-P_b).
\ee   
The total probability for a set of $N$ data points $X=\{x_1,...,x_N\}$ is then
\be
p(X|\theta,\theta_b,P_b,\sigma_x) & = & \prod_{i=1}^{N} p(x_i|\theta,\theta_b,P_b,\sigma_{x,i}).
\ee
To infer the model parameters, one uses the Bayes theorem 
and computes 
\be
p(\theta,\theta_b,P_b|X,\sigma_x) & \propto & p(X|\theta,\theta_b,P_b,\sigma_x)p(\theta,\theta_b,P_b).
\ee
Here, $p(\theta,\theta_b,P_b)$ represents our prior
knowledge about the parameters. We discuss
this in detail in the next section. 

\begin{figure}
\centering \includegraphics[width=0.95\textwidth]{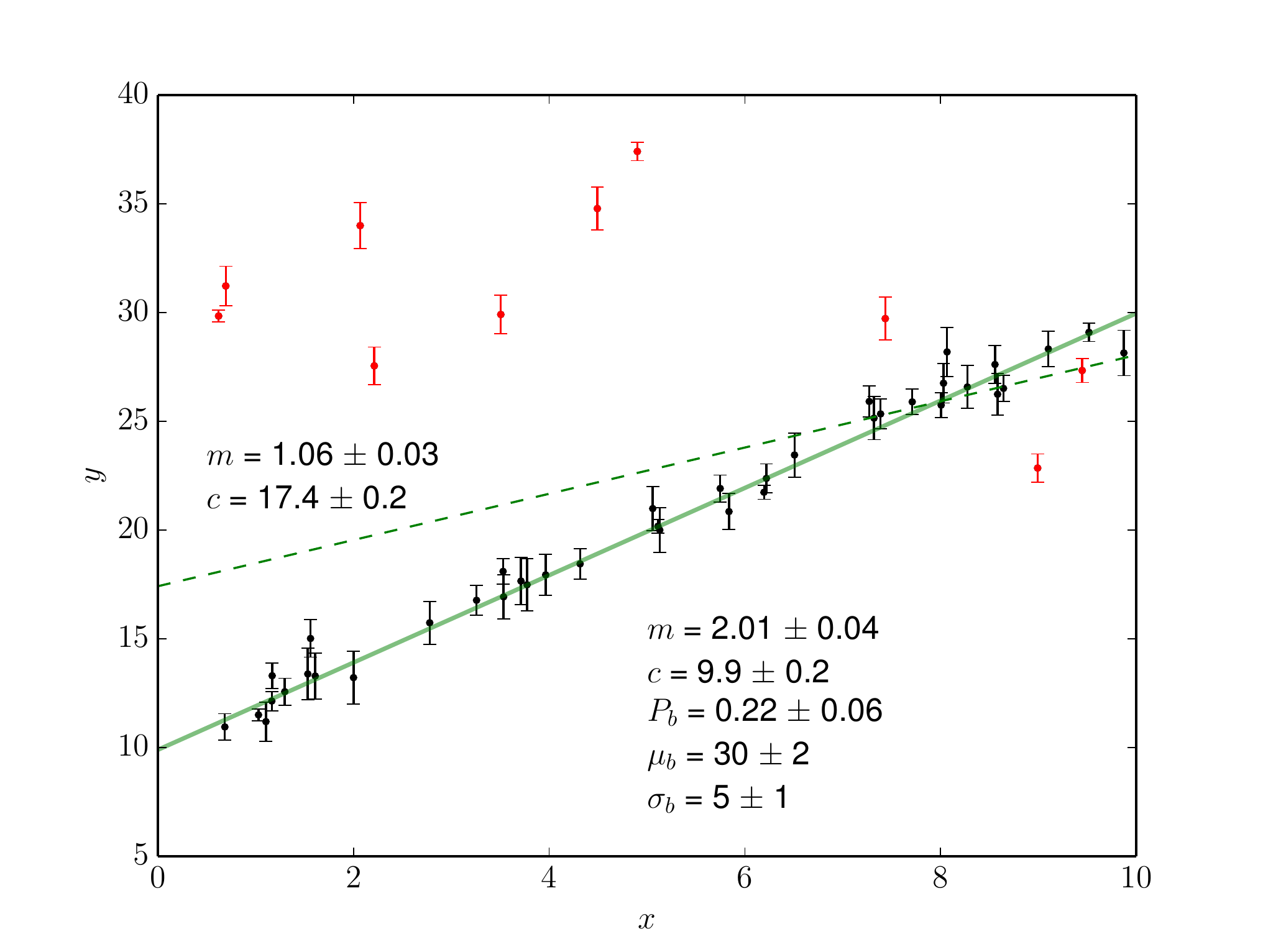}
\caption{Fitting a straight line to data with outliers.  The
  outliers are shown as red points. The dashed line is the
  best fit line when an outlier model is not used. The solid
  line is the best fit line with an outlier model. The data
  was generated with model parameters $m=2$ and $c=10$. 20\%
  of the points were set as outliers and sampled from $\mathcal{N}(30,5^2)$.
\label{fig:stline_b1}}
\end{figure}

We consider the problem
of fitting a straight line
with equation $y=mx+c$ to some data points
$X=\{x_1,...,N\}$ and $Y=\{y_1,...,N\}$, with uncertainty 
$\sigma_{y,i}$ on the $y$ ordinate. We generated 50 data
points with $m=2.0$ and $c=10.0$;  20\% of the data points 
were set as outliers and were sampled from
$\mathcal{N}(30,5^2)$. 
To simulate random uncertainty 
the $y$ ordinate was scattered with a Gaussian function 
having dispersion in range $0.25<\sigma_y<1.25$.  
The data along with the results of our fitting exercise are 
shown in \fig{stline_b1}. 
The image shows the outliers and 
data sampled from a straight line.
We first fitted a simple model without taking the 
outliers into account (dashed line). 
Here, $\theta=\{m, b\}$ and the generative model of the data
is 
\be
p(y_i|m,c,x_i,\sigma_{y,i})=\frac{1}{\sqrt{2\pi}\sigma_{y,i}}\exp\left(-\frac{(y_i-mx_i-b)^2}{2 \sigma_{y,i}^2}\right)
\ee 
It can be seen that the ``best fit'' line is not a good
description for the data points that were sampled from
a straight line.
Next, we extended the model by adding a model for the outliers as 
\be
p(y_i|\mu_b,\sigma_b,x_i,\sigma_{y,i})=\frac{1}{\sqrt{2\pi(\sigma_{y,i}^2+\sigma_b^2)}}
\exp\left(-\frac{(y_i-\mu_b)^2}{2 (\sigma_{y,i}^2+\sigma_b^2)}\right)
\ee 
The full model being
\be
p(Y|m,c,P_b,\mu_b,\sigma_b,X,\sigma_y)=\prod_{i=1}^N
[p(y_i|\mu_b,\sigma_b,x_i,\sigma_{y,i})P_b+p(y_i|m,c,x_i,\sigma_{y,i})(1-P_b)] \label{equ:outliers}
\ee
The best-fit line resulting from this model obtained 
by sampling the posterior distribution using a Markov Chain
Monte Carlo scheme  
is shown  in \fig{stline_b1}. The best-fit parameters of
the model resemble well the true parameters that were used
to create the synthetic data set (the example is implemented
in the software that we provide).

\subsection{Priors}
Priors are one of the most important ingredients of the
Bayesian framework. 
Priors express our present state of knowledge about the
parameters of interest, which we wish to constrain   
by analyzing new data. 
In a multi-dimensional parameter space, it is quite common 
to have degeneracies among the parameters. Here 
priors can play a crucial rule in restricting the posterior 
to a small region of the parameter space as compared to 
the much larger region allowed by the likelihood function. 
Priors can be broadly classified into two types, 
uninformative  and  informative.  
Uninformative priors express our state of ignorance 
and have very little restricting power. They are also known as 
ignorance prior.  Typically 
their distributions are diffuse. Informative 
priors on the other hand 
By contrast, informative priors
are typically very restricting.  
They might come from the analysis of some previous data.  
They are important when the data alone are  
not very informative and without 
external information the data cannot adequately constrain 
the parameters being investigated. 

Ignorance priors are used in cases where
we have very little knowledge about the parameters we want to
constrain, and we wish to express 
our ignorance by using uninformative priors. 
Certainly a prior with sudden jumps or oscillating features 
is too detailed for expressing ignorance! So smoothness is
certainly an important criterion for an ideal uninformative
prior. In fact, if the data are informative, 
almost any prior that is sufficiently smooth in the region
of high likelihood will lead to very similar conclusions. 
Is there a formal and unique way to express our ignorance?  

A number of techniques exist for constructing  ignorance
priors. We here discuss a few simple and commonly used ones;  
for a detailed  review see \citet{kass1996selection}. 
The simplest is Laplace's principle of insufficient
reason which assigns equal probability to all possible
values of the parameter. If the parameter space 
consists of a finite set of points, then it is easy to 
apply this principle. However, for a continuous parameter 
space, the prior depends upon the chosen partitioning
scheme.   

Ignorance priors can also be specified using the 
invariance of the likelihood  
function, $p(x'|\theta'){\rm d}x'=p(x|\theta){\rm d}x$, under the action of a 
transformation group $(x',\theta')=h(x,\theta)$, e.g., translation, scaling or 
rotation of coordinates. If the priors are really
uninformative, consistency demands that 
we should make the same Bayesian inference, 
which implies that the priors should also be invariant to the transformation   
and satisfy $p(\theta'){\rm d}\theta'=p(\theta){\rm d}\theta$ 
\citep{jaynes2003probability}. 
For two special types of parameters, this leads to unique choices
for expressing ignorance. These are the {\it location
  parameters} and the {\it scale parameters}. An example is 
the mean $\mu$  and dispersion $\sigma$  
of a normal distribution $\mathcal{N}(x|\mu,\sigma^2)$ which 
are the location and the scale parameters respectively. 
The likelihood $\mathcal{N}(x|\mu,\sigma^2)$ is invariant 
under transformation $(x',\mu')=(x+b,\mu+b)$, demanding 
invariance for the prior leads to $p(\mu)={\rm constant}$. 
Similarly, $\mathcal{N}(x|\mu,\sigma^2)$ is also invariant 
under $(x'-\mu',\sigma')=(a(x-\mu),a\sigma)$, which leads 
to $p(\sigma) \propto 1/\sigma$. In general, $\mu$ and
$\sigma$ are 
location and scale parameters if likelihood is of the form 
$f((x-\mu)/\sigma)/\sigma$. 

Another commonly used technique to specify ignorance priors is the Jeffreys rule, 
\be
p(\theta) \propto \det (\mathcal{I}(\theta))^{1/2}
\mathrm{,\ where\ }
[\mathcal{I}(\theta)]_{ij}=\int p(x|\theta) \frac{\partial^2
}{\partial
  \theta_i \partial \theta_j} \ln p(x|\theta) {\rm d}x
\ee
is the Fisher information matrix and $\theta$ a vector of
parameters.  It is based on the idea that the prior should
be invariant to reparameterization of $\theta$. Applying it 
to the case where the likelihood is a normal
distribution $\mathcal{N}(x|\mu,\sigma^2)$, gives 
$p(\mu)= {\rm constant}$ (for a fixed $\sigma$) and  
$p(\sigma)= 1/\sigma$ (for a fixed $\mu$). However, 
when applied to both $\mu$ and $\sigma$ together, 
it gives $p(\mu,\sigma)=1/\sigma^2$. To avoid this
contradiction the rule was modified to 
\be
p(\mu_1,..\mu_k,\theta) \propto \det
(\mathcal{I}(\theta))^{1/2}, 
\ee
where $\mu_i$ are location parameters
and $\mathcal{I}(\theta)$ is calculated keeping them fixed.

The principle of maximum entropy 
\citep{Jaynes1957information} is also 
helpful for selecting priors.  Suppose we are interested 
in knowing the pdf of a variable, e.g., the probability of a given
face of a six-faced die landing up. 
Suppose we also have some macroscopic constraint 
available to us, e.g., the mean value obtained when the die is rolled a
large number of times. Such a constraint cannot uniquely
identify a pdf but can be used to rule out a number of
pdfs. The principle says that out of all possible
pdfs satisfying the constraint,
the most likely one is the one having maximum entropy, 
where the entropy is defined as $S=-\sum p_i\log p_i$.
We now use this principle to derive the most likely
distribution of a variable for two common cases.  
\begin{itemize}
\item If for a variable $x$ we know the expectation
value $\bar{x}$ and the fact that it lies in the range $[0,\infty]$ then 
the maximum entropy distribution of $x$ is
$p(x|\bar{x}) = \exp(-x/\bar{x})/\bar{x}. $
\item If $\bar{x}$ and variance 
$\sigma^2=\langle (x-\bar{x})^2\rangle$ 
are known, then 
$p(x|\bar{x},\sigma)=\frac{1}{\sigma\sqrt{2\pi}}\exp\left(-\frac{(x-\bar{x})^2}{2
\sigma^2}\right).$   
\end{itemize}

\subsection{Fitting a straight line}
We now consider the Bayesian solution for fitting a straight
line in detail \citep[see also][]{jaynes1991straight,2010arXiv1008.4686H}. 
We first discuss the solution for the general
case where we have uncertainties on both $x$ and $y$ coordinates 
and then discuss the case where the uncertainties are
unknown.  
Suppose we have a collection of points $(X=\{x_1,...,x_N\}$, 
$Y=\{y_1,...,x_N\})$, 
with uncertainties
$\bsym{\Sigma}=\{\bsym{\Sigma}_1,...,\bsym{\Sigma}_N\}$. 
Here $\bsym{\Sigma}_i$ is the covariance matrix defined as  
 \be
 \bsym{\Sigma}_i=   \left[ {\begin{array}{cc}
       \sigma_{x,i}^2 & \sigma_{xy,i}^2\\       
       \sigma_{xy,i}^2 & \sigma_{y,i}^2\\       
   \end{array} } \right].
\ee
We want to fit a line $y=ax+b$ to these data. 
For the time being, we assume $\bsym{\Sigma}_i$ to be a
diagonal matrix with $\sigma_{xy,i}=0$. 
Let $(x,y)$ be the true values
corresponding to the point $(x_i,y_i)$. 
Then the probability of measuring the point $(x,y)$ at
$(x_i,y_i)$ is 
\be
p(x,y|x_i,y_i,\sigma_{x,i},\sigma_{y,i}) & = & \frac{1}{2 \pi
  \sigma_{x,i}\sigma_{y,i} }\exp\left(-\frac{(x-x_i)^2}{2\sigma_{x,i}^2}-\frac{(y-y_i)^2}{2\sigma_{y,i}^2}\right).
\ee

Let us consider a generative model for the line. 
We consider the pdf of
a line to be described by a Gaussian with width $\sigma_p$ 
along a direction perpendicular to the line and width $\sigma_h$
along the line.  Here, $\sigma_p$ can be thought of as 
an intrinsic scatter about the linear relation that 
we wish to investigate.  
In the limit $\sigma_h \to \infty$, the 
probability of a point $(x,y)$ to be sampled from this 
generative model is  
\be 
p(x,y|a,b,\sigma_p) & = & \frac{1}{\sqrt{2 \pi}
  \sigma_p}\exp\left(-\frac{(y-(ax+b))^2}{(1+a^2)2
  \sigma_p^2}\right).
\ee
Hence, the probability of $(x_i,y_i)$ being sampled from the
generative model of the line is 
\be
 p(x_i,y_i|a,b,\sigma_{x,i},\sigma_{y,i},\sigma_p)  && =   
\int \int p(x,y|a,b,\sigma_p) 
p(x,y|x_i,y_i,\sigma_{x,i},\sigma_{y,i}) \ {\rm d}x{\rm d}y
  \\
&& =  \frac{1}{\sqrt{2 \pi
  (\sigma_{\perp,i}^2+\sigma_p^2)}}\exp\left(-\frac{d_i^2}{2(\sigma_{\perp,i}^2+\sigma_p^2)}\right)
\ee
where
$\sigma_{\perp,i}=\sqrt{(\sigma_{y,i}^2+a^2\sigma_{x,i}^2)/(1+a^2)}$
is the component of the error vector perpendicular to the
line,  and $d_i=(y_i-a  x_i-b)/\sqrt{1+a^2}$ is the perpendicular
distance of the point from the line.   
For a general matrix $\bsym{\Sigma}_i$,
$\sigma_{\perp,i}=\hat{\bsym{u}}^T\bsym{\Sigma}_i
\hat{\bsym{u}}$
for which $\hat{\bsym{u}}=(-a/\sqrt{1+a^2},1/\sqrt{1+a^2})$ is a unit vector
  perpendicular to the line. 
For the full sample, 
\be
p(X,Y|\bsym{\Sigma},a,b,\sigma_p) & = & \prod_i^N
 \frac{1}{\sqrt{2 \pi(\sigma_{\perp,i}^2+\sigma_{p}^2)}}\exp\left(-\frac{d_i^2}{2(\sigma_{\perp,i}^2+\sigma_{p}^2)}\right).
\ee
\begin{figure}
\centering \includegraphics[width=0.75\textwidth]{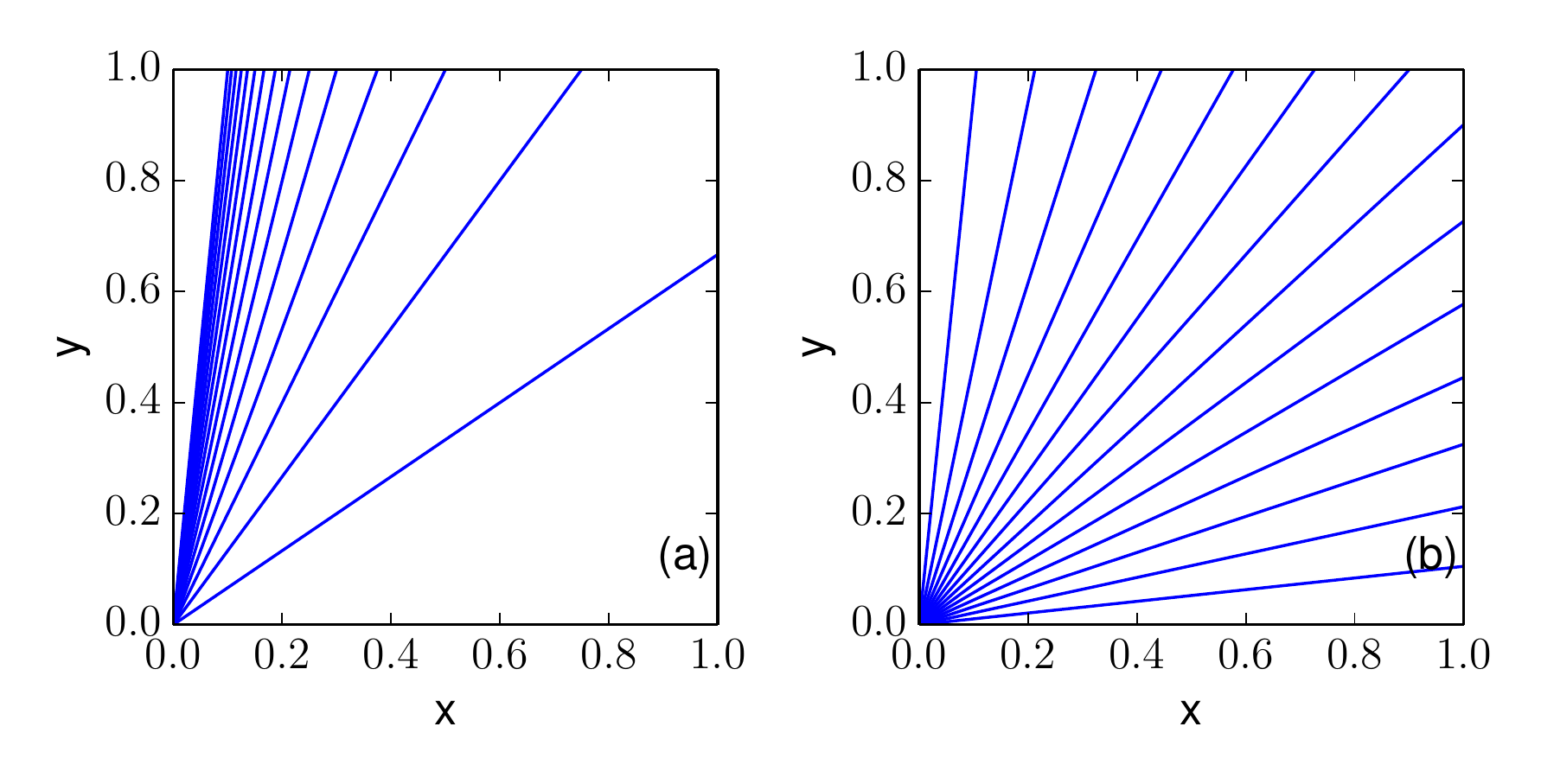}
\caption{Prior for slope of a straight line. The left panel
  represents a prior uniform in slope $a$, straight lines
  with constant interval $\Delta a$. The right panel
  represents a prior symmetric with respect to rotation, 
  straight lines  with constant interval $\Delta \theta$ and
  slope $a=\tan(\theta)$. 
\label{fig:stline_prior}}
\end{figure}
If we desire to compute $a$ and $b$, then 
\be
p(a,b|X,Y,\bsym{\Sigma}) & = &
A p(a,b) p(X,Y|\bsym{\Sigma},a,b). 
\ee
Henceforth, $A$ is a normalization constant which may be
different in different equations. 
The $p(a,b)$ is the prior distribution of parameters of the
line. The two common
choices for the prior are the uniform (flat) and Jeffreys
prior.  Neither is appropriate. 
Given the rotational symmetry in the problem, a
sensible choice is to have priors that are  
symmetric with respect to rotation. Let $\theta=\tan^{-1}a$
be the angle made by the
line with $x$ axis, and $b_{\perp}=b\cos(\theta)$ be the
distance of the line from the origin. 
A uniform prior on $\theta$ and $b_{\perp}$ is symmetric
with respect to  rotation. This leads to 
\be
p(a,b)\:{\rm d}a\:
{\rm d} b=\frac{{\rm d}\theta}{\pi}\frac{{\rm
    d}b_{\perp}}{2B_{\perp}}=\frac{1}{(1+a^2)^{3/2}}\frac{{\rm d}a\:{\rm  d}b}{2B_{\perp}\pi}
\ee
In \fig{stline_prior}, we graphically show how a prior 
uniform in $a$ differs from a prior uniform 
in $\theta=\tan^{-1}(a)$. In the left panel, we show straight lines 
uniformly spaced in $a$.  The 
lines tend to crowd at high value of $a$, and this can 
bias the estimate of the slope $a$. In the right panel, 
the lines are uniformly spaced in $\theta$, 
and there is no crowding effect.  
    
The log-likelihood of the full solution after taking
the prior into account is 
\be
\ln L & =& \ln p(a,b|\{x_i\},\{y_i\},\bsym{\Sigma}_i,\sigma_p) \\
& = & K -\frac{3}{2}\ln (1+a^2) -\sum_{i=1}^{N}\frac{1}{2}\ln (\sigma_{\perp,i}^2+\sigma_{p}^2)-\sum_{i=1}^{N}\frac{d_i^2}{\sigma_{\perp,i}^2+\sigma_{p}^2}. 
\ee
 
We now study the case where $\bsym{\Sigma_i}$ is unknown and
$\sigma_p=0$. 
For simplicity, we assume the uncertainty is the same for all data
points, i.e., $\sigma_{\perp,i}=\sigma_{\perp}$. 
\be
p(a,b,\sigma_\perp|\{x_i\},\{y_i\}) & = &
p(a,b,\sigma_\perp) p(\{x_i\},\{y_i\}|\sigma_{\perp},a,b) \\
& = & p(a,b)p(\sigma_\perp|a,b) \frac{1}{(2 \pi
  \sigma_{\perp}^2)^{N/2}}\exp\left(-\frac{\sum_i d_i^2}{2\sigma_\perp^2}\right)
\ee
and integrate over 
$\sigma_\perp$ using Jeffreys prior $p(\sigma_\perp|a,b)=1/\sigma_{\perp}$ to arrive at
\be
p(a,b|\{x_i\},\{y_i\}) & = & A p(a,b)\left(\sum_i d_i^2\right)^{-N/2}
\ee
So, if we ignore the prior factor, the best fit line is 
simply the line that minimizes the sum of the squared
perpendicular distances of points from the line.

\subsection{Model comparison}\label{sec:modelcomp}
When we have multiple models to explain data, we are faced
with the question of which model is better. There is no 
unique definition of better and depending upon 
what we mean by better we can come up with different
criteria to compare models. We have two main schools of 
thought, a) to compare the probability of the model given
the data and b) to compare the expected predictive accuracy of the
model for the future data. The former is inherently 
a Bayesian approach and is known as Bayesian model
comparison. The latter is inspired by frequentist ideas but 
can also be argued from a Bayesian perspective 
\citep{vehtari2012survey, gelman2014understanding_waic}.  

\subsubsection{Bayesian model comparison}
In the Bayesian formulation, the usefulness of a model is
indicated by the probability of a model $M$ given the data $D$,  
\be
p(M|D)=\frac{p(D|M)p(M)}{p(D)}.
\ee
The prior model probability $p(M)$ is generally assumed to
be unity. Note in some cases it may not be so, and we might have 
more reason to believe one model over the other. 
The $p(D)$ is the same for all models, so it is irrelevant 
when comparing models. Thus the main thing we need to compute 
is the evidence $p(D|M)$ (also know as marginal likelihood). 
Hence, for two models  $M_1$ and $M_2$, the odds ratio in favor of 
$M_2$ compared to $M_1$ is mainly determined by the 
ratio of their evidences, $B_{21}$, also known as the
``Bayes factor'' \citep[for a review and a guide to
  interpreting the Bayes factor, see][]{kass1995bayes}.
\be
\frac{p(M_2|D)}{p(M_1|D)} & = & \frac{p(D|M_2)p(M_2)}{p(D|M_1)p(M_1)}=\frac{p(D|M_2)}{p(D|M_1)}=B_{21}
\ee
For some given data $D$ and a model $M$ parameterized by $\theta$, we have  
\be
p(\theta|D,M)=\frac{p(D|\theta,M)p(\theta|M)}{p(D|M)}.
\label{equ:bayesfac1}
\ee
The evidence $p(D|M)$ appears as the denominator on the right
hand side and can be obtained by integrating
both sides of \equ{bayesfac1} over all $\theta$. For
properly normalized quantities, the left
hand side integrates to unity, leading to  
$p(D|M)=\int p(D|\theta,M)p(\theta|M) {\rm d}\theta$.

Note, the Bayes factor depends upon the adopted range of the
prior which leads to some 
conceptual difficulties \citep[see the paradox in][]{lindley1957statistical}. 
The range of prior 
is not an issue for parameter estimation but it is for model
selection; we cannot use improper priors. 
In most cases, we do have a reasonable sense of the range of
priors and they are unlikely to extend to infinity. 
To better understand the role of priors, consider two models 
$M_1$ and $M_2$, where $M_2$ has a free parameter $\theta$,  
while $M_1$ has no free parameter (with $\theta$
being fixed to $\theta_0$).  
Let $\Delta
\theta_{\rm likelihood}$ be the characteristic width of the likelihood
distribution and $\Delta \theta_{\rm prior}$ the range of a uniform
prior which encloses the likelihood peak. 
The Bayes factor 
in favor of model $M_2$ as compared to $M_1$ is then
\be
B_{21} &=& \frac{p(D|M_2)}{p(D|M_1)} = \frac{\int p(D|\theta)p(\theta) {\rm d}\theta}{L(\theta_{\rm
    0})}  
=  \frac{\int p(D|\theta)
{\rm d}\theta}{L(\theta_{\rm
    0}) \Delta \theta_{\rm prior}}  =  \frac{L(\theta_{\rm
    max})}{L(\theta_{\rm 0})} \frac{\Delta \theta_{\rm likelihood}}{\Delta \theta_{\rm prior}}.
\ee
The first term on the right hand side will in general be 
greater than one and will favor $M_2$, as the simpler model $M_1$ is a
special case of $M_2$. However, the second term penalizes $M_2$   
if it has a large range in priors. 

The conceptual difficulty associated with the dependence of
the Bayes factor on the adopted prior range is alleviated if one thinks 
of hypothesis as a specification of a model as well as the prior on
its parameters. A model $M_2$ with a larger range in priors allows 
for a larger number of possible data sets consistent with
the hypothesis as compared to a simpler model $M_1$ with narrow range
of prior. Hence $p(D|M_2)$, being a normalized probability
over possible data sets, will be lower as compared to
$p(D|M_1)$ \citep{mackay2003information}. 
Also, $M_1$ is more precise as a hypothesis as compared with $M_2$.

If we have more free
parameters in a model, the penalty term in the Bayes factor 
will be 
higher, being of the form $\prod_{i=1}^{d} \Delta
\theta_{\rm likelihood,i}/\Delta \theta_{\rm prior,i}$. 
In this sense, the Bayes factor has a 
built-in safeguard to prevent overfitting 
(a model with a large number of free parameters will 
fit a given set of data better but will perform poorly when
presented with new data). 

\begin{marginnote}
\entry{BIC}{Bayesian information criterion} 
\entry{WBIC}{Widely applicable Bayesian information criterion}
\end{marginnote}
Computing the Bayes factor or the Bayesian evidence 
is computationally challenging. 
Generally, the likelihood is 
peaked and confined to a narrow region in the 
prior range, but has long tails whose contributions cannot be 
neglected. Some commonly employed numerical techniques are 
(1) simulated annealing, (2) nested sampling,  
(3) Laplace's approximation, (4) Lebesgue integration
theory \citep{weinberg2012computing}, and (5) the
Savage-Dickey density ratio \citep{verdinelli1995computing}. 
Two useful approximations of the Bayes free energy $\mathcal{F}=-\ln
p(D|M)$ are 
\begin{eqnarray*}
{\rm BIC}/2  =& -\ln p(Y|\hat{\theta})+(d \ln n)/2 &\textrm{\citep{schwarz1978estimating}}\\
{\rm WBIC}/2 =& \mathbb{E}_{\theta}^{\beta}[- \ln p(Y|\theta)]
\mathrm{\ where\ }
\beta=\frac{1}{\ln n} &\textrm{\citep{watanabe2013widely}}, 
\end{eqnarray*} 
Here, ${\rm E}_{\theta}^{\beta}$ denotes expectation taken over the
posterior distribution $p(\theta|Y) \propto p(Y|\theta)^{\beta}p(\theta)$
of $\theta$. The case of $\beta=1$
corresponds to the Bayesian estimation of the posterior. The
posterior can be sampled using an MCMC algorithm
Assuming weak priors and that the posterior is
asymptotically normal we have $\mathcal{F}=\mathrm{BIC} +O(1)$.
WBIC is an improved version of BIC, which is also
applicable for singular statistical models where BIC fails.
A model is singular if the Fisher information matrix is  
not positive definite, which typically occurs when the
model contains hierarchical layers or has hidden variables.  

\subsubsection{Predictive methods for Model comparison} 
A statistical model $p(x|\theta)$ can be thought of as an approximation 
of the true distribution $q(x)$ from which the observed data 
$Y=\{y_1,y_2,...,y_n\}$ were generated. 
$Y$ represents a set of independently
  observed data points such that  $p(Y|\theta)=\prod_{i=1}^{n}p(y_i|\theta)$.
The Bayesian predictive distribution can then be defined as
$p(x|Y)={\rm E}_{\theta}[p(x|\theta)]$, while the 
maximum likelihood estimate is given by
$p(x|\hat{\theta}(Y))$. 
Predictive methods judge models by their ability to 
fit future data $X=\{x_1,x_2,...,x_n\}$, e.g., via
the log-likelihood function $-\ln p(X|Y)$. 
Given that we do not have future data, the
idea is to measure out-of-sample-prediction error from the
sample at hand. Cross validation is a natural way to do
this, where we divide the current data set into training and 
testing samples. But this is computationally costly. 
Hence, alternate criteria have been developed.
We start by computing the training error 
$T_e  =  -\frac{1}{n}\sum_{i=1}^{n} \ln p(y_i|Y)$.  
However, this is a biased estimator of  $\mathbb{E}_x[-\ln
  p(x|Y)]$ as the data are used twice, once
to estimate the model and once more to compute the log likelihood 
of the data. If we have more parameters in the model, it will 
certainly fit the given data better but will also give rise
to larger variance in the estimator, and  we need to penalize 
the model for this.  This variance,  
which represents the effective degrees of freedom in the
model, can be calculated from the data and the model.   
A list of some useful information criteria 
based on the above idea are given below. They can be 
easily computed using samples of $\theta$ obtained by an 
MCMC simulation of the posterior $p(\theta|Y)$. 
\begin{marginnote}
\entry{AIC} {Akaike information criterion} 
\entry{DIC}{Deviance information criterion}
\entry{WAIC}{Widely applicable Bayesian information criterion}
\end{marginnote}
\begin{eqnarray*}
{\rm AIC}/2 =& -\ln p(Y|\hat{\theta})+d & \textrm{\citep{akaike1974new}}\\
{\rm DIC}_{1}/2 =&   -\ln p(Y|{\rm E}_{\theta}^{1}[\theta])+2\left( \ln
p(Y| {\rm E}_{\theta}[\theta])-{\rm  E}_{\theta}^{1}[\ln  p(Y| \theta)]
 \right) &\textrm{\citep{spiegelhalter2002bayesian}} \\
{\rm DIC}_{2}/2 =&   -\ln p(Y|{\rm E}_{\theta}^{1}[\theta])+2 {\rm Var}_{\theta}^{1}[\ln
   p(Y|\theta)] &\textrm{\citep{spiegelhalter2002bayesian}}\\
{\rm WAIC}_{1}/2 =& -\sum_i^{n} \ln {\rm
  E}_{\theta}^{1}[p(y_i|\theta)] +2\sum_i^{n} \ln {\rm 
  E}_{\theta}^{1}[p(y_i| \theta)]-{\rm E}_{\theta}^{1}[\ln
  p(y_i| \theta)] &\textrm{\citep{watanabe2010asymptotic}}\\
{\rm WAIC}_{2}/2 =& -\sum_i^{n} \ln {\rm
  E}_{\theta}^{1}[p(y_i|\theta)] +\sum_i^{n} {\rm
  Var}_{\theta}^{1}[\ln p(y_i|\theta)] &\textrm{\citep{watanabe2010asymptotic}}
\end{eqnarray*}
Here, ${\rm  Var}_{\theta}^{1}$ denotes variance 
taken over the posterior distribution
$p(Y|\theta)p(\theta)$ of $\theta$.
The first term is a measure of how well the model fits the
observed data while the second term is a penalty for 
the degrees of freedom $d$ in the model. 

In general, the predictive criteria have a well-defined 
information-theoretic interpretation
\citep{burnham2002model, watanabe2010asymptotic}. 
Specifically, the expected value of AIC and WAIC, 
is equivalent to the expected Kullback-Leibler divergence
$\int q(x )\ln [(q(x)/
p(x|Y)] dx $ of the predictive distribution from the true
distribution, the expectation is taken over the random realizations 
of the observed data set $Y$, which samples the true
distribution $q(x)$. 
Also, in the asymptotic limit of large sample
size, both AIC and WAIC are equivalent to leave-one-out 
cross-validation (LOOCV).  

An extra parameter in a model need not necessarily
contribute to extra variance in the predictive density,
e.g., if we have informative priors on the parameter, 
the likelihood has a very weak dependence on the parameter or if the 
model is hierarchical then multiple parameters
might be restricted. The use of AIC can be problematic in
such cases. DIC and WAIC overcome this problem by estimating 
the effective degrees of freedom
directly from the likelihood function of the data and samples 
of $\theta$ obtained from the posterior $p(\theta|Y)$. 

WAIC offers some additional advantages as compared to 
AIC and DIC. 
AIC and DIC use a point estimate for $\theta$ when
computing predictive density, whereas WAIC uses the Bayesian
predictive density. 
If a model is singular, criteria such as AIC, DIC and BIC do not work well. 
In contrast, 
WAIC works for such cases, and in the asymptotic limit of large sample
size, WAIC is always equivalent to Bayesian LOOCV. 

It is instructive to study the differences between BIC and
AIC, as they represent two very different approaches to the 
problem of model selection  
\citep[for a detailed discussion, see][]{burnham2002model}. 
Due to the presence of the $\ln n$ term, for $n>7$ the 
BIC penalizes free
parameters more heavily as compared to AIC. 
So BIC is more parsimonious or cautious when it comes 
to admitting new parameters in a model.  
In situations where two models 
can give rise to the same predictive distribution, BIC 
will favor the model with fewer degrees of freedom while 
AIC will treat them equally.  
An example is nested models, where a simpler model can be
considered as a special case of a complex model but with 
few of its parameters being fixed. 
Interestingly, AIC can also 
be argued to be using the approach of Bayes factors, but with a prior
whose variance decreases with sample size $n$, whereas
BIC would correspond to the choice of a weak prior with
fixed variance \citep{smith1980bayes}. 

To conclude, the Bayesian and the predictive
methods both have their strengths and weaknesses. 
If the choice of priors is well justified, then the 
methods based on Bayes factor are best suited for model 
selection. However, if our aim 
is best predictive accuracy for future data, predictive 
methods like WAIC are a better choice. 

\section{Monte Carlo methods for Bayesian computations}\label{sec:mcmc}
Having discussed how to set up problems in the Bayesian
framework, we now discuss methods to perform the 
inference, i.e., how to estimate the pdf 
of parameters given the data. Except for some simple 
cases, closed form analytical solutions are in general 
not available. So one makes use of Monte Carlo based 
methods to sample from the desired distribution.  
The most popular method to do this today is the Markov Chain 
Monte Carlo (MCMC) method. MCMC is a class of methods for 
sampling a pdf using a Markov chain
whose equilibrium distribution is the desired distribution. 
Once we have a sample 
distributed according to some desired distribution, we can 
compute expectation values and integrals of various 
quantities in a process analogous to Monte Carlo
integration.
The word Monte Carlo in MCMC comes from the use of random numbers to
drive the Markov process and the close analogy to 
Monte Carlo integration schemes. Note in conventional  
Monte Carlo integration, the random samples are statistically 
independent whereas in MCMC they are correlated.  
We first broach the theory
behind Markov chains and then discuss specific MCMC 
methods based on it.

\subsection{Markov Chain}\label{sec:markovchain}
A Markov chain is a sequence of random variables  $X_1,
..., X_n$ such that, given the present state, the
future and past are independent. It is formally written as 
\be
{\rm Prob}(X_{n+1}=x| X_1=x_1,X_2=x_2,...,X_n=x_n) & = &
{\rm Prob}(X_{n+1}=x|
X_n=x_n) 
\ee
In other words, the conditional distribution of $X_{n+1}$ in future,  
depends only upon the present state $X_n$.  
If the probability of transition is independent of $n$, it is
a time-homogeneous chain. Such a chain is defined by
specifying the probabilities of transitioning
from one state to another.  
To simplify mathematical notation, we sometimes 
consider the state space to be continuous and sometimes 
discrete. 
But the presented results are equally valid for either type
of spaces. 
For a continuous state space where a probability density can
be defined we can write the transition probability as 
\be
K(x,y)={\rm Prob}(X_{n+1}=y|X_n=x)
\ee
For a discrete state space the transition probability 
is a matrix and is written as $K_{xy}$. 
On a given state space,  a time-homogeneous Markov chain
has a stationary distribution (invariant measure) $\pi$ if 
\be
\pi(y)=\int dx\ \pi(x)  K(x,y)
\ee
A Markov chain is irreducible if it can go from any state
$x$ of a discrete state space to any
other state $y$ in a finite number of steps, i.e., 
there exists an integer $n$
such that $K^n_{xy}>0$. 
If a chain having a stationary distribution is irreducible,
the  stationary distribution is unique, and the chain is 
positive recurrent.  
For an aperiodic, positive recurrent chain with stationary
distribution $\pi$, the distribution is limiting
(equilibrium distribution). 
This means if we start 
with any initial distribution $\lambda$ (a row vector 
specifying probability over states of a discrete state space)
and apply the transition 
operator $K$ (a matrix) many times, the final distribution will 
approach the stationary distribution $\pi$ (a row vector), 
\be
\lim_{n \rightarrow \infty}\|\lambda K^n-\pi\|=0.
\ee 
For an irreducible Markov chain with a unique 
stationary distribution $\pi$, there is a law of large numbers
which says that the expectation value of a function $g(x)$
over $\pi$  approaches the average taken over the output of a Markov
chain, 
\be
E_{\pi}[g(x)] = \int g(x) \pi(x) dx = \lim_{n \rightarrow \infty}\frac{1}{n}\sum_{i=1}^{n} g(x_i).
\ee
This property allows one to compute Monte Carlo estimates of specific 
quantities of interest from a Markov chain. Techniques that
do this are known as Markov chain Monte Carlo or MCMC.

A chain having a stationary distribution is said to be
reversible if the chain starting from a stationary
distribution looks the same when run forward or
backward in time. In other words, if $X_n$ has distribution $\pi$ 
then the pair $(X_{n},X_{n+1})$ has the same joint
distribution as $(X_{n+1},X_{n})$. 
\be
{\rm Prob}(X_n,X_{n+1})={\rm Prob}(X_{n+1},X_{n})
\ee
For the transition kernel $K$ this means 
\be
\pi(X_n)K(X_{n},X_{n+1})=\pi(X_{n+1})K(X_{n+1},X_{n})
\ee
and is known as the {\it condition of detailed balance}.  
For a Markov chain, it is not necessary to satisfy reversibility 
in order to have a stationary distribution. However,
reversibility guarantees the existence of a stationary
distribution, and is thus a stronger condition. This 
is the reason that most MCMC algorithms are designed to 
satisfy detailed balance.  

\subsection{Metropolis Hastings algorithm}\label{sec:mhalgo}
The most general MCMC algorithm is the Metropolis-Hastings (MH)
algorithm 
\citep{metropolis1953equation,hastings1970monte}. 
Suppose we are interested in sampling a
distribution $f(x)$ on a state space $E$, with $x \in E$.  
To construct a transition kernel $K(x,y)$ to go from 
$x$ to $y$, MH algorithm uses a two
step process: 
\begin{itemize}
\item Specify a proposal distribution
$q(y|x)$.   
\item Accept draws from 
$q(y|x)$ with acceptance ratio $\alpha(x,y)=
{\rm min}\left[1,\frac{f(y)q(x|y)}{f(x)q(y|x)}\right]$. 
\end{itemize}
So the transition kernel is given by $K(x,y)=q(y|x)\alpha(x,y)$.
The full algorithm is as follows
\begin{algorithm}
\SetKwInOut{Input}{Input}
\SetKwInOut{Output}{Output}
\Input{Starting point $x_1$, function $f(x)$,
  transition kernel function $q(y|x)$}
\Output{An array of $N$ points ${x_1,x_2,...x_{N}}$}
\For{ $t=1$ \KwTo $N-1$}
	{
          Obtain a new sample $y$ from $q(y|x_{t})$ \;
          Sample a uniform random variable U \;
          \lIf{$U<\frac{f(y)q(x_{t}|y)}{f(x_{t})q(y|x_{t})}$}
          {
            $x_{t+1}=y$ 
          }
          \lElse
          {
            $x_{t+1}=x_t$             
          } \;
	}
\caption{Metropolis Hastings Algorithm \label{MH1}}
\end{algorithm}

The transition kernel of the MH algorithm is reversible and satisfies
detailed balance, $f(x)K(x,y)=f(y)K(y,x)$. 
Note the reversibility condition by
itself does not lead to a unique form for the acceptance ratio
$\alpha(x,y)$ and alternatives exist \citep{barker1965monte}.
However, it has been shown that the acceptance ratio of the
MH algorithm results in a chain with the fastest
mixing rate \citep{peskun1973optimum}.  

There are multiple ways to construct the proposal
distribution $q$ each leading to a new version of the MH algorithm. 
\begin{itemize}
\item {\bf Symmetric Metropolis:} $q(y|x)=q(x|y)$ which
  simplifies the acceptance probability to ${\rm min}\left\{1,f(y)/f(x)\right\}$; this is the version that was
  proposed by Metropolis and colleagues.
\item {\bf Random walk Metropolis-Hastings (RWMH):}
  $q(y|x)=q(y-x)$; the direction and distance of the new
  point from the current point is independent of the current
  point. Common choices are $N(x,\sigma^2)$ and ${\rm Uniform}(x-\sigma,x+\sigma)$. 
\item {\bf Independence sampler:} $q(y|x)=q(y)$; i.e., the new state is drawn independent of the current
  state. The acceptance probability is given by ${\rm
    min}\left\{1,\frac{f(y)q(x)}{f(x)q(y)}\right\}$, a generalization of the accept-reject algorithm. The quantity
    $q(x)$ should resemble $f(x)$ but with longer
    tails.  
\item {\bf Langevin algorithm:} $q(y|x) \sim
  N(x+\frac{\sigma^2}{2}\nabla \log f(x),\sigma^2)$; this is
  useful when the gradient is available. 
\end{itemize}

Except when $f(y)=f(x)$ (uniform target density), the mean
of the acceptance ratio $\alpha$ is always 
less than unity.  Decreasing $\sigma$ in the RWMH algorithm 
increases $\alpha$ but
lowers the independence of the sampler. Increasing $\sigma$ 
improves the independence but lowers $\alpha$.  
In the Langevin algorithm, one makes use of the information
in the gradient to allow faster mixing of the chain.

\subsection{Gibbs sampling}\label{sec:gibbs}
The Gibbs sampler introduced by \citet{geman1984stochastic} 
is one of the most popular computational 
methods for doing Bayesian computations.
Suppose we want to sample $f(x)$ where $x \in \chi \subseteq
\mathcal{R}^d$. In Gibbs sampling, the transition kernel $K(x,y)$ is split into
multiple steps. In each step, one coordinate is advanced
based on its conditional density with respect to other
coordinates. The algorithm is as follows:
\begin{algorithm}
\SetKwInOut{Input}{Input}
\SetKwInOut{Output}{Output}
\Input{Starting point $x^1$, Function $f(x)$}
\Output{An array of $N$ points ${x_1,...,x_{N}}$}
\For{ $t=1$ \KwTo $N-1$}
    {
      Sample $x^1_{t+1}$ from $f(x^1|x^2_{t},...x^d_{t})$ \;
      Sample $x^2_{t+1}$ from $f(x^2|x^1_{t+1},x^3_{t}...x^d_{t})$  \;
      Sample $x^d_{t+1}$ from $f(x^d|x^1_{t+1},...x^{d-1}_{t+1})$ \;
    }
\caption{Gibbs Sampling Algorithm \label{GS1}}
\end{algorithm}

\noindent
The full transition kernel is written as,  
\beq 
\kappa_{1 \to d}(x_{t+1}|x_{t})=\prod_{i=1}^{d}
f(x^i_{t+1}|x^1_{t+1},...x^{i-1}_{t+1},x^{i+1}_{t},...,x^d_{t+1}). 
\eeq
Similarly one can define a reverse move, 
\beq
\kappa_{d \to 1}(x_t|x_{t+1})=\prod_{i=d}^{1}
f(x^i_{t}|x^1_{t+1},...x^{i-1}_{t+1},x^{i+1}_{t},...,x^d_{t+1}). 
\eeq
It can be easily shown that 
\beq
f(x_{t})\kappa_{1\to
  d}(x_{t+1}|x_{t}) =  f(x_{t+1})\kappa_{d\to
  1}(x_{t}|x_{t+1}).
\eeq 
Integrating both sides leads to 
\beq
\int f(x_{t})\kappa_{1\to  d}(x_{t+1}|x_{t}) dx= f(y). 
\eeq
Thus, $f$ is the stationary distribution of the Markov
chain formed by the transition kernel  $\kappa_{1\to
  d}(x_{t+1}|x_{t})$. Note the Gibbs sampler as given above
({\it systematic} scan) is not reversible. However, the
reversible ones can easily
be produced, e.g., at each iteration picking a random
component to update (random-scan).  The random-scan 
Gibbs sampler can be viewed as a special case of 
MH sampler with acceptance ratio ${\rm
  min}(1,\frac{f(y)q(x|y)}{f(x)q(y|x)})=1$. 
It follows that 
\be
f(y)q(x|y) &= & f(y^i|y_{-i})f(y^{-i})f(x^i|y^{-i}) =
f(y^i|x^{-i})f(x^{-i})f(x^i|x^{-i}) \nonumber \\ 
& = & f(x^i|x^{-i})f(x^{-i})f(y^i|x^{-i}) = f(x)q(y|x). 
\ee
Here,
$x^{-i}=\{x^1,...,x^{i-1},x^{i+1},...,x^{d}\}$ and
$y^{-i}=x^{-i}$, as only the $i$-th component is changed in
each step.   

\subsection{Metropolis within Gibbs}\label{sec:mwg} 
One problem with the Gibbs sampler is that it requires 
one to sample from the conditional distributions  
which can be difficult. 
In such cases, one can replace the sampling of
conditional densities with the MH step. This then becomes 
the Metropolis within Gibbs (MWG) scheme
\citep[see ][]{muller1991generic}, which is shown in Algorithm
\ref{MWG} (it is implemented in the code that we provide).
\begin{algorithm}
\SetKwInOut{Input}{Input}
\SetKwInOut{Output}{Output}
\Input{Starting point $x^1$, Function $f(x)$}
\Output{An array of $N$ points ${x^1,x^2,...x^{N}}$}
\For{ $t=1$ \KwTo $N-1$}
    {
      \For{ $i=1$ \KwTo $d$}
          {
            Generate $x^i_{*}$ from $q_i(x^i|x^1_{t+1},...,x^{i-1}_{t},x^{i+1}_{t}...x^d_{t})$  \;
            Calculate
            $r=\frac{f_i(x^i_{*}|x^1_{t+1},...,x^{i-1}_{t},x^{i+1}_{t}...x^d_{t})}{f_i(x^i_{t}|x^1_{t+1},...,x^{i-1}_{t},x^{i+1}_{t}...x^d_{t})}
            \frac{q_i(x^i_{t}|x^1_{t+1},...,x^{i-1}_{t},x^{i+1}_{t}...x^d_{t})}{q_i(x^i_{*}|x^1_{t+1},...,x^{i-1}_{t},x^{i+1}_{t}...x^d_{t})}$\;
          \lIf{$U<{\rm Min}(1,r)$}
              {
                $x^{i}_{t+1}=x^{i}_*$ 
              } \lElse
              {
                $x^i_{t+1}=x_t$             
              }\;
          }
    }
\caption{Metropolis-Within-Gibbs Algorithm \label{MWG}}
\end{algorithm}
Rather than updating all the variables step by step, 
one can also choose to update a subset of variables
together, leading to block updates. 
The fact that the full sampling of a complicated
distribution can be broken up into a sequence 
of smaller and easier samplings, is the main strength 
of the Gibbs sampler and has resulted in its widespread use \citep[e.g.][]{2012MNRAS.427.2119S,2014ApJ...793...51S}.

\subsection{Adaptive Metropolis}\label{sec:adapmetro}
\begin{figure}
\centering \includegraphics[width=0.75\textwidth]{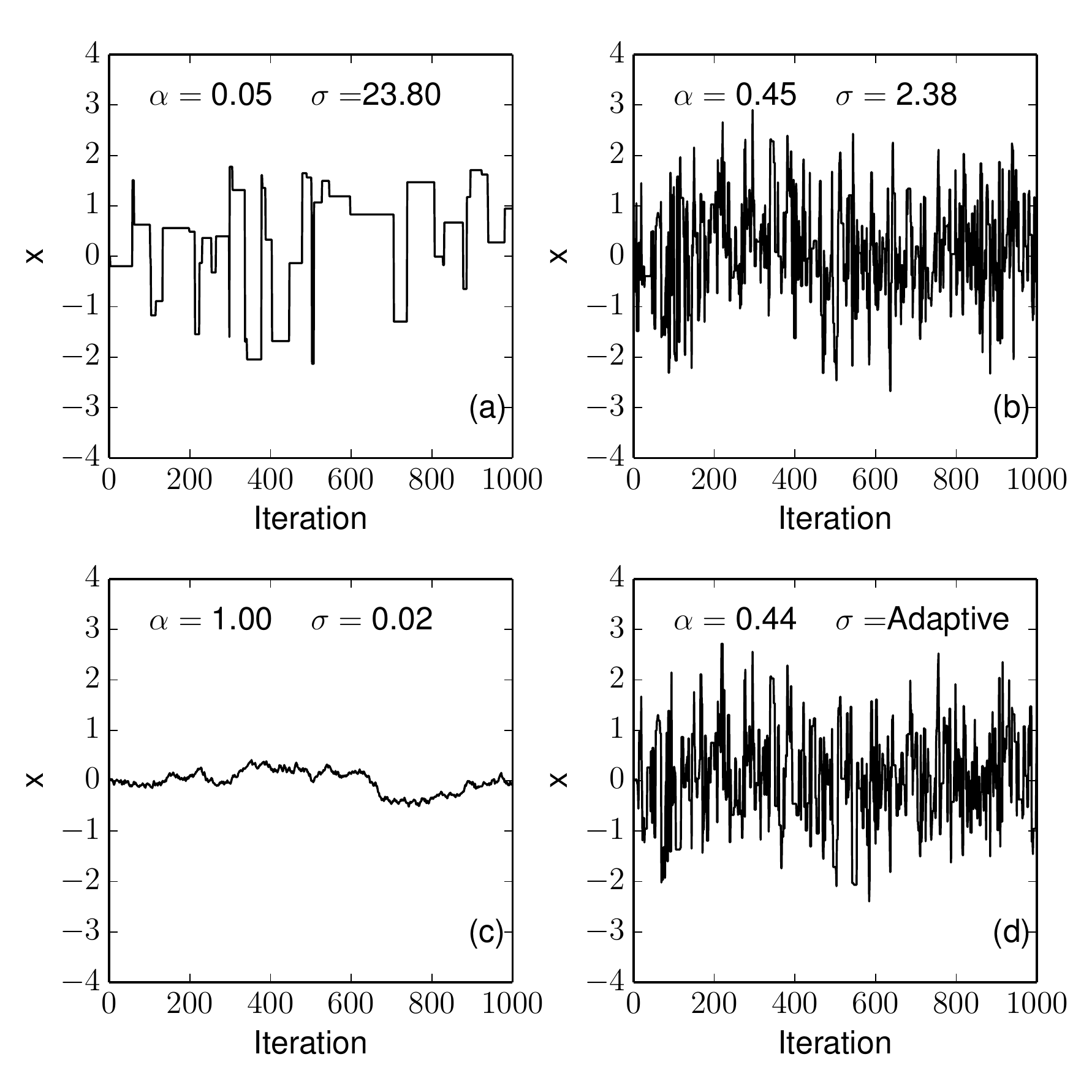}
\caption{MCMC chains for different widths of the
  proposal distribution. The variable $x$ is sampled from a
  Gaussian distribution $N(0.0,1.0)$ using MCMC with 
  different proposal distributions. The proposal
  distributions are also normal and are characterized by width 
  $\sigma$. 
  The ideal width is $\sigma=2.38$ and the chain for
  this is shown in panel (b). 
  In panel (a), $\sigma$ is ten times larger while in panel
  (c) it is one hundred times smaller. Panel (d) shows the
  chain when an adaptive scheme is used to adjust $\sigma$. 
  The performance of this is same as for the ideal case shown 
  in panel (b).  
\label{fig:adaptive1}}
\end{figure}

The efficiency of the MH algorithm depends crucially upon
the proposal distribution. By efficiency we typically mean
how independent are the samples. If the samples are not
independent then they have high correlation. For Markov chains,
the correlation falls off with distance between samples. If
the correlation is large, this means the mixing in the 
chain is slow.  If the width of the proposal
distribution is too small, the acceptance ratio is high 
but the chain mixes very slowly. If the width of the proposal
distribution is too large, the acceptance ratio is too small 
and the chain again mixes slowly (see \fig{adaptive1} for an
illustration of this effect). 
\citet{gelman1996efficient} showed that optimal
covariance matrix $\Sigma$ for the RWMH algorithm using the multivariate
normal distribution is $(2.38^2/\mathcal{D})\Sigma_{\pi}$, where $\mathcal{D}$
is the dimensionality of the space and $\Sigma_{\pi}$ is the covariance
matrix of the target distribution $\pi$. The optimal acceptance
ratio $\alpha_{\rm opt}$ is 0.44 for dimension $\mathcal{D}=1$ and then 
falls off with increasing number of dimensions reaching an
asymptotic value of $0.23$ for $\mathcal{D} \to \infty$. 
The convergence
is quite fast  
($\alpha=$[0.441, 0.352, 0.316, 0.279, 0.275, 0.266] for
$\mathcal{D}=$[1, 2, 3, 4, 5, 6]).
The efficiency as compared to independent samples is
$0.331/\mathcal{D}$.

These results suggest a possible way to choose the optimal 
proposal distribution.
Estimate the covariance matrix $\Sigma_{\pi}$ by a trial 
run and then use it for the actual run. Even doing this 
is cumbersome as it is unclear how long 
the trial run should be. To circumvent this, 
\citet{haario2001adaptive} proposed an adaptive scheme in
which $\Sigma$ is updated on the fly using past values. 
Naively, any scheme that uses proposals that depend upon the 
full past history violates the Markovian property, i.e., 
the future should only depend on the present and should be
independent of the past. The trick is to adapt the proposal 
distribution in such a way that it converges to the optimal
one. The resulting chain then also converges to the target distribution.  
\citet{andrieu2001controlled} showed that such a scheme 
can be described as part of a more general adaptive
framework. 

At the heart of most adaptive algorithms is the 
\citet{robbins1951stochastic} recursion. They proposed an iterative
stochastic algorithm to find roots of functions that are 
stochastic, i.e., their algorithm solves $M(x)=\alpha$, where instead of
$M(x)$ the function available is $N(x)$, which is
stochastic and is such that $\langle N(x) \rangle=M(x)$. 
Starting with some initial value $x_0$ 
the algorithm to get the $n+1$th iterate is  
\be
x_{n+1}=x_n+\gamma_n(\alpha-N(x_n))
\ee
Here $\gamma_1,\gamma_2,...$ is a sequence of positive
steps. The $x_n$ then converge to the true solution 
provided the sequence $\gamma_n$ satisfies 
\be
\sum_{n=0}^{\infty} \gamma_n=\infty \textrm{ and }
\sum_{n=0}^{\infty} \gamma_n^2 < \infty
\ee
The first condition makes sure that irrespective of where we
start, the solution can be reached in a finite number of
steps. The second condition makes sure that we do converge. 
A possible choice of $\gamma_n$ is
$\gamma_n=\gamma/n^{\beta}$ where $0<\beta<1$.  

A nice description of various adaptive 
algorithms is given by \citet{Andrieu2008}. 
Below we discuss Algorithm 4 from their paper
which is quite general and is implemented in the software that we provide.
 \begin{algorithm}
\SetKwInOut{Input}{Input}
\SetKwInOut{Output}{Output}
\Input{Starting point $x_0$,$\mu_0$, $\Sigma_0$, $\alpha^{*}$, function $f(x)$}
\Output{An array of $N$ points ${x_0,x_2,...x_{N-1}}$}
\For{ $i=1$ \KwTo $N-1$}
	{
          Obtain a new sample $y$ from
          $N(x_{i},\lambda_i\Sigma_{i})$ \;
          Set $\alpha_{i}(x_i,y)=\frac{f(y)}{f(x_{i})}$ \;
          Sample a uniform random variable U \;
          \eIf{$U<\alpha_{i}(x_i,y)$}
              {$x_{i+1}=y$ \; }
              {$x_{i+1}=x_{i}$ \;}
          $\log \lambda_{i+1}=\log
         \lambda_{i}+\gamma_{i+1}(\alpha_{i}(x,y)-\alpha^*)$\;
          $\mu_{i+1}=\mu_{i}+\gamma_{i+1}(x_{i+1}-\mu_{i})$ \;
          $\Sigma_{i+1}=\Sigma_{i}+\gamma_{i+1}[(x_{i+1}-\mu_{i})(x_{i+1}-\mu_{i})^{\rm
              T}-\Sigma_{i}]$ \;
	}
\caption{Adaptive Symmetric Random Walk Metropolis Hastings Algorithm \label{AMH1}}
\end{algorithm}
If $\beta$ is too small, the convergence is too slow; if 
$\beta$ is too large the convergence is too fast and  
the simulation can quickly lean towards a wrong solution 
and will take a long time to get out of it. 
For adaptive MCMC, we find a choice of $\beta=0.6$ 
to be satisfactory for most test cases. \fig{adaptive1}d 
shows an adaptive MCMC chain obtained using Algorithm
\ref{AMH1}. 
The adaptive chain looks very similar to the ideal case
shown in \fig{adaptive1}b, and this demonstrates the
usefulness of the adaptive MCMC scheme. 

\subsection{Affine invariant sampling}\label{sec:affine}
An elegant solution to the problem of tuning the proposal
density is to use the idea of ensemble samplers
\citep{gilks1994adaptive}. Here multiple chains (walkers)
are run in parallel but allowed to interact in such a way 
that they can adapt their proposal densities. 
\citet{Goodman2010} provide a general purpose algorithm 
to do this, known as the affine invariant sampler \citep[see
  also][]{christen2010general}. A python implementation 
of this
(emcee: the MCMC hammer, \url{http://dan.iel.fm/emcee/current/}) 
is provided by \citet{2013PASP..125..306F}
and is widely used in astronomy. We now 
describe this algorithm.

We saw in the previous section that adapting the proposal density can
violate the Markovian property of a chain. The trick lies 
in using the information available in the ensemble but 
in a way that does not violate the Markovian property. 
This is achieved by using the idea of partial resampling which is a
generalized version of the Gibbs sampling procedure. 
Let us consider an ensemble of walkers
$X=(x_1,x_2,...,x_L)$ and a Markov chain that walks 
on a product space with distribution
$\Pi(X)=\pi(x_1)\pi(x_2)...\pi(x_L)$.  
Then if $x_i$ is updated conditional on other walkers
$x_{[-i]}=\{x_1,...,x_{i-1},x_{i+1},...,x_L\}$ 
(complementary set of walkers), but satisfying
detailed balance $p(y_i|x_i,x_{[-i]})=p(x_i|y_i,x_{[-i]})$, 
then each walker samples from $\pi(x)$.   
  
One way to do this is to choose a point $x_j$ from $x_{[-i]}$
and a scalar $r$ with density $g(r)$, and propose a new point $y$ as 
\be
y=x_i+(r-1)(x_i-x_j)=x_j+r(x_i-x_j).
\ee 
The inverse transformation is given by $x_i=x_j+(y-x_j)/r$.
Now if we want the proposal to be symmetric then  
$q(y_i|x_i,x_{[-i]})=q(x_i|y_i,x_{[-i]})$, and this implies  
$g(1/r)=rg(r)$.
A good choice of such a function is 
\be
g(r)=\frac{1}{\sqrt{r}} \textrm{ for } r \in
\left[\frac{1}{a},a\right] \textrm{ and } a>1.
\ee
To satisfy detailed balance, the acceptance probability is
given by
$\textrm{min}\left[1,r^{n-1}\frac{\pi(Y)}{\pi(X_i)}\right]$.
The factor $r^{n-1}$ is because the proposal is restricted
along a line and not the full hypersphere over the actual
space. This means an appropriate Jacobian has to
be calculated; for details see \citet{gilks1994adaptive} and 
\citet{roberts1994convergence} (the proof is much easier
when using the reversible jump MCMC formalism of \citet{green1995reversible}). 

Moves other than the stretch move can also be constructed, e.g. a proposal $y=x_i+W$, where $W$ has 
a covariance computed from a subset of walkers in the 
complementary sample. It is also possible to construct 
algorithms which use a combination of both the stretch and
the walk move. Although the \citet{Goodman2010} affine invariant algorithm  
elegantly solves the problem of choosing a 
suitable proposal distributions, it has one drawback.  
The computational 
cost of warm-up scales linearly with the number of walkers.  
Note, like most other MCMC algorithms, multimodal distributions 
(distributions with many well separated peaks) 
also pose a problem for this algorithm. 

\subsection{Convergence Diagnostics}\label{sec:convergence}
Having studied MCMC methods in order to sample from distributions, we now 
discuss how to detect convergence; i.e., how long should 
we run an MCMC chain. Several convergence diagnostics 
have been proposed in the literature. 
\citet{cowles1996markov} provide a good review of 13
convergence diagnostics. Other reviews include 
\citet{brooks1998general} and \citet{robert2013monte}. 
Unfortunately, because there is no method to detect convergence, 
we can only detect failure to converge. 
So convergence diagnostics are necessary conditions but not
sufficient. Below we present two schemes 
to monitor convergence. The first scheme makes use of the
correlation length of the chain to compute the effective 
number of independent samples in a chain. The second 
scheme makes use of multiple chains to see if they 
are converging. 

\subsubsection{Effective sample size}
Let us begin by estimating how many independent 
samples we need to get reliable estimates of mean 
and variance of a quantity. 
For a posterior of some variable $x$ with standard deviation
$\sigma_x$,  the Monte Carlo standard error goes 
as $\sigma_x/\sqrt{N}$ for sample of size $N$. So to measure
the mean of a quantity with about 3\% error as compared to
the overall uncertainty $\sigma_x$ we need $N=1000$.  
\citet{raftery1992many} showed that to measure $0.025$
quantile to within $\pm0.005$ with probability 0.95 requires 
about 4000 independent samples.

However, the MCMC is not an independent sampler. As we
have seen, the points in an MCMC chain are
correlated. Autocorrelation provides a measure of
this. Autocorrelation $\rho_{xx}(t)$ for a sequence is the correlation
between two points separated by a fixed distance $t$; i.e.
\be
\rho_{xx}(t)=\frac{\mathbb{E}[(x_i-\bar{x})(x_{i+t}-\bar{x})]}{\mathbb{E}[(x_i-\bar{x})^2]}
\ee
An automatic windowing procedure is discussed by  
\cite{sokal1997} for the computation of integrated 
autocorrelation \citep[see also][]{goodman1989multigrid,Goodman2010}. 
Typically the autocorrelation falls off exponentially 
as $\sim \exp^{-t/\tau_{x}}$ and $\tau_{x}$ is known as the
correlation time (or correlation length).  
The integrated autocorrelation
is defined as $\tau_{\rm int,x}=(1/2)
\sum_{t=-\infty}^{\infty} \rho_{xx}(t)$. The variance of
the mean of $x$ for a sample of size $N$ can be shown to be
\be
{\rm Var}(\bar{x})=(2 \tau_{\rm int,x}) \frac{\mathbb{E}[(x_i-\bar{x})^2]}{N}
\ee
So for correlated samples the variance is $2
\tau_{\rm int,x}$ times larger than the variance of
independent samples. Using $\tau_{\rm int,x}$, one can
measure the number of effective independent samples in a correlated
chain $-$ also known as the {\it effective sample size} (ESS) $-$
as $N/(2\tau_{\rm int,x})$ and then use it to decide if we 
have enough samples (e.g., $1000<\:$ESS$\:<4000$). 

\subsubsection{Variance between chains}
The most widely used criterion for studying convergence was first 
presented
by \citet{gelman1992inference}.
Let us suppose we have $M$ chains each consisting of $2N$
iterations out of which we use only the last $N$
iterations. For any given scalar parameter of interest
$\theta$, let      
\be
\bar{\theta}_{j}=\frac{1}{n}\sum_{i=1}^{n}\theta_{i,j}
\textrm{\ \ \ \  and\ \ \ \ }
\bar{\theta}=\frac{1}{m}\sum_{j=1}^{m}\bar{\theta_j} 
\ee
The index $i$ runs over points in a chain, and the index $j$ 
runs over the chains.
Then the between chain variance and the mean within chain variance 
can be written as
\be
B=\frac{1}{m-1}\sum_{j=1}^m(\bar{\theta}_j-\bar{\theta})^2
\textrm{\ \ \ \  and\ \ \ \ }
W=\frac{1}{m}\sum_{j=1}^m \frac{1}{n-1}\sum_{i=1}^{n}(\theta_{i,j}-\bar{\theta_j})^2.
\ee
The total variance $\hat{\sigma}^2$ for 
the estimator $\bar{\theta}$ can be written as a weighted
average of $W$ and $B$, $\hat{\sigma}^2=W(n-1)/n+B$.
If we account for the sampling variability of the estimator
$\bar{\theta}$, then this yields a pooled variance of 
\be
V=\hat{\sigma}^2+\frac{B}{m}=\frac{n-1}{n}W+\frac{m+1}{m}B
\ee 
for the mixture of chains.
If the initial
distribution is over-dispersed, then $B>\sigma^2$ and $V$ always 
overestimates the true variance $\sigma^2$. 
For any finite $n$, $W$ is expected to be 
less than $\sigma^2$, as individual sequences 
in a chain would 
not have had the time to explore the full target distribution.  
So, initially we expect $V/W>1$.  
However, in the limit $n\to \infty$,
the variance $B$ between chains, which is expected to fall off
as $1/n$, goes to 0 and $W$ will approach the true variance
$\sigma^2$, making $V/W$ approach 1.
Therefore the ratio $\hat{R}=\sqrt{V/W}$, also known as the
potential scale reduction factor, can be used to monitor the convergence. 

\subsubsection{Thinning}
For making inferences from an MCMC chain, 
some algorithms use only the $k$-th iteration of each sequence such that successive draws are approximately independent, a process known as {\it thinning}. 
However, there is no additional advantage of thinning 
other than savings in storage. Since we are throwing away information, 
an estimate from a thinned chain can never 
be better than the original chain 
\citep{geyer1992practical,maceachern1994subsampling}. 
Moreover, it is difficult to choose an appropriate $k$ 
without studying the autocorrelation of the full chain. 
So thinning is useful only in situations where the
autocorrelation is known a priori and is known to be 
large. Here again $k$ should be chosen such that it is 
smaller than the autocorrelation length, to retain 
as much information as possible.

\subsection{Parallel Tempering}\label{sec:partemp}
Multimodal distributions in general pose problems 
for all MCMC algorithms. Parallel tempering 
is one way to address this problem.  
It is a type of ensemble sampler where 
multiple chains are simulated in parallel but are allowed to
exchange information. Each chain has a target distribution 
different from the other and is controlled by a parameter 
$T$ known as the temperature. Let $\pi(x)=\exp(-H(x))$ be the 
actual target distribution, then a ladder of distributions 
\be
\pi_i(x)=\exp(-H(x)/T_i), i=1,....n
\ee 
is created, controlled via the parameter $T_i$, such that 
$T_1>T_2>...>T_{n}$. $T_n$ is set to 1. Hence,  
$\pi_n$ represents the target distribution.  
The temperature broadens the target distribution 
and allows a wider exploration of the parameter space
which makes it useful to explore multimodal distributions. 
To exchange information between the chains, a state swapping 
procedure is used. A swap is proposed between a randomly 
chosen chain $i$ 
and its neighbors $i-1$ and $i+1$ with probability 
$q_{i,i-1}=q_{i,i+1}=0.5$ and $q_{1,2}=q(n,n-1)=1$. 
Naively, accepting the swap will violate the detailed
balance condition. So the swap proposal is accepted 
with probability 
\be
p_{ij}= {\rm
  min}\left(1,\frac{\pi_i(x_j)\pi_j(x_i)}{\pi_i(x_i)\pi_j(x_j)}\right)
={\rm
  min}\left(1,\exp\left(\left[H(x_i)-H(x_j)\right]\left[\frac{1}{T_i}-\frac{1}{T_j}\right]\right)\right)
\ee
which satisfies detailed balance. 

In parallel tempering 
the temperature ladder needs to be chosen carefully. 
If the neighboring temperatures are too far apart, the 
acceptance rate will be diminished leading to slow mixing. 
If the neighboring temperatures are too close, a large number of elements 
in the ladder will be required to explore a wide range in
parameter space, and this can increase the
computational cost significantly. However, by exploiting the trial runs, a suitable 
ladder can be constructed \citep{liang2011advanced}. 

The idea of parallel tempering can be generalized to
construct evolutionary algorithms that incorporate 
features of genetic algorithms into the framework of 
MCMC. The basic idea is to have parallel chains as in 
parallel tempering and allow exchange of information 
while satisfying detailed balance on the product space 
defined by the chains. The exchange of information is based 
on ideas of mutation and crossover from genetic algorithms 
\citep{liang2001evolutionary,liang2001real}.

\subsection{Monte Carlo Metropolis Hastings}\label{sec:mcmh}
\label{sec:MCMH}
\begin{algorithm}
\SetKwInOut{Input}{Input}
\SetKwInOut{Output}{Output}
\Input{$\tilde{p}(\theta|x,Y),\ q(\theta'|\theta),\ T$ and $x$}
\Output{A set of points $({\theta_1,\theta_2,...\theta_{N}})$
  sampled approximately from $p(\theta|x)$}
\For{ $t=1$ \KwTo $T-1$}
	{
         Generate $\theta'$ from $q(.|\theta)$\;
         Generate $m$ auxiliary samples $Y=(y_1,..y_m)$
         conditioned on $\theta$\;
         Obtain a Monte Carlo estimate
         $\tilde{r}(x,\theta,\theta',Y)$ of the MH ratio
         $\tilde{r}=\tilde{p}(\theta'|x,Y)/\tilde{p}(\theta|x,Y)$  \;
         \lIf{$U<{\rm Min}(1,\tilde{r})$}
             {
               $\theta_{t+1}=\theta'$ 
             } \lElse
             {
               $\theta_{t+1}=\theta$             
             }\;
	}
\caption{Monte Carlo Metropolis Hastings\label{MCMH}}
\label{alg:mcmh}
\end{algorithm}
In MCMC based Bayesian inference, we are concerned with
simulating samples from some pdf 
$p(\theta|x)=p(x|\theta)p(\theta)$. However, there are 
situations when $p(x|\theta)$ cannot be easily evaluated 
or is not available in an analytically tractable form. 
In such situations one can make
use of Monte Carlo based  techniques to approximately evaluate 
$p(\theta|x)$. More generally, the Metropolis 
Hastings ratio $r=p(\theta'|x)/p(\theta|x)$ is used 
to update an MCMC chain. In such techniques, typically, one generates 
a set of auxiliary samples $Y=\{y_1,...,y_m\}$ conditioned on 
$\theta$ and then uses them to compute  
$\tilde{p}(\theta|x,Y)$ (an approximation of $p(\theta|x)$)
or $\tilde{r}$ (an approximation of the ratio $r$).  
However, Monte Carlo based
estimates are stochastic and special care is needed when 
working with them in an MCMC scheme.  
An algorithm to
make use of Monte Carlo based estimates inside a Metropolis
Hastings algorithm is given in \alg{mcmh} 
(it is implemented in the software that we provide). 

There are many variants of \alg{mcmh}, depending
upon how and at what stage the auxiliary sample is generated $-$ see
Chapter 4 in \citet{liang2011advanced}. 
The invariant
stationary distribution of such Markov chains is not necessarily the 
target density $p(\theta|x)$. The 
characteristics of such chains and their convergence
properties are discussed by \citet{beaumont2003estimation}
and \citet{andrieu2009pseudo}. 
In \alg{mcmh}, the auxiliary sample is refreshed in each iteration and 
the same sample $Y$ is used to estimate both $\tilde{p}(\theta'|x,Y)$ 
and $\tilde{p}(\theta|x,Y)$. This makes \alg{mcmh} 
more robust compared to other similar alternatives. 
In classical MCMC, one can reuse the previous estimate of 
$p(\theta|x)$ when computing $r$. However, when the
Metropolis Hastings ratio $\tilde{r}$ is
stochastic, if $\tilde{p}(\theta|x,Y)$ is not evaluated in each
iteration using a fresh sample of $Y$, then the
MCMC chain tends to get stuck at a stochastic
maxima of the estimated likelihood
\citep{2014ApJ...793...51S}. 
The smaller the size of the auxiliary sample, or the more inaccurate the
Monte Carlo estimate of $\tilde{r}$, the worse is this
problem. Using the same sample $Y$ to estimate 
both $\tilde{p}(\theta'|x)$ and $\tilde{p}(\theta'|x)$ 
leads to lower noise in the estimated ratio of $\tilde{r}$.
This property was also noticed by \citet{2013MNRAS.433.1411M} 
in the context of fitting models of the gravitational 
potential of the Milky Way to spatio-kinematic data 
of stars orbiting inside it. Two specific cases where the 
above algorithm can be used are given below. 

\subsubsection{Unknown  normalization constant}
In fitting a model to data,  
we are interested in sampling $p(\theta|x)=p(x|\theta)p(\theta)$. 
To do this, the function $p(x|\theta)$ should be 
properly normalized over the data space, in the sense that 
$\int p(x|\theta)dx=1$. However, on many occasions, we have
\be
p(x|\theta) =\frac{1}{Z(\theta)}\exp(-U(x,\theta))=\frac{1}{Z(\theta)}f(x|\theta),
\ee
where $f(x|\theta)$ is known but the normalization 
constant $Z(\theta)$ is not known. An example is  
the problem of fitting a density profile
$\rho(r|\theta)$ ($r$ being the Galactocentric distance) 
to a sample of stars with Galactic latitude
$b>30^{\circ}$, longitude $l>30^{\circ}$ and heliocentric
distance $s<50$ kpc. Here we have
$Z(\theta)=\int_{b=\pi/6}^{\pi/2} db \int_{0}^{50} ds \int_{\pi/6}^{2\pi}
dl \rho(l,b,s|\theta) s^2 cos(b)$. 

Our aim is to compute the Metropolis Hastings ratio 
$r=p(\theta'|x)/p(\theta|x)=[Z(\theta)/Z(\theta')][f(x|\theta')/f(x|\theta)]$
that is used to advance an MCMC chain, and  it is the ratio 
$R=Z(\theta)/Z(\theta')$ that is unknown.  
If one can sample exactly from $f(x|\theta)$,  
then it is possible to cancel the normalization constant 
using ingenious algorithms by \citet{moller2006efficient} and 
\citet{murray2012mcmc}. However exact sampling is not
always feasible. In such cases 
a Monte Carlo estimate of the ratio of the unknown normalization
constant $R=Z(\theta)/Z(\theta')$ can be done using samples
$Y=(y_1,..., y_m)$\ generated from density $f(y|\theta)$, such that
\be
\tilde{R}(\theta,\theta';Y)=\frac{1}{m}\sum_{i=1}^{m} \frac{f(y_i|\theta')}{f(y_i|\theta)}.
\ee
This sampling can be done by various means, e.g., exact
sampling, MCMC, and rejection sampling. If $f(y|\theta')$ is
difficult to sample
from, one can use so-called
``importance sampling'' by drawing samples from
a distribution $g(y|\theta)$ that is easy to sample
from. The required ratio of normalization constants is then given by 
\be
\tilde{R}(\theta,\theta';Y)= \frac{\frac{1}{m}\sum_{i=1}^{m}
  f(y_i|\theta')/g(y|\theta)}{\frac{1}{m}\sum_{i=1}^{m} f(y_i|\theta)/g(y|\theta)},
\ee
and the MH ratio is given by $\tilde{r}=\tilde{R}(\theta,\theta';Y)[f(x|\theta')/f(x|\theta)]$.

\subsubsection{Marginal inference}
Here we are interested in the marginal density
$p(\theta|x)=\int p(\theta,y|x) dy$, but the integral may not be analytically tractable and 
may also be difficult to do by deterministic schemes. In such
situations, the integration can be done by Monte Carlo 
importance sampling,  using auxiliary samples $Y$ generated from some density 
$g(y|\theta)$ that is easy to sample from. Thus we have
\be
\tilde{p}(\theta|x,Y)=\frac{1}{m}\sum_{i=1}^{m} \frac{p(\theta,y_i|x)}{g(y_i|\theta)}.
\ee

\subsection{Hamiltonian Monte Carlo}\label{sec:hmc}
One of the attractive features of MCMC 
for sampling pdfs 
is its better performance for higher
dimensions. However, for very large dimensions, traditional
MCMC algorithms start running into problems. 
While for lower dimensions, a typical set of the posterior
(e.g. region encompassing 99\% of the total probability) 
lies close to the center, for higher dimensions, a typical  
set lies in a shell that has a very large volume. Since, a
shell cannot be traversed with large step sizes, it takes a long time to explore the posterior.    

Hamiltonian Monte Carlo (HMC) tries to address this problem by
introducing an auxiliary variable called momentum $u$ for 
each real variable $x$ called position \citep{1987PhLB..195..216D,neal1993probabilistic}. The log of posterior 
(target density) $\pi(x)$ is assumed to define the potential
energy $U(x)=-\ln \pi(x)$, 
and the momenta define the kinetic energy $K(u)$. Together
they define the Hamiltonian $H(x,u)=U(x)+K(u)$, where
$K(u)=u^2/2$. The distribution to be explored is 
\be
p(x,u)=\exp\left[-H(x,u)\right]=\exp\left[-\ln\pi(x)-\frac{1}{2}u^2\right]
\ee

Next, principles of Hamiltonian dynamics are used to advance 
a given point to a new location. The point is then 
accepted or rejected based on the MH algorithm. The use 
of Hamiltonian dynamics to advance a given point allows 
the point to travel to locations which are 
far from its current location. This allows faster 
exploration of the parameter space. 

There are two major obstacles involved with using HMC, and this has
prevented its widespread use. 
First, it requires the gradient of the target density. 
Secondly, it requires two extra parameters to be tuned by the user:
a step size $\epsilon$ to advance from the current state 
and the number of steps over which
to evolve the Hamiltonian system.  Considerable progress has
been made to address both these issues.  

The automatic/algorithmic differentiation can be used 
to accurately compute the derivatives of a given function 
without any user
intervention \citep{griewank2008evaluating}. 
The idea is that 
any function written as a computer program can be described as 
a sequence of elementary arithmetic operations, and then 
by applying the chain rule of derivatives repeatedly on
these operations, the derivatives can be computed. 
Alternatively, one can create analytical functions to 
approximate the target density and use these to compute the
derivatives.  
This is because the exact Hamiltonian is only required 
when computing the acceptance probability 
and this does not require derivatives.    
For simulating the trajectory, one needs derivatives 
and here one can use an approximate Hamiltonian  \citep{neal2011mcmc}. 
An application of HMC for fitting cosmological
parameters is given by   
\citet{hajian2007efficient} and
\citet{2008MNRAS.389.1284T}. 
\citet{homan2014no} provide additional algorithms for
automatic tuning of step size $\epsilon$ and the number 
of steps $L$, known as the No-U-Turn Sampler. 
This is used in the open-source Bayesian inference 
package Stan (available at \url{http://www.mc-stan.org}).

\subsection{Population Monte Carlo}
Population Monte Carlo is an iterative importance sampling
technique that adapts itself at each iteration and 
produces a sample approximately simulated from
the target distribution. The sample along with its
importance weights 
can be used to construct unbiased estimates of quantities 
integrated over the target distribution. 
Suppose $h(x)$ is a quantity of interest.  
One of the major applications for 
MCMC applications is to compute integrals like 
$J = \int h(x) \pi(x) dx$.
In importance sampling, this is replaced by 
\be
J = \frac{1}{N}\sum_{i=1}^{N} h(x_i) \frac{\pi(x_i)}{q(x_i)},  
\ee
where $(x_1,...,x_n)$ are sampled from a distribution $q(x)$
which is easier to sample than $\pi(x)$. The closer the
importance function to the target distribution, the better
the quality of the estimate (lower variance).
In practise it is difficult to guess a good importance
function. 

The main idea in population Monte Carlo is to
start with a reasonable guess of the importance function
$q_0$ and then iteratively improve $q_t$ by making use 
of the past set of samples $(x_1^{t-1},...,x_N^{t-1})$. 
The importance function can adapt not only in time (with
each iteration), but also in space, and can be 
written in general as $q_{t}(.|x_{i}^{t-1})$. Suppose 
$X^t=\{x_1^t,...,x_N^t\}$ are the set of points at iteration 
$t$. Let $x_i^t$ be produced from  importance distribution 
$q_t(x|x_i^{t-1})$.  
An estimate of $J$ is then given by 
\be
J^{t}=\sum_{i=1}^{n} w_i^t h(x_i^t) \textrm{ where } \rho_i^t=\frac{\pi(x_i^t)}{q_{it}(x_i^t)} \textrm{ and }
    w_i^t=\frac{\rho_i^t}{\sum_{i=1}^{n} \rho_i^t}. 
\ee
Thus the expectation
value of any function $h(x)$ computed using importance
sampling is unbiased, i.e.
\be
\mathbb{E}\left[h(X^t)\frac{\pi(X^t)}{q_t(X^t|X^{t-1})}\right]
& = & \int h(x) \frac{\pi(x)}{q_t(x|y)}q_t(x|y) g(y) {\rm
  d}x {\rm d}y =  \int h(x) \pi(x) dx 
\ee
Here $g$ is distribution of $X^{t-1}$ and the equality is
valid for any $g$. 

A simple choice for the importance function is to have 
set it as a mixture
of normal or $t$-distributions, e.g., $q^t(x)=\sum_{d=1}^{D}
\alpha_d^t \mathcal{N}(x|\mu_d^t,\Sigma_d^t)$
\citep{cappe2008adaptive}. This has been used 
for cosmological parameter estimation
\citep{2009PhRvD..80b3507W}  and model comparison
\citep{2010MNRAS.405.2381K}.

\subsection{Nested Sampling}\label{sec:nested}
In \sec{modelcomp}, we saw that computing the evidence is
computationally challenging. Nested sampling
\citep{skilling2006nested} is designed to
ease this computation. 
To compute the evidence,  
we are interested in computing quantities like 
\be
Z = \int L(\theta)\pi(\theta)\: {\rm d}\theta =\int L(\theta)\: {\rm d} \pi (\theta).
\ee
Integration is basically chopping up the full space into
small volume elements and summing the contribution of the
integrand over these cells. We are free to chop up the volume 
and order or label the cells as we wish. So we divide the
space by iso-likelihood contours and define a variable 
$X$ to label them. A convenient choice is the 
prior probability mass enclosed by an iso-likelihood
contour, i.e. 
\be
X(L)=\int_{L(\theta)>L} \pi(\theta) d\theta 
\ee
If the the prior probability is normalized, then it 
ranges from 0 for the highest likelihood, to 1 for the lowest 
likelihood. Given the above definition, we can also define 
an inverse function $L(X)$, which is the likelihood that 
encloses a probability mass of $X$. So the integral for $Z$ 
can now be written as $Z=\int L(X)dX$.  

Suppose we generate $N$ samples uniformly from the prior
distribution. Next, we sort them in decreasing sequence of $L$
to give prior mass $X_i=i/N$. Then using trapezoidal rule,
one can easily perform the numerical integration. 
However, a significant contribution to the integral comes from a region 
with small prior mass $X$. So, the integral should be done
in equal steps in $\ln(X)$ rather than $X$. This can
be done using an iterative procedure. We start 
with a set $A$ of $N$ points drawn from the prior.  
At each iteration, let $L_i$ be the point with lowest $L$;
we replace it in set $A$ with a new point drawn uniformly from the
prior but satisfying $L>L_i$. This generates a sequence of 
$L_{i}$ for which the expected $X_i=\exp(-i/N)$. 

Nested sampling is widely used for cosmological 
model selection and parameter estimation. Three publicly  
available packages based on nested sampling are CosmoNest
\citep[][see \url{https://github.com/dparkins/CosmoNest}]{2006PhRvD..73l3523P,2006ApJ...638L..51M}, 
MultiNest \citep[][see \url{https://ccpforge.cse.rl.ac.uk/gf/project/multinest}]{2009MNRAS.398.1601F} and 
DNEST \citep[][see \url{https://github.com/eggplantbren/DNest4}]{brewer2011diffusive}.

\section{Bayesian hierarchical modelling (BHM)}\label{sec:bhm}
In the simplest setting, we have some observed data $Y$ generated 
by some model having parameters $\theta$ which can be
inferred using the Bayes theorem as 
\be
p(\theta|Y) \propto p(Y|\theta)p(\theta), 
\ee 
where $p(\theta)$ denotes our prior knowledge or belief 
about $\theta$.
If the model parameters $\theta$ depend upon another
set of parameters $y$ through $p(\theta|\phi)p(\phi)$, then 
$\theta$ and $\phi$ can be inferred using 
\be
p(\theta,\phi|Y) \propto p(Y|\theta)p(\theta|\phi)p(\phi).
\ee 
The variable $\phi$ is known as the hyperparameter and $p(\phi)$, 
the distribution of the hyperparameter, as a hyperprior.  
Alternatively, the observed data $Y$ may depend upon another set of
hidden variables $X$, which in turn depend on $\theta$. The
inference of $\theta$ and $X$ can then be established
using  
\be
p(\theta,X|Y) \propto p(Y|X)p(X|\theta)p(\theta).
\ee 
Such situations lead to hierarchies and Bayesian models of
this type are known as hierarchical models.
It turns out that hierarchies are quite common in real world
applications, often where more than two levels
exist, and Bayesian hierarchical modelling provides 
a framework for capturing this. 

\begin{figure}
\centering \includegraphics[width=0.95\textwidth]{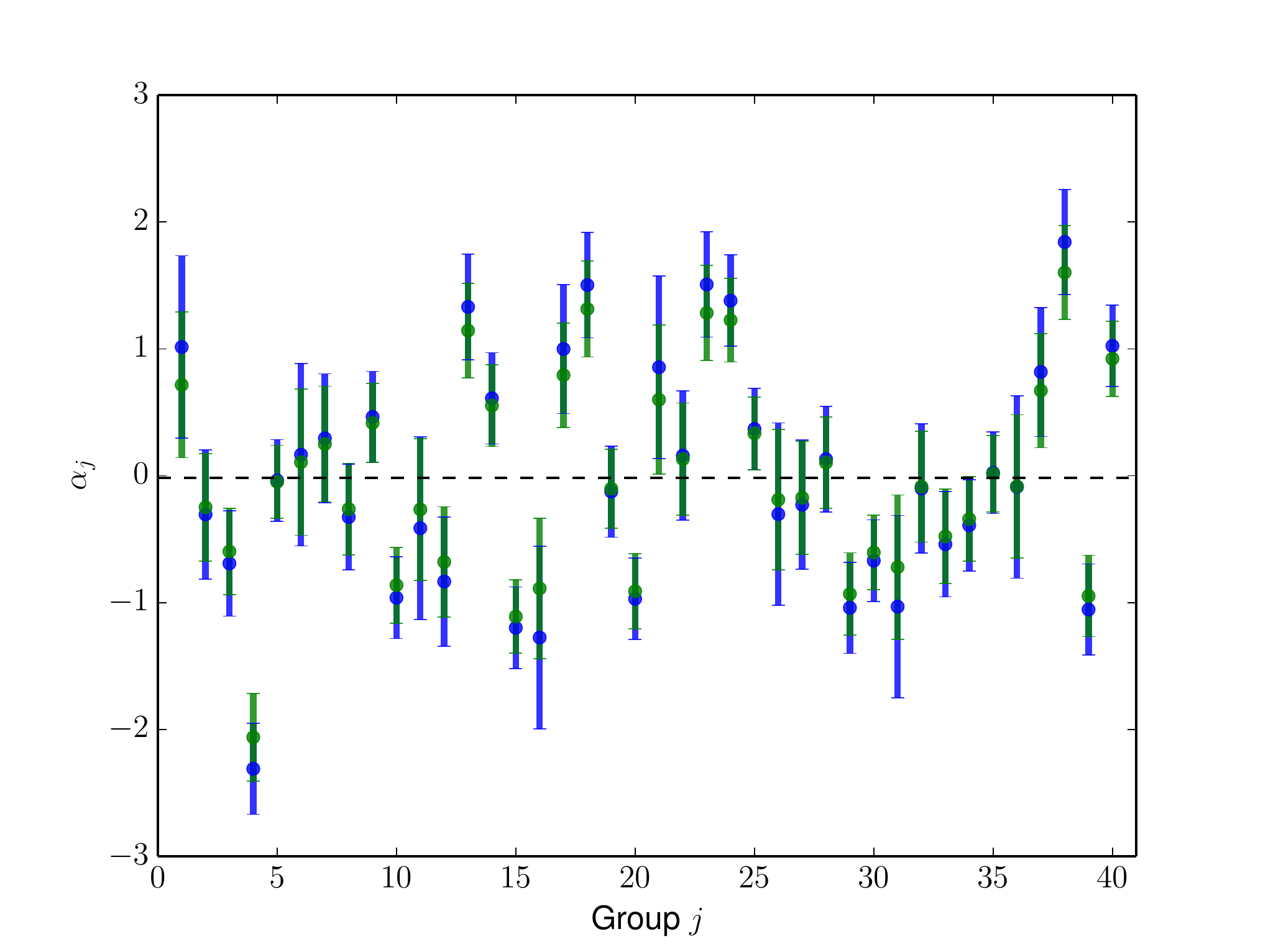}
\caption{Analysis of group mean using a hierarchical
  Bayesian model. The dashed line is the global mean from
  all data points. The blue dots are group means
  computed from the data points in the group. The green
  points are estimates of group mean using a hierarchical
  Bayesian model, which makes use of the full
  information available. The green points have smaller error bars and are systematically 
  closer to the global mean than the blue points. 
\label{fig:hbayes_demo1}}
\end{figure}

Let us consider a simple example, for details see
\citet{gelman2013bayesian}. Suppose
we observe some data $Y$ (a set of measurements of some
variable $y$) with uncertainty 
$\sigma$, and we are interested in the mean
$\alpha=\bar{y}$. Now suppose that the data
$Y=\{y_{ij}|0<j<J, 0<i<n_j\}$ 
are grouped into $J$ independent groups, and we have reason to believe 
that the group mean $\alpha_j$ varies from group to group. 
For observations within a group $j$, our model is  
\be
p(y_{.j}|\alpha_j,\sigma)=\mathcal{N}(y_{.j}|\alpha_j,\sigma^2), 
\ee
where we denote by $y_{.j}$ an observation belonging to 
group $j$. We now compute the group mean $\bar{y}_{.j}$, instead of 
global mean $\bar{y}$, to capture 
the variation of mean across groups. 
A global mean is certainly an inaccurate description 
of data, whenever the group mean is far away from the 
global mean. However, if the number of data points in a group 
is very small, e.g., $n_j=2$, then the uncertainty in the group
mean is large and it is much better to
trust the global mean than the group mean.   

Bayesian hierarchical modelling provides a natural way 
to handle the above problem of group means. 
It can act like a middle ground 
between the two extremes, global mean versus group mean. 
To demonstrate this,  
we set up the above problem using a Bayesian
hierarchical model. 
Suppose the group means are distributed according 
to a normal distribution 
\be
p(\alpha_j|\mu,\omega)=\mathcal{N}(\alpha_j|\mu,\omega^2)
\ee
where $\mu$ and $\omega$ are unknown parameters of the
model. The $\mu$, $\omega$ and the group means  
$\alpha=\{\alpha_1,...,\alpha_J\}$ can then be inferred 
from data $Y$ using 
\be
p(\alpha,\mu,\omega|Y) & \propto &
p(Y|\alpha,\sigma)p(\alpha|\mu,\omega)p(\mu,\omega)
 \propto  p(\mu,\omega) \prod_{j=1}^J p(\alpha_j|\mu,\omega)\prod_{i=1}^{n_j} p(y_{ij}|\alpha_j,\sigma)
\ee
We generated synthetic data with 
$\mu=0$, $\omega=1$, $J=40$, $\sigma=1$ and $2<n_j<10$; 
we then estimated $\alpha$, $\mu$ and $\omega$ (assuming
flat priors for $\mu$ and $\omega$).  
The results are shown in \fig{hbayes_demo1}. 
The BHM based group mean estimates are systematically
shifted with respect to standard group mean estimates (
computed from the data points in a group). The BHM estimates
are closer to the global mean than the standard
estimates. The shift between the two estimates is more for
cases where the error bars are large. The BHM estimates also
have smaller error bars. This is because , when estimating
the group mean, in addition to points within a group the BHM
model also makes use of information available from other
groups.

\subsection{Expectation maximization, data
  augmentation and Gibbs sampling} \label{sec:gsem}
The easiest way to analyze a Bayesian hierarchical 
model is via Gibbs sampling,
 and the motivation for doing this was provided by the  
the expectation maximization (EM) algorithm. 
In fact, the EM algorithm 
led to the development of the DA  
algorithm, which in turn provided the idea to use 
Gibbs sampling to solve Bayesian hierarchical 
models. 

Hence we begin by exploring the 
EM algorithm \citep{dempster1977maximum} which   
is one of the most influential algorithm in the field of
statistics. Let us suppose that we have some observed data 
$x=\{x_1,...,x_N\}$ generated by some model $p(x|\theta)$ 
having parameters $\theta$.  
We want to
compute the most likely parameters of the model given the
data, i.e., $\hat{\theta} = {\rm
  argmax}_{\theta}[p(x|\theta)]$. The full model is specified
by  $p(x,z|\theta)$ with $p(x,z|\theta)=\prod_{i=1}^{N}
p(x_i,z_i|\theta)$, where $z$ are variables which are
either missing or hidden or unobserved. 
The EM algorithm solves this problem as follows. 
The algorithm has two steps. It starts with a
fiducial value of $\theta_0$, then does the following 
at every iteration $t$.
\begin{itemize}
\item E-step: Compute $Q(\theta|\theta_t,x)= \int dz\  p(z|\theta_t,x) \log
  p(x,z|\theta)$. In other words,  it computes the
  expectation  of the log likelihood $ \log p(x,z|\theta)$
  with respect to  $p(z|\theta_t,x)$.
\item M-step: Find the value of $\theta$ that maximizes  
$Q(\theta|\theta_t,x)$ and set $\theta_{t+1}={\rm argmax}_{\theta}[Q(\theta|\theta_t,x)]$.
\end{itemize}
These steps are repeated iteratively until $\theta_{t+1} \sim \theta_{t}$.  
The proof that the EM algorithm increases the likelihood
$p(x|\theta)$ at each stage is as follows. 
The conditional density of the missing data $z$ given the 
observed data $x$ and the model parameter $\theta$ is given
by 
\be
p(z|\theta,x)=\frac{p(x,z|\theta)}{p(x|\theta)}. 
\ee
Taking the $\log$ and then the expectation with respect to
$p(z|\theta_t,x)$, we get 
\be
\log p(x|\theta) & =  & \int dz\  p(z|\theta_t,x) \log
  p(x,z|\theta) -\int dz\  p(z|\theta_t,x) \log
  p(z|\theta,x), \\
 & = & Q(\theta|\theta_{t},x)+S(\theta|\theta_{t},x),
\ee
which is valid for any $\theta$. Using this result, we can compute
the difference 
\be
\log p(x|\theta_{t+1}) -\log p(x|\theta_{t}) = Q(\theta_{t+1}|\theta_{t})-Q(\theta_{t}|\theta_{t})+S(\theta_{t+1}|\theta_{t})-S(\theta_{t}|\theta_{t})
\ee
Due to the M-step,
$Q(\theta_{t+1}|\theta_{t},x)-Q(\theta_{t}|\theta_{t},x) \geq
0$. Also, from Gibbs' inequality,
$S(\theta_{t+1}|\theta_{t})-S(\theta_{t}|\theta_{t})\geq
0$. 
This means that each EM iteration is guaranteed to increase the marginal
likelihood $p(x|\theta)$. 
This guarantees a convergence towards a maximum, 
but not necessarily a global maximum. The algorithm can still get stuck at a 
saddle point, or a local maximum.

The EM algorithm as presented above is deterministic. 
In general, it is not always easy to compute the expectation 
value, as it involves integrals over high dimensions.   
A general way to compute the $Q(\theta|\theta_t,x)$, would be to draw 
$m$ random samples of $z$ from distribution $k(z|\theta_t,x)$
and take its mean. We label this stochastic estimate 
$Q_S(\theta|\theta_t,x)$, which in the limit $m \to \infty$ 
is same as $Q(\theta|\theta_t,x)$. Having computed $Q_S$, the 
M-step can proceed as usual to maximize it and 
compute a new $\theta_{t+1}$. In fact $m$ can be set to 1.
This is the stochastic version of EM
(SEM) as given by \citet{celeux1985sem}. 
Because of stochasticity, one does not get a unique answer but 
instead a distribution. In fact, SEM  
generates a Markov chain, which under mild regularity
conditions converges to a stationary distribution. The 
algorithm has an additional advantage in that it is less 
likely to get stuck at a local maximum. 

If we now replace the M-step with a draw of $\theta$ 
from the $Q_S(\theta|\theta_t,x)$, 
this becomes a fully stochastic method
; this is, as previously mentioned 
the DA algorithm of \citet{tanner1987calculation}. 
This is equivalent to a 
two-step Gibbs Sampler for sampling from 
\be
p(\theta,Z|X) \propto p(X,Z|\theta)p(\theta)
\ee
\begin{enumerate}
\item Sample $Z_{t+1}$ from $p(Z|\theta_{t},X)$. 
\item Sample $\theta_{t+1}$ from $p(\theta|Z_{t+1},X)$. 
\end{enumerate}
From the properties of the
Gibbs sampler, we know that the sequence of  
$(\theta_{t},Z_{t})$ forms a Markov chain that samples 
$p(\theta,Z|X)$. Although Gibbs sampling requires sampling from the
conditional distribution, the inner step can be
replaced by MH sampling, leading to the
Metropolis-within-Gibbs method as discussed in \sec{gibbs}. 
This provides a completely general scheme for handling 
missing data. 

Finally, the DA algorithm is not limited to just 
missing variables of the data, but can also be applied to 
unknown parameters of the model, e.g., $\alpha$ in      
\be
p(\theta,\alpha|X) \propto p(X|\theta,\alpha)p(\theta)p(\alpha)
\mathrm{\ \ \ \ \ or\ \ \ \ \ }
p(\theta,\alpha|X) \propto p(X|\theta,\alpha)p(\theta|\alpha)p(\alpha).
\ee
Such dependencies are common in Bayesian hierarchical modeling. 
In general, the Bayesian hierarchical modeling 
provides a framework for handling 
marginalization in Bayesian data analysis, i.e., 
handling parameters or variables that are either unknown  
or missing but are necessary to model the data.  

\subsection{Handling uncertainties in observed data}
Marginalization is not limited to handling missing data. It
can also be used to handle data $X=\{x_i| i=1,...,N\}$ with uncertainty 
$\sigma_X=\{\sigma_{x,i}| i=1,...,N\}$. Consider 
\be
p(\theta|X,\sigma_X) & \propto & p(\theta) \prod_i \int
p(x|\theta)p(x_i|x,\sigma_{x,i}) dx \mathrm{\ \ and\ }
\label{equ:pthetax1} \\
p(\theta,X^t|X,\sigma_X) & \propto & p(\theta)\prod_i p(x_{i}^t|\theta)p(x_{i}|x_i^t,\sigma_{x,i}),
\label{equ:pthetax2}
\ee
where $X_t=\{x_{i}^t|i=1,...,N\}$ is the true values of
the observed data $X$. Here again, instead of doing an
integration, one treats the true values as unknowns and
sample them using the Gibbs scheme. We demonstrate this with a simple example
where $p(x_i^t|\theta) \sim \mathcal{N}(x_i^t|\mu,\sigma^2)$ is the model
that generates the data, and $\theta=(\mu,\sigma)$ are the 
unknowns which we wish to evaluate. The data has uncertainty
described by another Gaussian function
$p(x_i|x_i^t,\sigma_{x,i}) \sim \mathcal{N}(x_i|x_i^t,\sigma_{x,i}^2)$. 
\begin{figure}
\centering \includegraphics[width=0.75\textwidth]{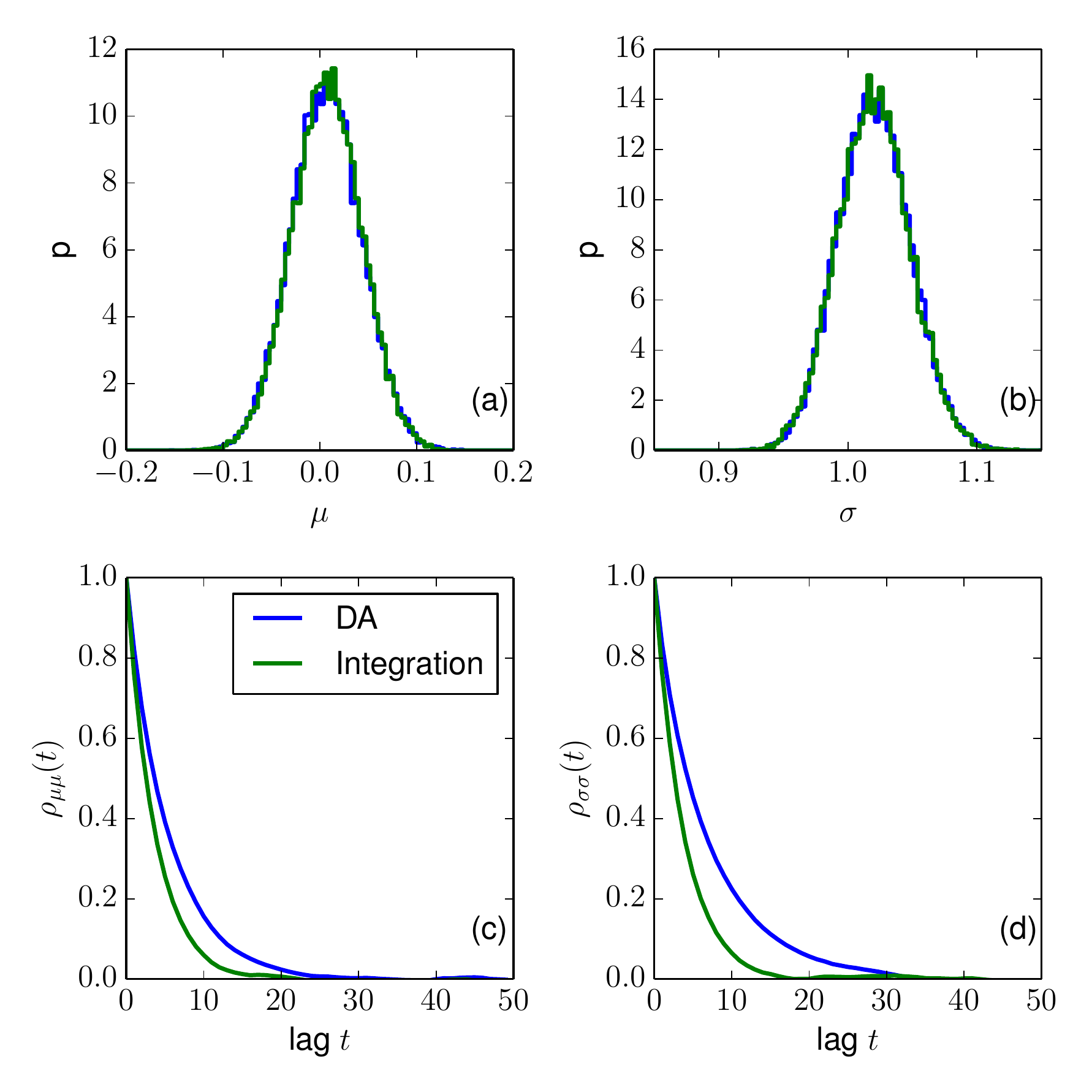}
\caption{Comparison of two methods to handle nuisance
  parameters. Here the nuisance parameter is the
  true coordinate which is related to the observed
  coordinate via a given uncertainty. In the DA algorithm, the
  nuisance parameter is sampled alongside other parameters
  using Gibbs sampling on a Bayesian hierarchical model.  
  In the other method, the nuisance
  parameter is marginalized via integration; an analytical form 
  of the marginalized likelihood is used. The estimated 
  parameters are $\mu$ and $\sigma$.  
  The panels (a) and (b) show the probability distribution
  function   of the parameters given the data.   
  The panels (c) and (d)
  show the autocorrelation function of the two parameters in
 their respective Markov chains. 
\label{fig:da_demo}}
\end{figure}
For this simple case, the integral in \equ{pthetax1} leads
to an analytical expression
\be
p(\theta|X,\sigma_X) \propto p(\theta) \prod_i \mathcal{N}(x_i|\mu,\sigma_{{\rm
  tot},i}^2), \textrm{ where } \sigma_{{\rm tot},i}=\sqrt{\sigma^2+\sigma_{x,i}^2}.
\label{equ:pthetax3}
\ee
We used  $(\mu,\sigma)=(0.0,1.0)$ and $\sigma_x=0.5$ to
generate test data and then estimated $\mu$ and $\sigma$ 
using two schemes: (1) DA algorithm which uses \equ{pthetax2} 
and treats $X_t$ as unknown and samples from it,  and (2)
explicit integration scheme which uses  \equ{pthetax3} 
where the variable $x_i^t$ has been integrated out of the 
equation. The Markov chain was run for 100,000 iterations.  
\fig{da_demo}$a,b$ shows the pdf of the 
estimates of the two parameters. Both schemes give 
identical results.  The autocorrelation function  
for the two parameters are shown in 
\fig{da_demo}$c,d$. The DA algorithm has a slightly higher
autocorrelation time $\tau$ as it has to sample an 
extra parameter for each data point.

\section{Case studies in astronomy}\label{sec:casestudy}
In this section, we study a range of cases in astronomy where 
MCMC based Bayesian analysis is making a significant impact. 
The emphasis is on showing how to set up 
a diverse range of problems within the Bayesian framework 
and how to solve them using MCMC techniques. 
The examples are intentionally chosen from different areas 
of astronomy so as to demonstrate the ubiquity of the techniques 
reviewed here. There is a long history of applying
 such techniques 
in the field of cosmology, and excellent reviews and books 
already exist here:
\citet{2008ConPh..49...71T,hobson2010bayesian,2013arXiv1302.1721P}.  

\subsection{Exoplanets and binary systems using radial
  velocity measurements}
\begin{figure}
\centering 
\begin{minipage}{6cm}
\centering 
\includegraphics[width=6cm]{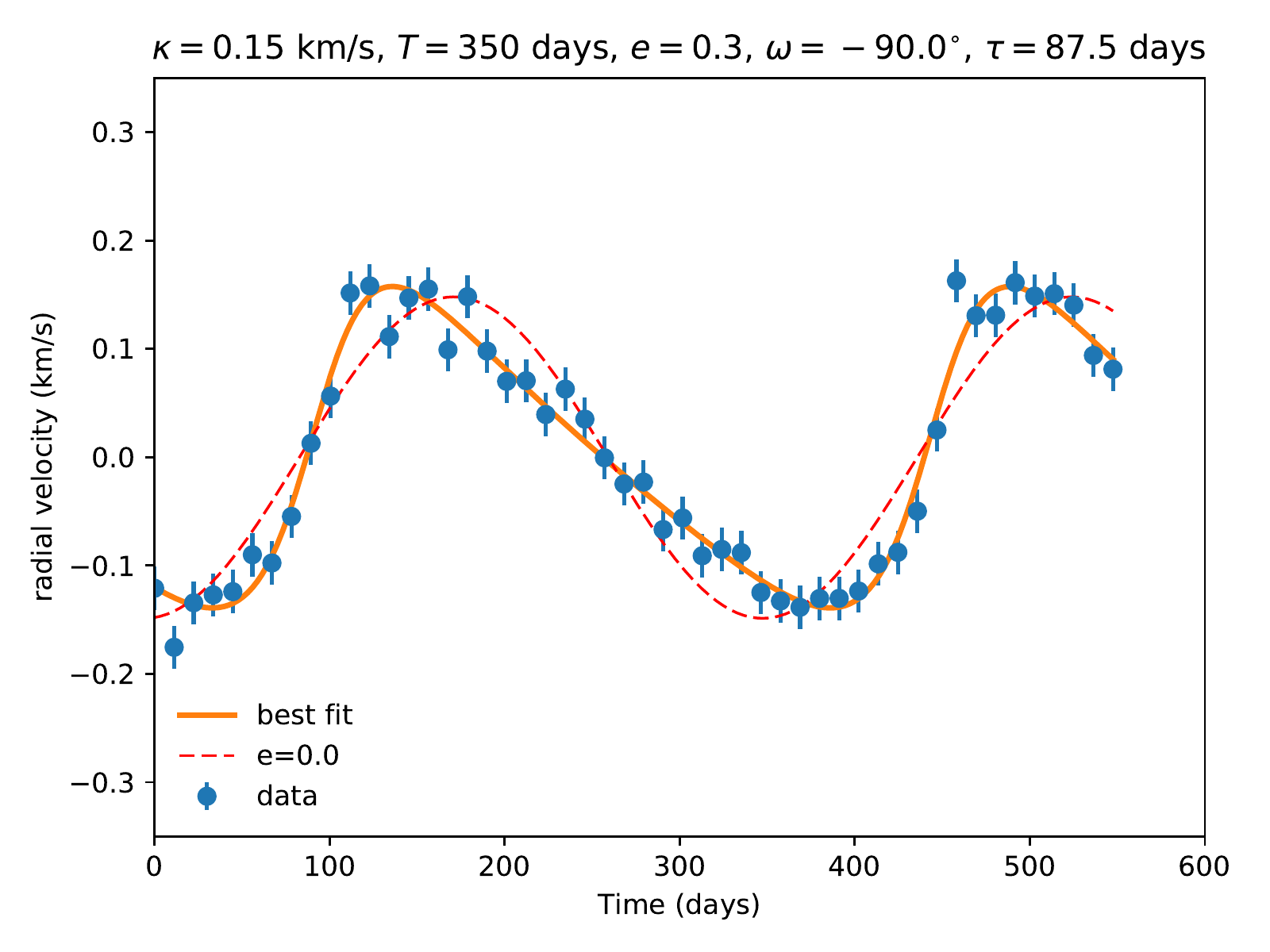}
\end{minipage}
\centering 
\begin{minipage}{6cm}
\centering 
\includegraphics[width=6cm]{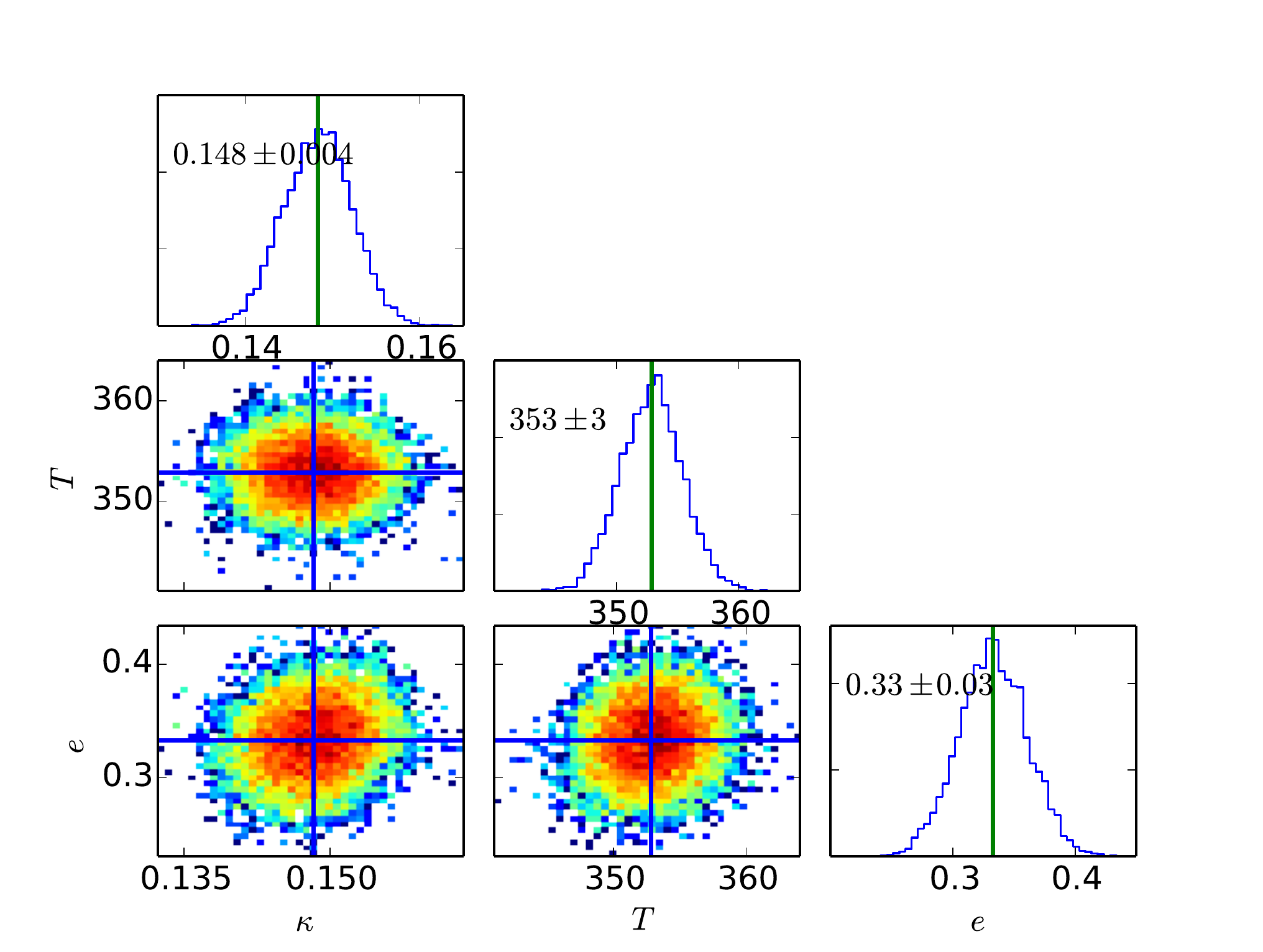}
\end{minipage}
\caption{Left: Radial velocity as a function of time for a star in
  a binary system. 
The parameters of the binary system are listed on the top.
The green line is the best fit solution 
  obtained using an MCMC simulation. The red and the green curves are generated from
  \equ{radial_vel} and differ only in the eccentricity $e$. 
  The plot shows that the shape of the radial velocity curve 
  depends sensitively upon the eccentricity $e$ of the
  orbit. Right: The posterior probability distribution of parameters
  obtained using the MCMC simulation.}
\label{fig:rv_mcmc}
\end{figure}

The presence of a planet or a companion star results in temporal
variations in the radial velocity of the host star. By 
analyzing the radial velocity data, one can draw inferences  
about the ratio of masses between the host and the companion, 
and orbital parameters like the period and eccentricity. 
We now describe how to set up the above inference problem in a 
Bayesian framework. We begin by describing the predictive
model for the radial velocity of a star in a binary system. 

The radial velocity of a star of mass $M$ in a binary system
with companion of mass $m$ in an orbit with time period
$T$, inclination $I$ and eccentricity $e$ is given by
\be
v(t)=\kappa \left[\cos(f+\omega)+e\cos
  \omega\right]+v_{0},\:\:{\rm with\ } \kappa=\frac{(2\pi G)^{1/3}m \sin I
}{T^{1/3}(M+m)^{2/3}\sqrt{1-e^2}}.
\label{equ:radial_vel}
\ee
The true anomaly $f$ is a function of time, but depends upon $e$,
$T$, and $\tau$ via, 
\be
\tan(f/2)=\sqrt{\frac{1+e}{1-e}}\tan(u/2), \quad  u-e\sin u=\frac{2\pi}{T}(t-\tau).
\ee
An example of radial velocity data is shown in 
\fig{rv_mcmc} which shows the radial velocity for two 
binary systems (the green and the red line) that differ in $e$ 
but have same values for all other parameters
$\kappa,T,\tau,\omega$ and $v_0$. The figure demonstrates that the radial 
velocity is sensitive to the eccentricity of the orbit.  
\begin{marginnote}
\entry{$v_{0}$}{the mean velocity of the center of mass of
the binary system}
\entry{$I$}{the inclination of the orbital plane with
respect to the sky (angle between orbital angular momentum and
line of sight)}
\entry{$\omega$ }{the angle of the pericenter
measured from the ascending node (the point where the orbit
intersects the plane of the sky)} 
\entry{$\tau$ }{time of passage through the pericentre}
\end{marginnote}

The actual radial velocity data will differ from the 
perfect relationship given in \equ{radial_vel} due to 
observational uncertainty (variance $\sigma_v^2$) and 
intrinsic variability of a star (variance $S^2$) and we can
model this by a Gaussian function
$\mathcal{N}(.|v,\sigma_v^2+S^2)$. 
For radial velocity data $D$ defined as a set of radial velocities $\{v_1,...,v_M\}$ at various
times $\{t_1,...,t_M\}$, one can fit and constrain seven 
parameters,  $\theta=(v_{0}, \kappa, T, e, \tau, \omega, S)$,
using the Bayes theorem as shown below   
\be
p(\theta|D) \propto p(D|\theta) p(\theta) \propto p(\theta) \prod_{i=1}^{M} \mathcal{N}(v_i|v(t_i;\theta),\sigma_v^2+S^2).
\ee
We generated test data using \equ{radial_vel} and then,
using the above equation, we tried to recover the parameters
$\theta$ (available in the supplied software). 
The posterior distribution $p(\theta|D)$ was
sampled using MCMC, and the results are shown in
\fig{rv_mcmc}. Panel $a$ shows the test data along with 
the best fit curve. It also shows
the radial velocity for the case with $e=0$. Panel $b$
shows the posterior distribution of the parameters
$\kappa, T$ and $e$.

If we have data for a large number of binary systems,
we can use it to explore the distribution of 
orbital parameters. A naive way to do this would be 
to get a ``maximum {\it a posteriori}'' (MAP) estimate of the 
orbital parameters for each star and then 
study the population distribution by constructing 
histograms out of it. Such a scheme will give incorrect 
estimates of the population distribution as the 
uncertainty associated with the parameter estimates
is ignored. In addition to this, 
as discussed by \cite{2010ApJ...725.2166H}, the MAP 
estimates are in general biased. In the context of 
radial velocity data, the estimates of $e$ are biased 
high. The problem is especially 
acute if the uncertainty associated with the parameters 
is large, which is often the case with 
radial velocity data from barycentric motions.

All of these problems can be avoided by setting up the 
problem of estimation of population distributions as 
a hierarchical Bayesian model. 
Let us suppose we have radial velocity
data for $N$ binary star systems, and denote by $y_i$ the radial 
velocity data set for the $i$-th system. Let $x_i=(v_{0i},
\kappa_i, T_i, e_i, \tau_i, \omega_i, S_i)$ be the orbital
parameters for the $i$-th system. Finally, let $\alpha$ 
be the set of hyperparameters that govern the 
population distribution of the parameters $x$. The problem 
to determine $\alpha$ can be set up as 
\be
p(\alpha,\{x_i\}|\{y_i\}) \propto p(\alpha) \prod p(y_i|x_i)
  p(x_i|\alpha) 
\ee
This is a BHM and  
can be sampled using the Metropolis-within-Gibbs scheme 
discussed in \sec{gsem}. The parameters $x_i$ can be
estimated alongside $\alpha$, and to get the marginal
distribution $p(\alpha|\{y_i\})$, one can simply ignore 
the computed $x_i$.  

However, the above scheme is not well suited 
to explore a variety of population models, 
especially if sampling from $p(y_i|x_i) p(x_i|\alpha)$ is  
computationally demanding. 
We now show a computationally efficient 
scheme by \citet{2010ApJ...725.2166H} 
that can in general be applied to BHMs of two levels. 
The marginal distribution of hyperparameters that 
we are interested in is given by 
\be
p(\alpha|\{y_i\}) \propto p(\alpha) \prod \int dx_i
p(y_i|x_i) p(x_i|\theta)
\ee
The integral on the right hand side can be estimated using 
a Monte Carlo integration scheme as follows:
\be
\int dx_i p(y_i|x_i) p(x_i|\alpha) = \int dx_i
p(y_i|x_i) p(x_i) \frac{p(x_i|\alpha)}{p(x_i)} =
\frac{1}{K}\sum_{k=1}^{K}
\frac{p(x_{ik}|\alpha)}{p(x_{ik})}, 
\ee
with $x_{ik}$ sampled from $p(x_i|y_i) \propto p(y_i|x_i)
p(x_i)$, which can be done by an MCMC scheme.

\subsection{Data driven approach to estimation of stellar parameters from a spectrum}
\begin{figure}
\centering 
\includegraphics[width=0.9\textwidth]{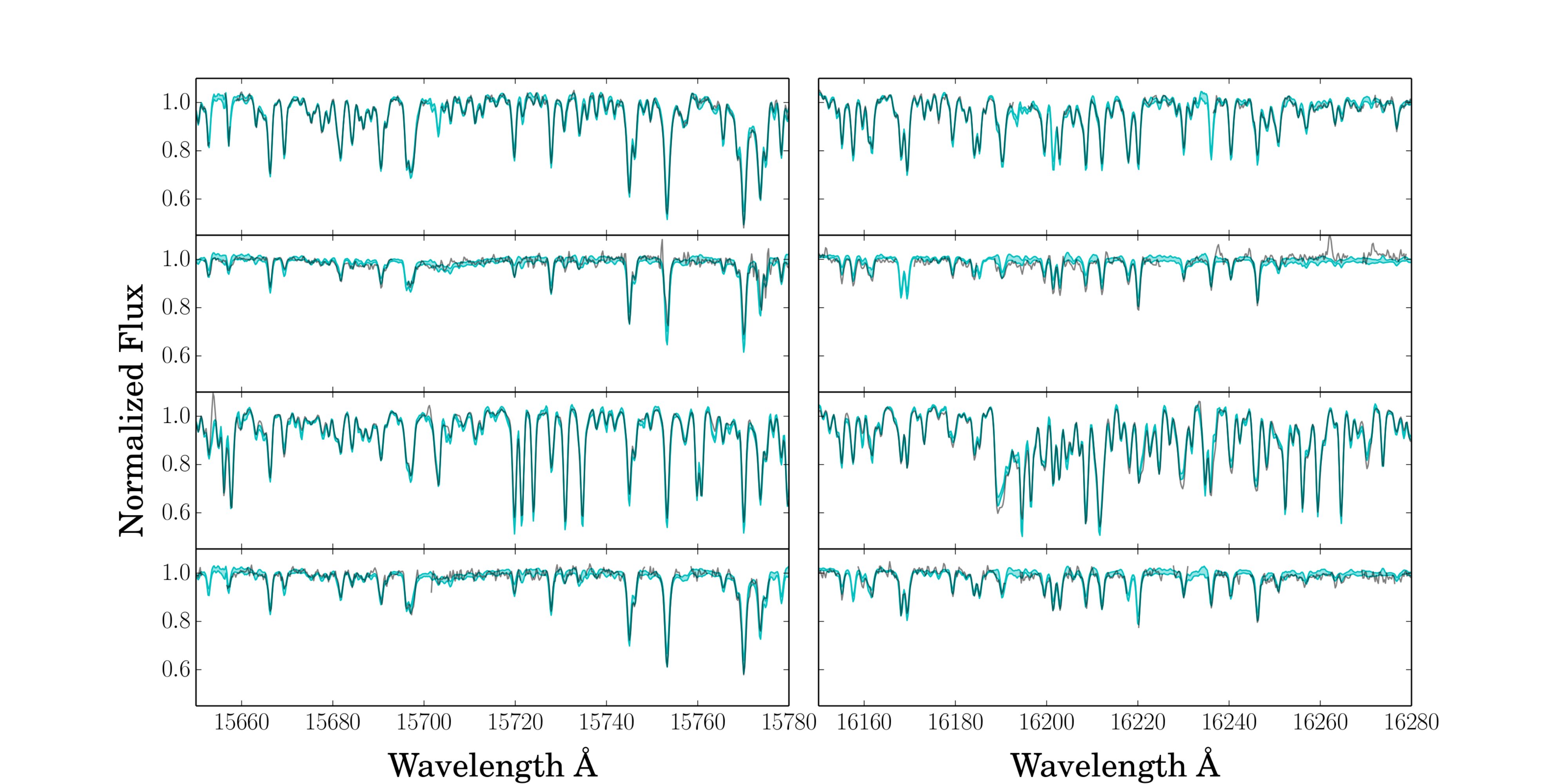}
\caption{The APOGEE spectra of four stars (black line) along with the best
  model spectra generated by {\it The Cannon} algorithm
  (cyan line) along with scatter around the fit. Each row
  shows the spectra of a single star in
  two wavelength intervals (left and right). Image from \citet{2015ApJ...808...16N}.}
\label{fig:data_model_cannon}
\end{figure}
The spectrum of a star contains information about its 
properties like temperature, gravity and the abundance of
different chemical elements that make up the star. 
Decoding information about stellar parameters 
from a stellar spectrum is a problem of great significance
for astronomy. With the advent of large spectroscopic
stellar surveys having several hundred thousand spectra, 
the need for fast and accurate methods to analyze the 
stellar spectra has gained prominence. 
Let us denote the stellar parameters 
(e.g., $T_{\rm eff},\log g,{\rm [Fe/H]},$ and ${\rm [X/Fe]}$)
by label vector ${\bf x}=(x_1,...,x_K)$ and 
the observed spectrum by vector ${\bf
  y}=\{y_1,...,y_L\}$, denoting normalized flux at 
specific wavelengths indexed by ${\bf \lambda}=(1,...,L)$  
(see \fig{data_model_cannon}). 
The problem is to find ${\bf x}$ given ${\bf y}$, which
using the Bayes
theorem can be written down as
\be
p({\bf x}|{\bf y},\theta) \propto p({\bf y}|{\bf x},\theta)p({\bf
  x}).
\label{equ:cannon_bayes}
\ee
Here
$p({\bf y}|{\bf x},\theta)$ denotes a probabilistic generative model 
for the data, with $\theta$ being the parameters of the
model. If we denote by $f_{\lambda}({\bf x},\theta_{\lambda})$ the flux
predicted by the model at wavelength $\lambda$ and by
$s_{\lambda}^2$ the variance or scatter about this relation 
(assuming Gaussian noise), then the probabilistic generative 
model for the full spectrum can be written as  
\be
p({\bf y}|{\bf x},\theta)=\prod_{\lambda=1}^L p(y_{\lambda}|{\bf x},\theta_{\lambda},s_{\lambda})=\prod_{\lambda=1}^L\mathcal{N}(y_{\lambda}|f_{\lambda}({\bf x},\theta_{\lambda}),s_{\lambda}^2)
\label{equ:cannon_genmod}
\ee 
Traditionally, $f_{\lambda}({\bf x},\theta_{\lambda})$ is calculated
from first principles using a physical theory for  the 
formation of spectral lines in a stellar atmosphere 
specified by stellar parameters ${\bf x}$. Frequently, 
$f_{\lambda}({\bf x},\theta_{\lambda})$ is evaluated on a grid defined
on $x$ and then interpolation is used to get the spectrum 
for any arbitrary value of ${\bf x}$. 
The $f_{\lambda}({\bf x},\theta_{\lambda})$ can also be computed 
by interpolating over a library of empirical spectra with predefined 
stellar parameters. A more refined data 
driven approach to the problem 
using machine learning techniques was presented in 
\citet{2015ApJ...808...16N}. 
In this approach, $f_{\lambda}({\bf
  x},\theta_{\lambda})$ is approximated by a simple 
(linear or quadratic) function of label vector ${\bf x}$. Therefore
\be
f_{\lambda}({\bf  x},\theta_{\lambda})=\theta_{\lambda
  0}+\sum_{i=1}^{K} \theta_{\lambda i}x_{i} +\sum_{i=1}^{K}\sum_{j=1}^{K}
\theta_{\lambda ij}x_ix_j.
\ee
Let us consider a training set of $N$ stars 
with label vectors $X=\{{\bf x}^1,...,{\bf x}^N\}$ 
and corresponding set of fluxes at wavelength $\lambda$ 
by $Y_{\lambda}=\{y_{\lambda}^1,...,y_{\lambda}^N\}$. 
One can estimate $\theta_{\lambda}$ by sampling within
MCMC such that 
\be
p(\theta_{\lambda},s_{\lambda}|X,Y_{\lambda}) \propto
p(Y_{\lambda}|X,\theta_{\lambda},s_{\lambda})p(\theta_{\lambda})p(s_{\lambda})
\propto p(\theta_{\lambda})p(s_{\lambda})\prod_{i=1}^{N}
p(y_{\lambda}^{i}|{\bf x}^{i},\theta_{\lambda}s_{\lambda}). 
\ee
Having obtained the model parameters 
$\theta=\{\theta_1,...,\theta_L,s_1,...,s_L\}$, one can now
estimate stellar parameters  ${\bf x}$ of a new star with 
given spectrum ${\bf y}$ using \equ{cannon_bayes}. 
This is the basis of {\it The Cannon} algorithm
(Ness et al. 2015) which is already widely used by
the stellar community.
The ability of the algorithm to model the spectra is
demonstrated  in \fig{data_model_cannon} which shows the 
spectra of fours stars along with the best-fit spectra for
each of them.

\subsection{Solar-like oscillations in stars}
\begin{figure}
\centering 
\includegraphics[width=0.9\textwidth]{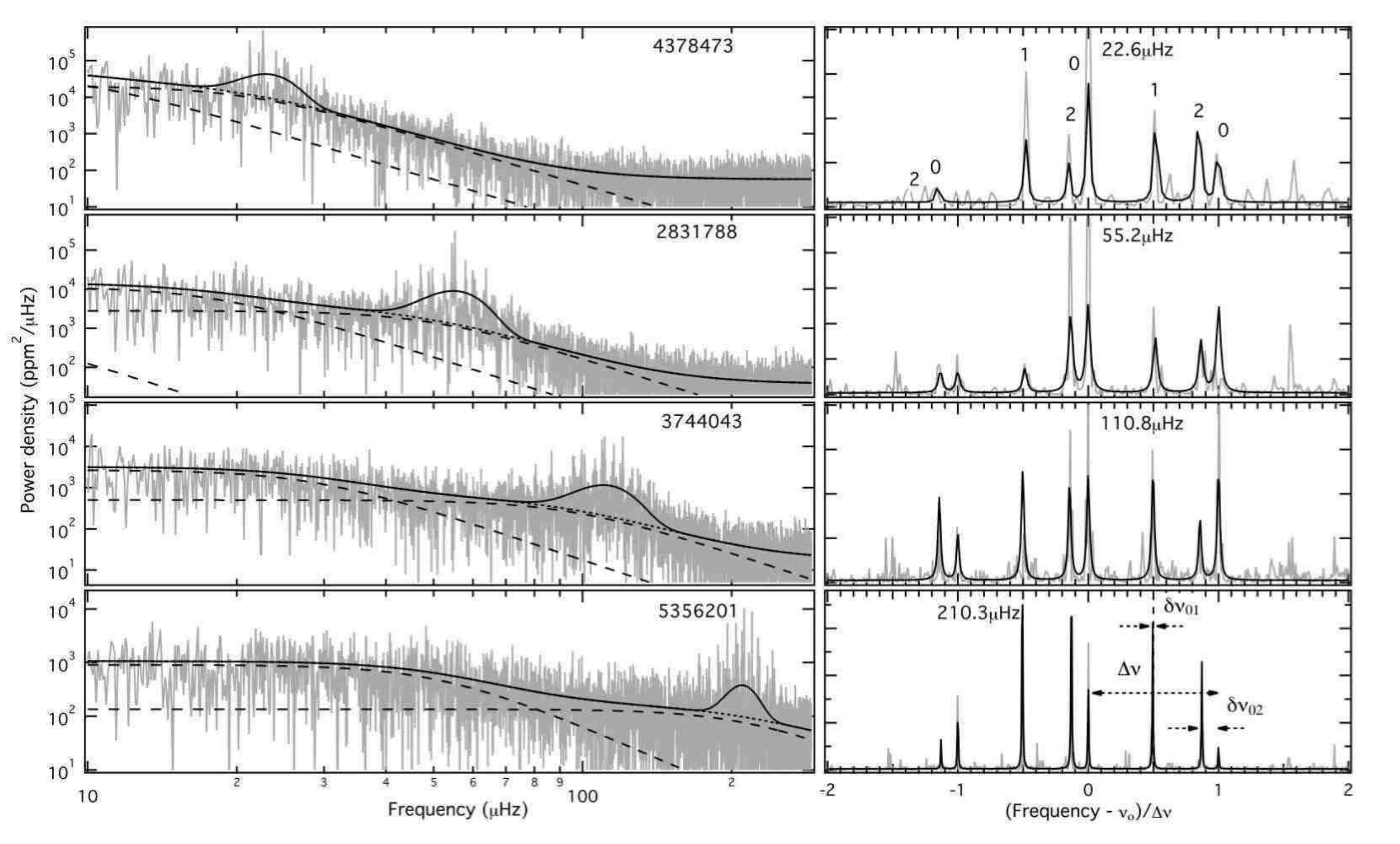}
\caption{Left: The power density of four stars observed with
  {\it Kepler} showing solar-like
  oscillations along with best fit model (solid black). The
  hump is the approximate Gaussian-like envelope that
  modulates the power spectrum. The dashed lines
  are the individual super-Lorentzian profiles. The dotted
  line is the background model without the Gaussian component. Right: The
  spectrum region around central frequency $\nu_0$ after subtracting the
  background model. Individual modes are clearly visible. The degree $l$
  of the modes is labelled in the top panel. The frequency
  separations $\Delta \nu, \delta_{01}$ and $\delta_{02}$ 
  are shown using dotted lines in the bottom panel. Image
  from \citet{2010A&A...522A...1K}.}
\label{fig:kallinger}
\end{figure}

Solar-like oscillations, which are excited and damped in
the outer convective envelopes of a star, are seen in stars 
like the Sun and red giants. With the advent of space-based 
missions like {\it Kepler} and COROT that provide high 
quality photometric data over a long time series, it has now 
becomes feasible to detect solar-like oscillations in 
tens of thousands of stars \citep{2013ApJ...765L..41S, 2015ApJ...809L...3S}.
Typically, the power spectrum of a star with solar-like
oscillations (\fig{kallinger}) shows a regular pattern of 
modes, characterized by a large 
frequency separation $\Delta \nu$. The overall amplitude 
is modulated by a Gaussian envelope and this is 
characterized by the frequency of maximum oscillation $\nu_{\rm max}$. 
Theory suggests that $\Delta \nu$ for a given star is related to its density
\citep{1986ApJ...306L..37U}, whereas the $\nu_{\rm
  max}$ is related to its surface gravity and
temperature \citep{1991ApJ...368..599B,1995A&A...293...87K}. 
Using the above two relations, the mass and the radius of a
star can be constrained. 
The mass of a red giant is sensitive to its age 
and this makes asteroseismology very useful for understanding
Galactic evolution \citep{2011Sci...332..213C, 2016ApJ...822...15S}.
For further details on solar type oscillations see review 
by \citet{2013ARA&A..51..353C}. 

Bayesian-MCMC based techniques are increasingly being
adopted to extract seismic properties, e.g., $\Delta \nu$ 
and $\nu_{\rm max}$, by analyzing the power spectrum 
generated from the time series photometry of a star 
\citep{2009A&A...506.1043G,2010A&A...522A...1K,2011A&A...527A..56H}.  
The probability that an observed power spectrum
$\mathbf{\Gamma}=\{\Gamma_1,...,\Gamma_N\}$ at frequencies $\mathbf{\nu}=\{\nu_1,...,\nu_N\}$ is
produced by a model spectrum  $\Gamma(\nu;\theta)$ (specified by a set of
parameters $\theta$), is given by
\be
p(\mathbf{\Gamma}|\theta)=\prod_{i=1}^{N} \frac{1}{\Gamma(\nu_i;\theta)}\exp\left(-\frac{\Gamma_i}{\Gamma(\nu_i;\theta)}\right)
\ee
as shown by \citet{duvall1986solar}. This forms the basis
for the Bayesian treatment of the problem of estimation of
parameters $\theta$ by 
$p(\theta|\{\Gamma_i\})=p(\{\Gamma_i\}|\theta)p(\theta)$. 
The power density is modelled as a sum of super-Lorentzian
functions  
\be
\Gamma(\nu;\theta)=\Gamma_{\rm
  wn}+\sum_k\frac{A_k}{1+(2\pi\nu\tau_k)^{c_k}}+P_{\rm
  g}\exp\left(\frac{-(\nu_{\rm max}-\nu)^2}{2\sigma_{\rm g}^2}\right) 
\ee
To fit the individual modes, one assumes Lorentzian
profiles. Spherical harmonics are used to describe the
oscillations; the modes are characterized 
by three wave numbers, $n,l$ and $m$.
In  \citet{2010A&A...522A...1K}, 
eight main modes are fitted 
(three $l=0$ and $l=2$ and two $l=1$), parameterized by 
the mode lifetime $\tau$, the central frequency $\nu_0$, three
spacings  $\Delta \nu, \delta\nu_{01}$ and
$\delta\nu_{01}$, and the amplitudes $A_i,A_j$ and $A_k$.
\be
\Gamma(\nu)&=&P_{\rm wn}+\sum_{i=-1}^{1}
\frac{A_i^2\tau}{1+4[\nu-(\nu_0+i\Delta\nu)]^2(\pi\tau)^2} 
+\sum_{j=-1}^{1}\frac{A_j^2\tau}{1+4[\nu-(\nu_0+j\Delta\nu-\delta\nu_{02})]^2(\pi\tau)^2}\nonumber\\
& & +\sum_{k=-1,1} \frac{A_j^2\tau}{1+4[\nu-(\nu_0+k\Delta\nu/2-\delta\nu_{01})]^2(\pi\tau)^2}
\ee
\fig{kallinger} shows the result of fitting the above model
to power spectra of fours stars observed by the {\it Kepler}
mission.

\subsection{Extinction mapping and estimation of intrinsic
  stellar properties} 
Given the mass $m$ and initial composition (e.g., metallicity
[M/H]) of a star, we can use the 
theory of stellar evolution to predict its state and
composition at a later time (age $\tau$). 
However, the
intrinsic parameters like mass $m$, [M/H] and
$\tau$ are not directly observable. For most stars 
we only have photometric information, apparent 
magnitudes in different photometric bands (for example $J,
Ks, u, g, r$ and $i$). The photometry 
of a star depends upon temperature $T_{\rm eff}$, 
gravity $g$, [M/H], distance $s$ and extinction $E$
(proportional to the dust density integrated along the line
of sight to the location of the star). 
If we have spectroscopy, then we can get temperature $T_{\rm eff}$, 
$g$ and even composition, but with uncertainties. From asteroseismology,
we can get average seismic parameters like $\Delta \nu$ 
and $\nu_{\rm max}$, which are sensitive to the mass, radius 
and temperature of a star. Given this state of affairs, it 
is quite common to ask the question that, given a
certain set of observables of a star, what are the intrinsic
parameters of a star or even some other set of observables. 
For example, given the photometry of a star, what is the 
distance, temperature and gravity of a star; or given 
photometry and distance, what is the temperature and gravity 
of a star; or given photometry and spectroscopy, what is the 
distance? And so on. 
Knowing the intrinsic parameters of a star is also important 
for understanding the formation and evolution of the Galaxy, 
for example, the star formation rate, 
the age-metallicity relation and the distribution of dust
in the Galaxy. 

\begin{figure}
\centering 
\begin{minipage}{6cm}
\centering 
\includegraphics[width=6cm]{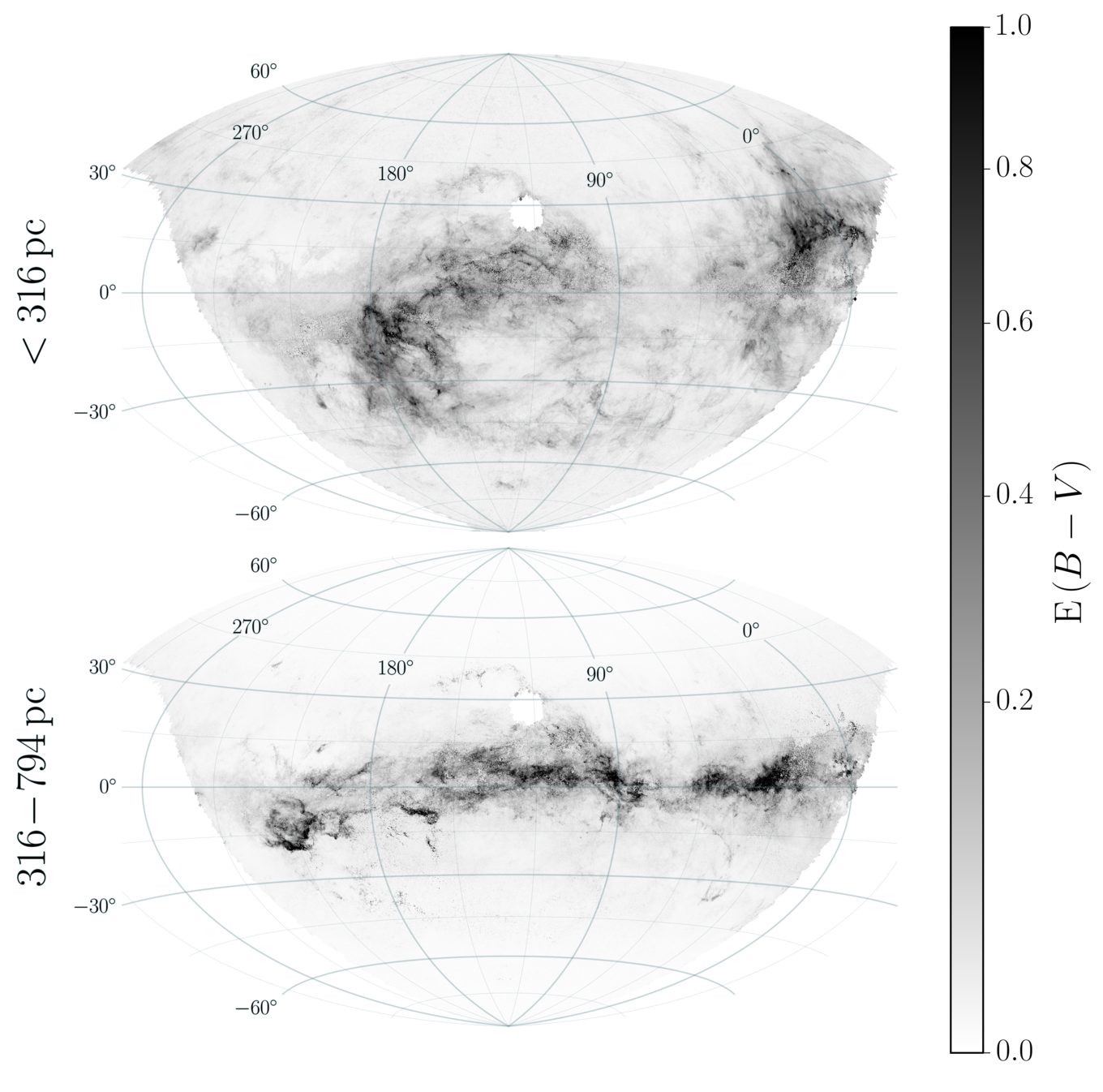}
\end{minipage}
\begin{minipage}{6cm}
\centering 
\includegraphics[width=6cm]{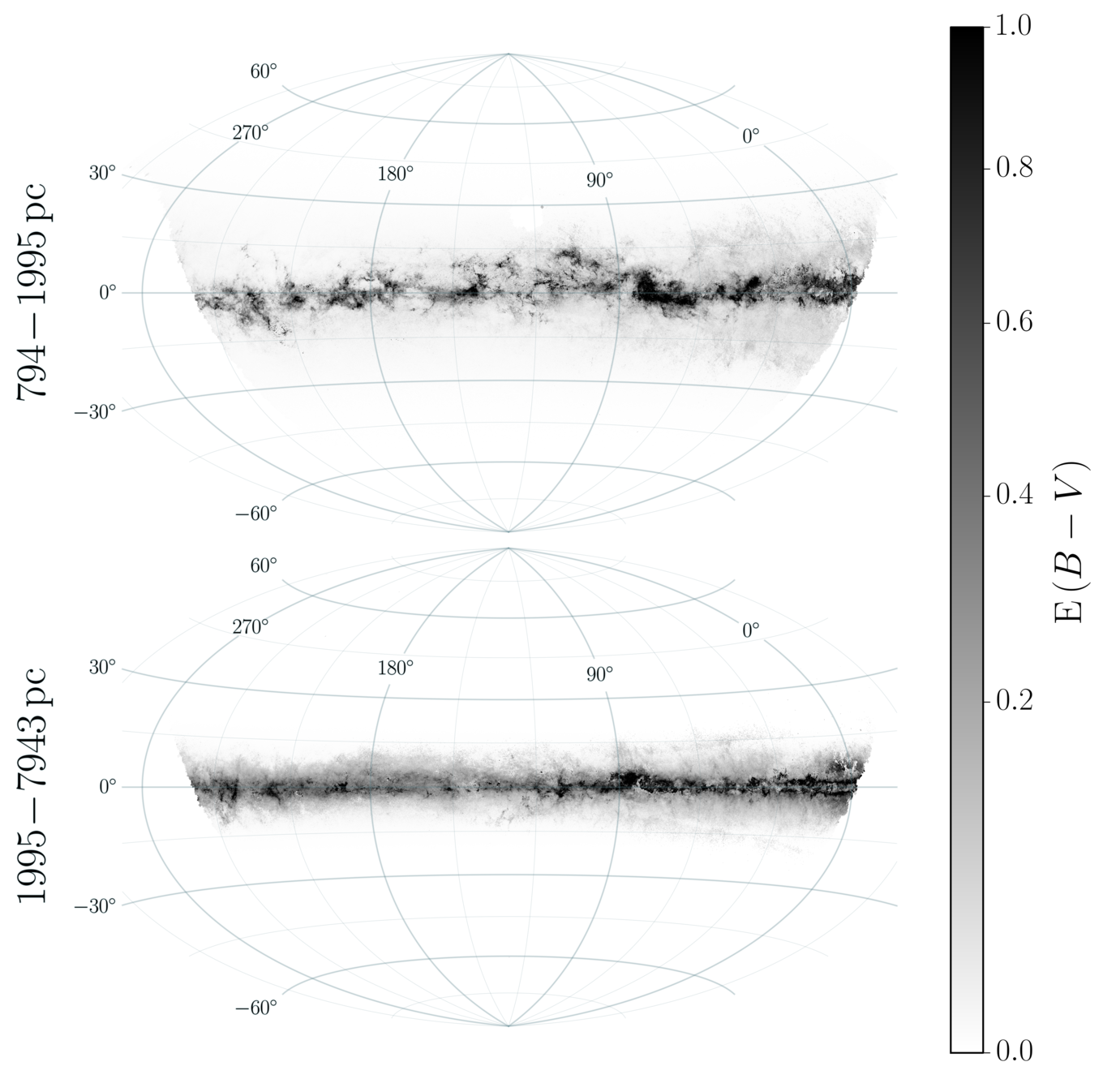}
\end{minipage}
\caption{A three-dimensional map of interstellar dust
  reddening in the Milky Way based on Pan-STARRS 1 and
  2MASS photometry. Shown are the mean differential reddening in different
  heliocentric distance ranges. The map is produced by \citet{2015ApJ...810...25G} and is available at http://argonaut.skymaps.info.
}
\label{fig:3d_ext_green}
\end{figure}

The problem of estimating intrinsic stellar parameters 
of a star given some observables can be formulated as
follows. Let 
${\bf y}=(J,J-Ks,J-H,T_{\rm eff},\log
g,[M/H]_{\rm obs},l,b)$
be a set of observables associated with a star
and $\sigma_{\bf y}$ their uncertainties.  
Let us denote the intrinsic variable of a star  
that we are interested in by
${\bf x}=([M/H],\tau,m,s,l,b,E)$.  
To specify prior probabilities on ${\bf x}$
we need a Galactic model, and we denote by $\theta$ the 
parameters of such a model. 
Typically, real catalogs have selection effects,
e.g., stars selected to lie in some apparent magnitude 
and color range, or a set of stars with parallax error 
less than 10\%, or stars with missing information in 
certain bands. To specify selection effects,  
we denote the event that a star exists in a catalog 
by $S$.
From theoretical isochrones we can predict ${\bf y}$ 
given ${\bf x}$, in other words a function 
${\bf y(x)}$ exists.
However, we are interested in the inverse problem of 
estimating ${\bf x}$ given ${\bf y}$. A Bayesian 
introduction to solving such a problem was given by 
\citet{2004MNRAS.351..487P} and \citet{2005A&A...436..127J} in the
context of estimating ages. The method was further
improved and refined by 
\citet{2010MNRAS.407..339B,Binney_Burnett_2011}
and \citet{2014MNRAS.437..351B} in the context of the estimation of
distances, with a better treatment of priors and selection
effects \citep[see also][]{2012MNRAS.427.2119S,2015MNRAS.452.2960S}.  
From the Bayes theorem we have 
\be
p({\bf x| y,\sigma_y},S,\theta) \propto p(S,{\bf
  y| x,\sigma_y}) p({\bf x}|\theta) \propto
p(S|{\bf y})p({\bf y}|{\bf x,\sigma_y})p({\bf x}|\theta).
\ee
We now explain each of the terms in detail.
\begin{enumerate}
\item $p({\bf x|y ,\sigma_y},S,\theta)$ is the {\it
  posterior} distribution of intrinsic parameters given the
  observables, the selection function and a Galactic model.  
\item $p(S|{\bf y,x,\sigma_y})$ is the {\it selection function}. This says
  given the observables what is probability that a star was
  observed. Typically this can be expressed as $p(S|{\bf
    y})p(S|{\bf x})$. The term $p(S|{\bf x})$ enters in
  situations where the value of an observable $y'$ is not known
  but constraints on it are. Then $p(S|{\bf x})=\int
  p(S|y')p(y'|x)dy'$. For example, a parallax of a star is 
  known to be greater than a certain limit, or the apparent 
  magnitude of a star may be missing in a band because the
  star is too bright or faint
  \citep{2010MNRAS.407..339B, 2012MNRAS.427.2119S}.     
\item $p({\bf y|x,\sigma_y})$ is the {\it likelihood} of the data given
  the uncertainty and the intrinsic parameters. This can be 
described by a Gaussian function $\mathcal{N}(y|y({\bf
  x}),\sigma_y^2)$ for each $y\in {\bf y}$. 
\item $p({\bf x}|\theta)$ is the {\it prior}. This describes the
  distribution of mass, metallicity, age and spatial distribution of
  stars in the Galaxy. More specifically it can be
  written as 
$p(x|\theta)=\sum_k p_k(m) p_k([M/H])p_k(\tau)p_k(r)$,
where the sum is over different Galactic components, e.g., 
thin disc, thick disc, bulge and stellar halo. 
\end{enumerate}
We now focus on the problem of estimating distance and
extinction. For simplicity, we ignore the selection effects;
for an in depth discussion, see    
\citet{2015MNRAS.452.2960S}. 
By marginalizing over stellar parameters $\tau,m$ and
$[M/H]$ one obtains $p(s,E|{\bf y,\sigma_y},\theta)$. If we have $N$ stars along a line of sight, we can estimate the distance-extinction relationship $E(s_i;\alpha)$ parameterized by $\alpha$ as 
\be
p(\alpha|\{{\bf y}\},\theta) & \propto & p(\alpha) \prod_{i=1}^{N}
\int d E_{i} ds_i\ p(s_i,E_i|{\bf y}_i,\sigma^i_{\bf y},\theta)\:
p(E_i|s_i,\alpha). 
\ee
The above method is used by 
\citet{2014ApJ...783..114G,2015ApJ...810...25G}, to
construct three dimensional maps of interstellar dust
reddening using Pan-STARRS 1 and 2MASS photometry
(\fig{3d_ext_green}). To estimate $p(s,E|{\bf
  y,\sigma_y},\theta)$, \citet{2015ApJ...810...25G} 
do a kernel density estimate  over samples generated by
MCMC, while \citet{2015MNRAS.448.1738S} 
present a method based on the Gaussian mixture model. 
As described in \citet{2012MNRAS.427.2119S}, we can 
also directly estimate $\alpha$ and intrinsic
parameters ${\bf x}$ of each star along a line of sight by setting 
up the problem as a BHM and sampling 
from the following posterior:
\be
p(\alpha,\{\bf x\}|\{{\bf y}\},\sigma_{\bf y},\theta) & \propto & p(\alpha) \prod_{i=1}^{N}
p({\bf y}_i|{\bf x}_i,\theta,\sigma_{\bf y}^i,\alpha) p({\bf
  x}_i|\theta) p(S|{\bf y}_i).
\ee  
The Metropolis-within-Gibbs scheme is used to
accomplish this sampling.

\subsection{Kinematic and dynamical modelling of the Milky Way} 
Understanding the origin and evolution of the Milky Way 
has received significant boost due to the emergence of large
data sets that catalog the properties of stars in the Milky
Way \citep{2011Prama..77...39B,2012MNRAS.419.2251M,2013MNRAS.433.1411M, 2013A&ARv..21...61R,  2013NewAR..57...29B,
  2016arXiv160207702B1}. 
Bayesian methods and MCMC based schemes are
now playing a prominent role in the analysis and interpretation 
of such large and complex data sets from, e.g.,  the GCS
survey \citep{2010MNRAS.403.1829S}, the SEGUE
survey \citep{2012ApJ...753..148B},  the APOGEE survey  
\citep{2012ApJ...759..131B,2013ApJ...779..115B},
and the RAVE survey \citep{2014ApJ...793...51S,2014MNRAS.445.3133P,
  2015MNRAS.449.3479S}. We focus on the problem of 
determining the mass distribution, or equivalently the
gravitational potential of the Milky Way, using 
halo stars \citep{2014ApJ...794...59K} and disc masers \citep{2017MNRAS.465...76M}. 

\begin{figure}
\centering 
\begin{minipage}{6cm}
\centering 
\includegraphics[width=5cm]{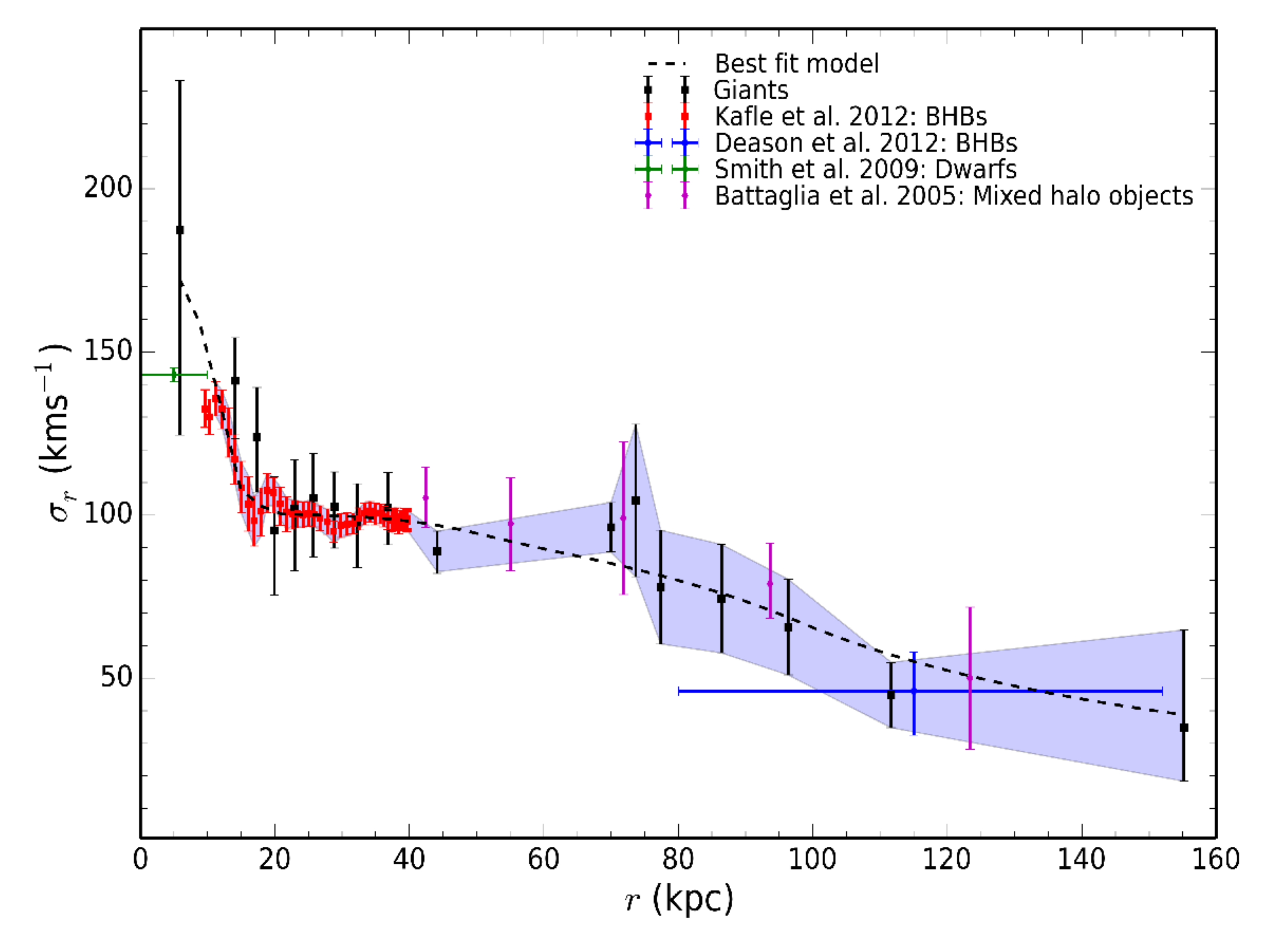}
\end{minipage}
\begin{minipage}{6cm}
\centering 
\includegraphics[width=5cm]{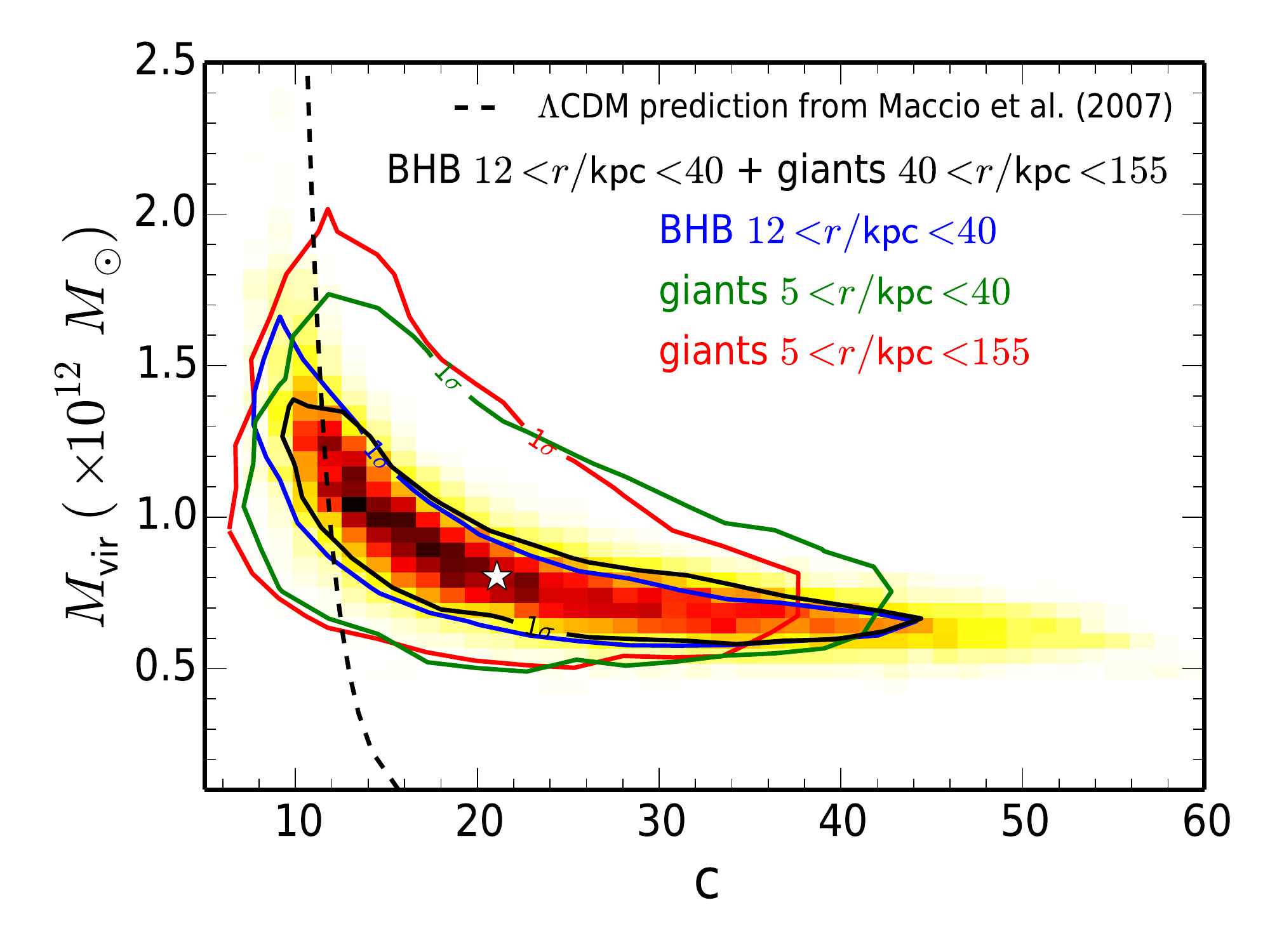}
\end{minipage}
\caption{$(a)$ Radial velocity dispersion as a function of radius
 for halo stars in the Milky Way. $(b)$
  Posterior distribution of virial mass and the 
  concentration parameter of the Milky Ways dark mater
  halo. Adapted from \citep{2014ApJ...794...59K} with
  permission. Abbreviation: BHB, blue horizontal branch}
\label{fig:mvir_c_kafle}
\end{figure}

The observational data of stars in the Milky Way is in 
heliocentric coordinates and is in the
form of angular positions on sky
(Galactic longitude $\ell$ and latitude $b$), heliocentric  
distance ($s$), heliocentric line of sight velocity 
($v_{\rm los}$), and proper motion (tangential motion on the
sky, $\mu_\ell$ and $\mu_b$). 
The velocity of halo stars can be described by a 
simple Gaussian model of the following form 
\be
p({\bf v}|\theta_v,\ell,b,s)=\mathcal{N}(v_r|0,\sigma_{vr})
\mathcal{N}(v_{r}|0,\sigma_{v\theta})
\mathcal{N}(v_{\phi}|v_{\rm rot},\sigma_{v\phi})
\ee
for which $\theta_v$ is the set of parameters that govern the 
velocity dispersion profiles $\sigma_{vr},\sigma_{v\theta}$
and $\sigma_{v\phi}$.   
The coordinates $(r, \theta, \phi)$ are in the Galactocentric
reference frame. The observed heliocentric coordinates 
can be converted to Galactocentric coordinates using 
prior estimates of the location and the motion of the sun.
For the stellar halo stars, tangential velocities  
cannot be accurately determined.  
The distance also has some uncertainty, $\sigma_s$. 
Hence we marginalize over unknown tangential velocities and 
true distance $s'$, to obtain
\be
p(v_{\rm los}|\theta_v,\ell,b,s,\sigma_s) = \int \int \int
p(v_\ell,v_b,v_{\rm los}|\theta_v,\ell,b,s') p(s'|s,\sigma_s)\: d v_\ell d v_b d
s' 
\label{equ:vlos_sh}
\ee 
The parameters $\theta_v$ can now be estimated using the data $D$ 
of multiple stars by  
\be
p(\theta_v|D) \propto \left( \prod_i p(v_{\rm los}|\theta_v,\ell_i,b_i,s_i,\sigma_{s,i})\right) p(\theta_v)
\ee
The  marginalization in \equ{vlos_sh} can be handled in
various ways. One can make use of deterministic numerical 
integration techniques (Gaussian quadrature) or one can 
achieve marginalization  
via Monte Carlo schemes making use of importance sampling. 
For Monte Carlo based integration one can make use of the
MCMH algorithm discussed in \sec{MCMH}. 
Alternatively, one 
can treat $v_\ell,v_b$ and $s$ as unknowns by setting them 
up as a BHM and estimate
them alongside $\theta$ by making use of the 
Metropolis-within-Gibbs scheme discussed in \sec{gsem}.  
The radial velocity dispersion profile of halo stars 
computed using blue horizontal branch and red giant stars 
in the SEGUE survey is
shown in \fig{mvir_c_kafle} \citep{2014ApJ...794...59K}.

We now proceed to estimating the potential $\Phi$. 
Given  $\Phi$, density of halo stars $\rho$ and anisotropy
$\beta=1-(\sigma_{v \theta}^2+\sigma_{v \phi}^2)/(2
\sigma_{v r}^2)$ as function of distance $r$ 
from the Galactic center, one can
solve for $\sigma_{vr}(r)$. 
Let $\theta$
be the set of parameters used to define the above profiles. 
So for given $\theta$, the model makes a prediction for 
radial velocity dispersion $\sigma_{vr}(r_i;\theta)$ at a
location $r_i$. This can be compared with the 
$\sigma_{vr}(r_i)$ estimated from the observed data. 
The probability of model parameters $\theta$ is then given
by 
\be
p(\theta|D) \propto p(\theta) \prod_{i=1}^{M} \mathcal{N}(\sigma_{vr}(r_i)|\sigma_{vr}(r_i;\theta),\gamma_i).
\ee
The posterior distribution for the virial mass and the concentration
parameter of the Milky Way halo using BHB and giant stars is shown in 
\fig{mvir_c_kafle} \citep{2014ApJ...794...59K}. 

We now discuss ways to incorporate prior 
information into the analysis. For example, 
the angular velocity of the Sun with respect to the
Galactic Center $\omega$ is well constrained to 
be within $30.24\pm0.12\: {\rm km\ s}^{-1} {\rm kpc}^{-1}$ \citep{2004ApJ...616..872R}. The vertical force at $1.1$ kpc above the Sun, in terms of surface mass
density, is given by $\Sigma_{1.1,\odot}=72\pm6$ \citep{1991ApJ...367L...9K}.
Let us denote such constraints by
$p(g_j(\theta)|\theta)$. 
Additional data sets $D_k$, constraining a certain subset of parameters 
can also exist. For example, the tangent point velocities 
or terminal velocities as a function of Galactic longitude 
$v_{\rm term}(\ell)$ help to constrain the shape of the circular velocity curve 
$v_{\rm circ}(R)=\sqrt{|Rd\Phi/dR|}$. The additional priors and data
all enter as multiplicative factors in the posterior, which   
is given by 
\be
p(\theta|D_1,...,D_K) \propto p(\theta) \prod_{k=1}^{K}
p(D_k|\theta) \prod_{j=1}^{J}p(g_j(\theta)|\theta).
\label{equ:mass_post}
\ee

The halo stars carry little information about the   
mass distribution close to the center and in the disc of the
Milky Way. 
Galactic masers associated with high mass 
star forming regions are very good tracers of the Milky Way
disc which makes them excellent candidates for studying the 
potential of the Milky Way
\citep{2009ApJ...700..137R,2011MNRAS.414.2446M,2014ApJ...783..130R,2017MNRAS.465...76M}. Due to extremely accurate 
astrometric information using very long baseline
interferometry, one has very accurate parallax ($\varpi$) and proper
motion measurements. When combined with line of sight
velocities from Doppler shift of spectral lines, one ends up
with full 6D phase space information for these
sources. Maser sources, are young and have
very little random motion which means their orbits are 
highly circularized. The distribution of velocities
can be described by a simple three dimensional Gaussian function, i.e.
\be
p(v_R,v_{\phi},v_{z}|\theta) = \mathcal{N}(v_{\phi}|v_{\rm
  circ}(R;\theta)+v_{\phi,{\rm
  M}},\sigma_{v{\rm M}}) \mathcal{N}(v_{R}|v_{R,{\rm
  M}},\sigma_{v{\rm M}}) \mathcal{N}(v_{z}|v_{z,{\rm
  M}},\sigma_{v{\rm M}})
\label{equ:maser_vel}
\ee
Here, ${\bf v}_{\rm M}=(v_{R,{\rm M}},v_{\phi,{\rm M}},v_{z,{\rm M}})$
is any systematic streaming velocity associated with the masers
and $\sigma_{v{\rm M}}$ is the velocity dispersion about the mean motion.
Now, we have
\be
p(\mu_{\alpha},\mu_{\delta},v_{\rm los}|\theta) &= & \int
d \varpi' p(\varpi'|\varpi) \int \int
\int   d\mu_{\alpha}'
\ p(\mu_{\alpha}'|\mu_{\alpha}) d\mu_{\delta}' p(\mu_{\delta}'|\mu_{\delta})
\nonumber \\ 
& & dv_{\rm los}' p(v_{\rm los}'|v_{\rm los})p(\mu_{\alpha}',\mu_{\delta}',v_{\rm los}'|\theta,\varpi')
\ee
The last term is evaluated using \equ{maser_vel}, by 
converting from heliocentric coordinates $(\mu_{\alpha}',\mu_{\delta}',v_{\rm
  los}',\alpha,\delta, \varpi')$ to Galactocentric
coordinates $(v_R,v_{\phi},v_{z},R,\phi,z)$. 
Let $D_1$ denote the full data of $N$ stars then
\be
p(D_1|\theta)=\prod_{i=1}^{N} p(\mu_{\alpha,i},\mu_{\delta,i},v_{\rm los,i}|\theta)
\ee 
This when put in \equ{mass_post} gives 
the posterior distribution of model parameters.

\section{Concluding remarks}
The power of the Bayesian probability theory lies in the  
fact that it is mathematically simple, being based on just 
two elementary rules, and yet it is broadly applicable. 
However, Bayesian calculations can be computationally 
demanding, and this has acted as a major bottleneck in the
past. But with the increase of computational 
power, we have witnessed a sharp increase in the adoption 
of Bayesian techniques. 
More recently, free availability of black-box computer
packages 
to efficiently sample from Bayesian posterior distributions 
has further accelerated the adoption of Bayesian
techniques in astronomy. 

Robust algorithms are now available to sample 
multidimensional and complex pdfs. 
The MH algorithm
is still the main workhorse of MCMC methods.  
Good solutions now exist for the issue of application 
specific tuning of the proposal distribution in the 
MH algorithm, e.g., adaptive Metropolis schemes  
and the affine invariant samplers. 
The MH algorithm 
when combined with parallel tempering allows one 
to sample a wide variety of commonly occurring 
distributions. Situations, in which the posterior 
is not analytically tractable, can also 
now be solved using the Monte Carlo version 
of the MH algorithm.    

Bayesian methods also provide a framework for 
model comparison via the use of Bayesian evidence. 
However, efficient computing of evidence still remains 
a challenge. Various alternate criteria for comparing 
models exist and importantly these can make use of the 
computed MCMC chain.  
 
Bayesian hierarchical models further increase the usefulness 
of the Bayesian framework. They can solve missing data
problems, marginalization over variables, convolution 
with observational uncertainties and so on. This makes 
a wide class of complex problems suddenly solvable. 
We showed that the Metropolis-within-Gibbs scheme is ideally 
suited for sampling posteriors generated by 
Bayesian hierarchical models and also provide 
a software for doing this.

Multimodal distributions still pose a problem for most 
MCMC algorithms. Parallel tempering can overcome them but 
requires more computational time and a careful choice of 
ladder. If dimensionality of the space being explored
is very high and the distribution is complex, efficient
exploration is not easy. Techniques are being developed to 
solve such problems that make use of derivatives of the
posterior distribution, e.g., Hamiltonian Monte Carlo. 
However, more work is required in this area. 
Efficient exploration of multi-level hierarchical models 
will play an increasingly important role in future
studies.

Communication of Bayesian results is also an area where we 
anticipate improvements. Traditionally, the estimates are
reported by means of confidence intervals. However,
there is much more information in the MCMC chain, in 
particular, 
the correlation between different variables.  
Also, there is an increasing need to feed results of one 
MCMC simulation into another. Such requirements are best
addressed by reporting the full pdfs or the thinned samples 
from it. Other alternatives that are economical in terms 
storage space are to approximate the pdf by 
analytical functions or to employ Gaussian mixture models.  
We also need better tools to visualize the Bayesian-MCMC
output, specially for  high dimensional and complex hierarchical models. 
Such tools will allow us to understand as to 
why a model fails and how we should improve it. 

There are key topics which we have not addressed
here. Non-parametric Bayesian methods are 
becoming increasingly important, e.g., Gaussian processes
\citep{beaumont2002approximate} and Dirichlet 
process mixture models \citep{neal2000markov}. 
\citet{2014MNRAS.437.2230M} uses this method
to estimate the gravitational potential 
of the Milky Way. 
 
Astronomy is no longer a data-starved science. With projects like
the Large Synoptic Survey Telescope and the Square
Kilometre Array, the quality 
and quantity of data are going to increase dramatically in the
coming years. 
Better quality and larger quantity of data means that we can
expect our data to answer more difficult questions,
which in
turn means more complex models (e.g. multi-level hierarchies
and a higher dimensional parameter space). 
Given that 
MCMC is a computationally expensive scheme, there will 
be an increasing demand for such techniques that can 
make full use of the vast quantity of data on offer 
and deliver results in an affordable amount of time.

Equivalently,
MCMC schemes that make use of computing environments 
with multiple processor and graphic processor units 
would also be useful. 
An MCMC chain is serial by nature and it 
requires special care to parallelize an MCMC algorithm, e.g., use of an ensemble of chains 
\citep{2013PASP..125..306F} or parallelizing the 
posterior computation by splitting up the data. 
Relaxing the condition of reversibility can lead to MCMC 
algorithms with faster mixing properties 
\citep{chen1999lifting,diaconis2000analysis,girolami2011riemann}. 
Finally, the development of approximate methods, 
both application specific and general, that can 
reduce the computational cost without significantly
compromising the quality of results also hold great promise 
for analyzing large data sets. Approximate Bayesian 
computation is one such framework
\citep{beaumont2002approximate}; see
\citet{2016ApJ...817...49B} for its use in astronomy 
to study the chemical homogeneity of stars in open clusters.

\section*{DISCLOSURE STATEMENT}
The author is not aware of any affiliations, memberships,
funding, or financial holdings that
might be perceived as affecting the objectivity of this review.

\section*{ACKNOWLEDGMENTS}
I am indebted to my colleague Joss Bland-Hawthorn for
suggesting this article and for supervising its 
development over the past year. I am thankful 
to James Binney, Jo Bovy, Brendon Brewer, Prajwal Kafle 
and Prasenjit Saha for numerous suggestions and 
discussions from which the review has benefited 
significantly. I am also thankful to David Hogg 
for words of encouragement on the draft. 
I acknowledge 
funding from a University of Sydney Senior Fellowship 
made possible by the office of the Deputy Vice
Chancellor of Research, and partial funding from 
Bland-Hawthorn's Laureate Fellowship from the
Australian Research Council. 

%
%

\bibliographystyle{ar-style2}

\end{document}